\documentclass{article}
\input{psfig}
\usepackage{amsmath}
\usepackage{amssymb}
\usepackage{amsfonts}

\title{\bf ORIGIN AND DYNAMICAL EVOLUTION OF COMETS AND THEIR RESERVOIRS}

\author{Alessandro Morbidelli\\
\small CNRS, Observatoire de la C\^pte d'Azur, Nice, France}
\begin{document}
\maketitle

\begin{abstract}
  This text was originally written to accompany a series of lectures that I
  gave at the `35th Saas-Fee advanced course' in Switzerland and at the
  Institute for Astronomy of the University of Hawaii. It reviews my current
  understanding of the dynamics of comets and of the origin and primordial
  sculpting of their reservoirs. It starts discussing the structure of the
  Kuiper belt and the current dynamics of Kuiper belt objects, including
  scattered disk objects. Then it discusses the dynamical evolution of Jupiter
  family comets from the trans-Neptunian region, and of long period comets
  from the Oort cloud. The formation of the Oort cloud is then reviewed, as
  well as the primordial sculpting of the Kuiper belt. Finally, these issues
  are revisited in the light of a new model of giant planets evolution
  that has been developed to explain the origin of the late heavy
  bombardment of the terrestrial planets.
\end{abstract}

\section{Introduction}
Comets are often considered to be the gateway for understanding Solar System
formation. In fact, they are probably the most primitive objects of the Solar
System because they formed in distant regions where the relatively cold
temperature preserved the pristine chemical conditions.  For this reason they
have been the target of very sophisticated and expensive space missions like
{\it Giotto}, {\it Stardust} and {\it Rosetta} for in-situ analysis or sample
return.  To best exploitthe information collected by ground based and
space based observations, however, it is necessary to know where comets come
from, where they formed, and how they evolved in the distant past. For
instance, did they form at 5, 30 or at 100 AU? Are they chunks of larger
objects that presumably underwent significant thermal and collisional
alteration or are they pristine planetesimals that could never grow larger?

In addition, the orbital structure of the comet reservoirs records information
of the dynamical processes that occurred when the Solar System was taking
shape.  For example, it carries evidence of the migration of the giant
planets, and/or of close encounters of our Sun with other stars. Modeling these
dynamical processes and comparing their outcomes with the observed
structures, gives us a unique opportunity to reconstruct the history of the
formation of the planets and of their primordial evolution.

The purpose of this chapter is to review our current understanding of comets
from the dynamical point of view and underline the open issues which still
need more investigation.  The first part is devoted to the current Solar
System. In Section\ref{KB} I describe the orbital and dynamical properties of
the trans-Neptunian population: the Kuiper belt and the scattered disk.
Section~\ref{comets} is devoted to the evolution of comets from their parent
reservoirs --the trans-Neptunian population or the Oort cloud-- to the inner
Solar System.  As we know the current Solar System quite well --the orbits of
the planets, its galactic environment-- the results discussed in this part are
quite secure.  In contrast, the second part of the chapter focusses on more
controversial topics, as it is devoted to the origin of the Solar System,
namely how the comet reservoirs formed and acquired their current shapes. More
precisely, section~\ref{Oort} is devoted to the formation of the Oort cloud,
section~\ref{sculpting} to the primordial sculpting of the trans-Neptunian
population and section~\ref{LHB} discusses a recently proposed connection
between these events and the Late Heavy Bombardment of the terrestrial
planets. In the final section I will speculate on a scenario of solar system
primordial evolution that would put all these aspects together in a coherent
scheme.

\section{The trans-Neptunian population}
\label{KB}

Our observational knowledge of the trans-Neptunian population\footnote{There
  is no general consensus on nomenclature, yet. In this work I call
  `trans-Neptunian population' the collection of small bodies with semi-major
  axis (or equivalently orbital period) larger than that of Neptune, with the
  exception of the Oort cloud (semi-major axis larger than 10,000~AU).} is
very recent. The first object, Pluto, was discovered in 1930, but
unfortunately this discovery was not quickly followed by the detection of
other trans-Neptunian objects. Thus, Pluto was thought to be an exceptional
body --a planet-- rather than a member of a numerous small body population, of
which it is not even the largest in size. It was only in 1992, with the advent
of CCD cameras and a lot of perseverance, that another trans-Neptunian object
--1992~QB$_1$-- was found \cite{QB1}. Now, 13 years later, we know more than
1,000 trans-Neptunian objects. Of them, about 500 have been observed for at
least three years. A time-span of 3 years is required in order to compute
their orbital elements with some confidence. In fact, the trans-Neptunian
objects move very slowly, and most of their apparent motion is simply a
parallactic effect. Our knowledge of the orbital structure of the
trans-Neptunian population is therefore built on these $\sim 500$ objects.

Before moving to discuss the orbital structure of the trans-Neptunian
population, in the next subsection I briefly overview the basic facts of
orbital dynamics. The expert reader can move directly to
section~\ref{orb-struct}. 

\subsection{Brief tutorial of orbital dynamics}
\label{orb-dyn}

Neglecting mutual perturbations, all bodies in the Solar System move relative
to the Sun in an elliptical orbit, the Sun being at one of the two foci of the
ellipse. Therefore, it is convenient for astronomers to characterize the
relative motion of a body by quantities that describe the geometrical
properties of its orbital ellipse and its instantaneous position on the
ellipse.  These quantities are usually called {\it orbital elements}.

\begin{figure}[t!]
\centerline{\psfig{figure=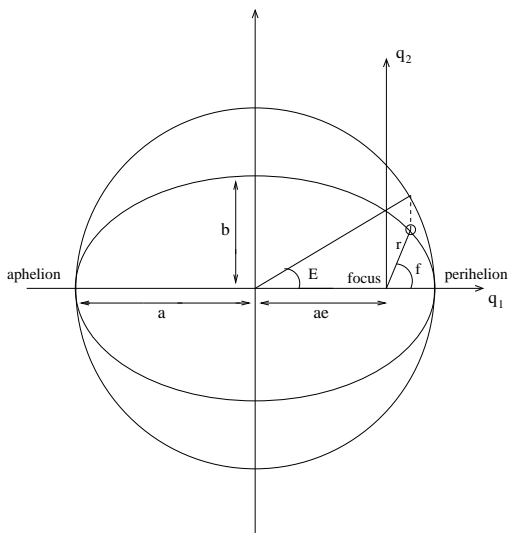,height=7.cm}}
\vspace*{0.cm}
\caption{Keplerian motion: definition of $a$, $e$ and $E$. 
%From Wertz (1980).
}
\vspace{0.3cm}
\label{ellisse_a}
\end{figure}

The shape of the ellipse can be completely determined by two orbital elements:
the semi-major axis $a$ and the eccentricity $e$
(Fig. \ref{ellisse_a}).
The name {\it eccentricity} comes from $e$ being
the ratio between the distance of the focus from the 
center of the ellipse and the semi-major axis of the ellipse. 
The eccentricity is therefore an indicator of
how much the orbit differs from a circular one: $e=0$ means that the 
orbit is circular, while $e=1$ means that the orbit is a segment of 
length $2a$, the Sun being at one of the extremes. Among all
``elliptical'' trajectories, the latter is 
the only collisional one,  if the
physical radii of the bodies are neglected. A semi-major axis of $a=\infty$ 
and $e=1$
denote parabolic motion, while the convention $a<0$ and $e>1$ is adopted for 
hyperbolic motion.
I will not deal with these kinds of unbounded motion in this chapter, 
hence I will concentrate, hereafter, on the elliptic case.
On an elliptic orbit, the closest point to the Sun is called the 
{\it perihelion}, and its heliocentric distance $q$ is equal to $a(1-e)$;
the farthest point is called the 
{\it aphelion} and its distance $Q$ is equal to $a(1+e)$.

To denote the position of a body on its orbit, it is convenient to
use an orthogonal reference frame $q_1,q_2$ with origin at the focus of the
ellipse occupied by the Sun and $q_1$~axis oriented towards the
perihelion of the orbit. Alternatively, polar coordinates $r,f$ can be used.
The angle $f$ is usually called the {\it true anomaly} of the body. From Fig.
\ref{ellisse_a}, with elementary geometrical relationships one has 
\begin{equation}
q_1=a(\cos E-e)\ ,\quad q_2=a\sqrt{1-e^2}\sin E
\label{cart-coord}
\end{equation}
and
\begin{equation}
r=a(1-e\cos E)\ ,\quad \cos f={{\cos E-e}\over{1-e\cos E}}
\label{pol-coord}
\end{equation}
where $E$, as Fig.~\ref{ellisse_a} shows, is the angle subtended at the center
of the ellipse by the projection --parallel to the $q_2$ axis-- of the
position of the body on the circle which is tangent to the
ellipse at perihelion and aphelion.  This angle is called the {\it eccentric
  anomaly}. The quantities $a$, $e$ and $E$ are enough to characterize the  
position of a body in its orbit.

From Newton equations, it is possible to derive \cite{damby} 
the evolution law of $E$ with respect to time,
usually called the {\it Kepler equation}:
\begin{equation}
E-e\sin E=n(t-t_0)\label{Kepler}
\end{equation}
where 
\begin{equation}
n=\sqrt{{\cal G}(m_0+m_1)}a^{-3/2}\label{mm}
\end{equation}
is the orbital frequency, or {\it mean motion}, of the body,
$m_0$ and $m_1$ are the masses of the Sun and of the body respectively and
${\cal G}$ is the gravitational constant;
$t$ is the time and $t_0$ is the time of passage at perihelion.

Astronomers like to introduce a new angle 
\begin{equation}
M=n(t-t_0)\label{m.anomaly}
\end{equation}
called the {\it mean anomaly},
 as an orbital element that changes linearly 
with time. $M$ also denotes the position of the body in its orbit, 
through equations (\ref{Kepler}) and (\ref{pol-coord}).

\begin{figure}[t!]
\centerline{\psfig{figure=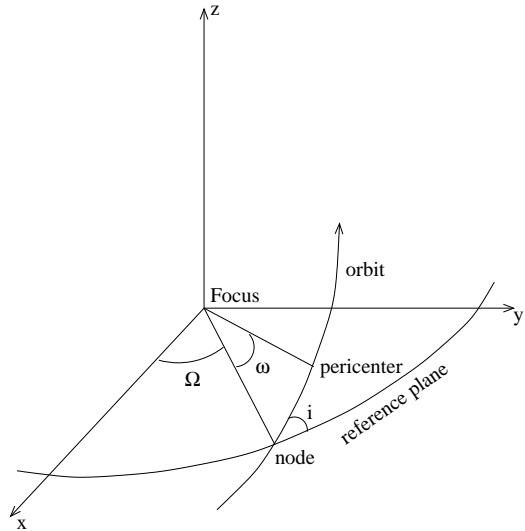,height=7.cm}}
\vspace*{0.cm}
\caption{Keplerian motion: definition of $i$, $\Omega$ and  $\omega$. 
%From Wertz (1980).
}
\vspace{0.3cm}
\label{ellisse_b}
\end{figure}

To characterize the orientation of the ellipse in space, with respect to an
arbitrary orthogonal reference frame $(x,y,z)$ centered on the Sun, one has to
introduce three additional angles (see Fig. \ref{ellisse_b}).  The first one
is the inclination $i$ of the orbital plane (the plane which contains the
ellipse) with respect to the $(x,y)$ reference plane.  If the orbit has a
nonzero inclination, it intersects the $(x,y)$ plane in two points, called the
{\it nodes} of the orbit. Astronomers distinguish between an {\it ascending}
node, where the body passes from negative to positive $z$, and a {\it
  descending} node, where the body plunges towards negative $z$.  The
orientation of the orbital plane in space is then completely determined when
one gives the angular position of the ascending node from the $x$~axis.  This
angle is traditionally called the {\it longitude of ascending node}, and is
usually denoted by $\Omega$.  The last angle that needs to be introduced is
the one characterizing the orientation of the ellipse in its plane. The {\it
  argument of perihelion} $\omega$ is defined as the angular position of the
perihelion, measured in the orbital plane relative to the line connecting the
central body to the ascending node.

In the definition of the orbital elements above, 
note that when the inclination is zero, $\omega$ and $M$ are 
not defined, because the position of the ascending node is  
not determined. Moreover, $M$ is not defined also when the 
eccentricity is zero, because the position of the 
perihelion is not determined.
Therefore, it is convenient to introduce  
the {\it longitude of perihelion}
 $\varpi=\omega+\Omega$ and the 
{\it mean longitude}
 $\lambda=M+\omega+\Omega$. The first angle is well 
defined when $i=0$, while the second one is well 
defined when $i=0$ and/or $e=0$.

In absence of external perturbations, the orbital motion is perfectly
elliptic. Thus, the orbital elements $a, e, i, \varpi, \Omega$ are fixed, and
$\lambda$ moves linearly with time, with frequency (\ref{mm}).  When a small
perturbation is introduced (for instance the presence of an additional
planet), two effects are produced. First, the motion of $\lambda$ is no longer
perfectly linear. Correspondingly, the other orbital elements have short
periodic oscillations with frequencies of order of the orbital frequencies.
Second, the angles $\varpi$ and $\Omega$ start to rotate slowly. This motion
is called {\it precession}. Typical precession periods in the Solar System are
of order of 10,000--100,000 years. Correspondingly, $e$ and $i$ have long
periodic oscillations, with periods of order of the precession periods.

The regularity of these short periodic and long periodic oscillations is
broken when one of the following two situations occur: (i) the perturbation
becomes large, for instance when there are close approaches between the body
and the perturbing planet, or when the mass of the perturber is comparable to
that of the Sun (as in the case of encounters of the Solar System with other
stars) or (ii) the perturbation becomes resonant.  In either of these cases
the orbital elements $a, e, i$ can have large non-periodic, irregular
variations.

A resonance occurs when the frequencies of $\lambda, \varpi$ or $\Omega$ of
the body, or an integer combination of them, are in an integer ratio with one
of the time frequencies of the perturbation. If the perturber is a planet, the
perturbation is modulated by the planet orbital frequency and precession
frequencies. We speak of {\it mean-motion resonance} when $k{\rm
  d}\lambda/{\rm d}t=k'{\rm d}\lambda'/{\rm d}t$, with $k$ and $k'$ integer
numbers and $\lambda'$ denoting the mean longitude with the planet.  We speak
of {\it linear secular resonance} when ${\rm d}\varpi/{\rm d}t={\rm
  d}\varpi'/{\rm d}t$ or ${\rm d}\Omega/{\rm d}t={\rm d}\Omega'/{\rm d}t$,
prime variables referring again to the planet. Other types of resonances exist
in more complicated systems (non-linear secular resonances, three--body
resonances, Kozai resonance etc.). I will return to discuss resonant motion
more specifically in subsection~\ref{KBdyn}, when reviewing the dynamical
properties of some trans-Neptunian sub-populations.

\subsection{The structure of the trans-Neptunian population}
\label{orb-struct}  

The trans-Neptunian population is ``traditionally'' subdivided in two
sub-\-po\-pu\-la\-tions: the {\it scattered disk} and the {\it Kuiper belt}.  The
definition of these sub-\-po\-pu\-la\-tions is not unique, with the Minor Planet
center and various authors often using slightly different criteria.  Here I
propose a partition based on the dynamics of the objects and their
relevance for the reconstruction of the primordial evolution of the outer
Solar System, keeping in mind that all bodies in
the Solar System must have been formed on orbits typical of an accretion disk
(e.g. with very small eccentricities and inclinations).

I call {\it scattered disk} the region of the orbital space that can be
visited by bodies that have encountered Neptune within a Hill's
radius\footnote{The Hill's radius is given by the formula
  $R_H=a_p(m_p/3)^{1/3}$, where $m_p$ is the mass of the planet relative to
  the mass of the Sun and $a_p$ is the planet's semi-major axis. It
  corresponds approximately to the distance from the planet of the Lagrange
  equilibrium points $L_1$ and $L_2$.}, at least once during the age of the
Solar System, assuming no substantial modification of the planetary orbits.
The bodies that belong to the scattered disk in this classification do not
provide us any relevant clue to uncover the primordial architecture of the
Solar System.  In fact their current eccentric orbits might have been achieved
starting from quasi-circular ones in Neptune's zone by pure dynamical
evolution, in the framework of the current architecture of the planetary
system.

I call {\it Kuiper belt} the trans-Neptunian region that cannot be visited by
bodies encountering Neptune. Therefore, the non-negligible eccentricities
and/or inclinations of the Kuiper belt bodies cannot be explained by the
scattering action of the planet on its current orbit, but they reveal that
some excitation mechanism, which is no longer at work, occurred in the past
(see section~\ref{sculpting}).

\begin{figure}[t!]
\centerline{\psfig{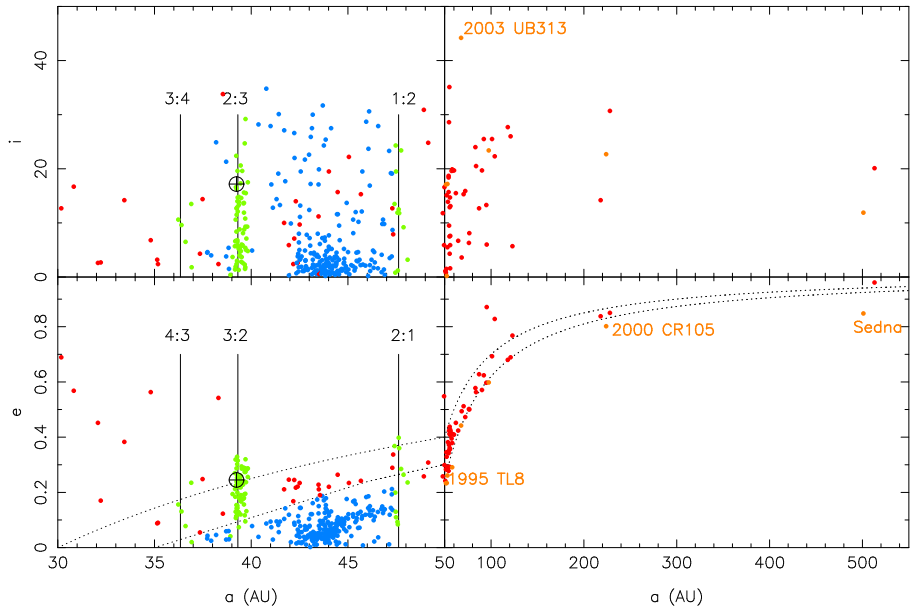}}
\caption{The orbital distribution of multi-opposition trans-Neptunian bodies,
  as of Aug. 26, 2005. Scattered-disk bodies are represented in red, extended
  scattered-disk bodies in orange, classical Kuiper belt bodies in blue and
  resonant bodies in green.  We qualify that, in absence of long term
  numerical integrations of the evolution of all the objects and because of
  the uncertainties in the orbital elements, some bodies could have been
  mis-classified.  Thus, the figure should be considered as an indicative
  representation of the various subgroups that compose the trans-Neptunian
  population.  The dotted curves in the bottom left panel 
  denote $q=30$~AU and $q=35$~AU; those in the bottom right panel $q=30$~AU
  and $q=38$~AU (right panel). The vertical solid lines mark the locations of
  the 3:4, 2:3 and 1:2 mean-motion resonances with Neptune.  The orbit of
  Pluto is represented by a crossed circle.  }
\label{aeai} 
\end{figure}

To categorize the observed trans-Neptunian bodies into scattered disk and
Kuiper belt, one can refer to previous works on the dynamics of
trans-Neptunian bodies in the framework of the current architecture of the
planetary system.  For the $a<50$~AU region, one can use the results by
\cite{DLB} and \cite{kuchner}, who numerically mapped the regions of the
$(a,e,i)$ space with $32<a<50$~AU that can lead to a Neptune encountering
orbit within 4~Gy.  Because dynamics are reversible, these are also the
regions that can be visited by a body after having encountered the planet.
Therefore, according to the definition above, they constitute the scattered
disk.  For the $a>50$~AU region, one can use the results in \cite{LD97} and
\cite{DL97}, where the the evolutions of the particles that encountered
Neptune in \cite{DLB} have been followed for another 4~Gy time-span.  Despite
the fact that the initial conditions did not cover all possible
configurations, one can reasonably assume that these integrations cumulatively
show the regions of the orbital space that can be possibly visited by bodies
transported to $a>50$~AU by Neptune encounters. Again, according to my
definition, these regions constitute the scattered disk.

Fig.~\ref{aeai} shows the $(a,e,i)$ distribution of the trans-Neptunian bodies
which have been observed during at least three oppositions. The bodies that
belong to the scattered disk according to my criterion are represented as red
dots. The Kuiper belt population is in turn subdivided in two sub-populations:
the {\it resonant population} (green dots) and the {\it classical belt} (blue
dots). The former is made of objects located the major mean-motion resonances
with Neptune (essentially the 3:4, 2:3 and 1:2 resonances, but also the 2:5
--see \cite{Chiang}, while the classical belt objects are not in any
noticeable resonant configuration.  Mean-motion resonances offer a protection
mechanism against close encounters with the resonant planet (see section
\ref{KBdyn}). For this reason, the resonant population can have perihelion
distances much smaller than those of the classical belt objects. Stable
resonant objects can have even Neptune-crossing orbits ($q<30$~AU) as in the
case of Pluto (see sect.~\ref{KBdyn}).  The bodies in the 2:3 resonance are
often called {\it Plutinos}, for the similarity of their orbit with that of
Pluto.  According to \cite{Trujillo}, the scattered disk and the Kuiper belt
constitute roughly equal populations, while the resonant objects, altogether,
make about 10\% of the classical objects.

Notice in Fig.~\ref{aeai} also the existence of bodies with $a>50$~AU, on
highly eccentric orbits, which do not belong to the scattered disk according
to my definition (magenta dots). Among them are 2000CR$_{105}$ ($a=230$~AU,
perihelion distance $q=44.17$~AU and inclination $i=22.7^\circ$), Sedna
($a=495$~AU, $q=76$~AU) and the currently size-record holder 2003~UB$_{313}$
($a=67.7$~AU, $q=37.7$~AU but $i=44.2^\circ$), although for some objects the
classification is uncertain for the reasons explained in the figure caption.
Following \cite{CR105}, I call these objects {\it extended scattered-disk}
objects for three reasons. {\it (i)} They are very close to the scattered-disk
boundary. {\it (ii)} Bodies of the sizes of the three objects quoted above
(300--2000~km) presumably formed much closer to the Sun, where the accretion
timescale was sufficiently short \cite{stern96}. This implies that they have
been transported in semi-major axis space (e.g. scattered), to reach their
current locations.  {\it (iii)} the lack of objects with $q>41$~AU and
$50<a<200$~AU should not be due to observational biases, given that many
classical belt objects with $q>41$~AU and $a<50$~AU have been discovered (see
Fig.~\ref{limits}). This suggests that the extended scattered-disk objects are
{\it not} the highest eccentricity members of an excited belt beyond 50~AU.
These three considerations indicate that in the past the true scattered disk
extended well beyond its present boundary in perihelion distance.  Why this
was so, is particularly puzzling. Some ideas will be presented in
sect.~\ref{sculpting}.

Given that the observational biases become more severe
with increasing perihelion distance and semi-major axis, the currently known
extended scattered-disk objects may be like the tip of an iceberg, e.g.  the
emerging representatives of a conspicuous population, possibly outnumbering
the scattered-disk population \cite{CR105}.

\paragraph{The excitation of the Kuiper belt.\,}
An important clue to the history of the early outer Solar System is
the dynamical excitation of the Kuiper belt.
While eccentricities and inclinations of
resonant and scattered objects are expected to have been 
affected by interactions
with Neptune, those of the classical objects should have suffered no such
excitation. Nonetheless, the confirmed classical belt objects have an
inclination range up to at least 32 degrees and an eccentricity range
up to 0.2, significantly higher than
expected from a primordial disk, even accounting for 
mutual gravitational stirring.

The observed distributions of eccentricities and inclinations in the Kuiper
belt are highly biased. High eccentricity objects have closer approaches to
the Sun and thus they become brighter and are more easily detected.
Consequently, the detection bias roughly follows curves of constant $q$.  At
first sight, this bias might explain why, in the classical belt beyond
$a=44$~AU, the eccentricity tends to increase with semi-major axis.  However,
the resulting $(a,e)$ distribution is significantly steeper than a curve
$q=$constant. Thus, the apparent relative under-density of objects at low
eccentricity in the region $44<a<48$~AU is likely a real feature of the
Kuiper belt distribution.

\begin{figure}[t!]
\centerline{\psfig{figure=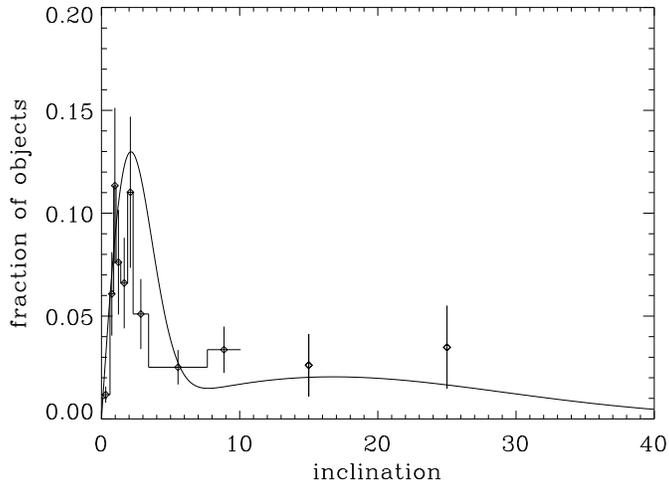,height=7.cm
,bbllx=7cm,bblly=7cm,bburx=15cm,bbury=20cm
%,bbllx=10cm,bblly=-20cm,bburx=20cm,bbury=0cm
}}
\caption{The inclination distribution (in deg.) of the classical
Kuiper belt, from \cite{MB}. The points with
error bars show the model-independent estimate constructed from a limited
subset of confirmed classical belt bodies, while the smooth line shows the 
best fit two-population model $f(i) di=\sin(i) [96.4\exp(-i^2/6.48)+3.6 
\exp(-i^2/288)] di$\cite{Brown}. In this model $\sim$60\% of the objects
have $i>4^\circ$.
} 
\label{incdist} 
\end{figure}

High inclination objects spend little time at low latitudes\footnote{Latitude
  (angular height over a reference curve in the sky) and inclination should be
  defined with respect to the local Laplace plane (the plane normal to the
  orbital precession pole), which is a better
  representation for the plane of the Kuiper belt than is the ecliptic
  \cite{elliot}.}  at which most surveys take place, while low inclination
objects spend zero time at the high latitudes where some searches have
occurred.  Using this fact, \cite{Brown} computed a de-biased inclination
distribution for classical belt objects (Figure \ref{incdist}).

A clear feature of this de-biased distribution is its bi-modality, with 
a sharp drop around 4 degrees and an extended, almost flat 
distribution in the $4$--$30$ degrees range,  
demanded by the presence of objects with large inclinations. This bi-modality 
can be modeled with two Gaussian functions and suggests the 
presence of two distinct classical Kuiper belt populations, called 
hot ($i>4$) and cold ($i<4$) after \cite{Brown}. 

\begin{figure}[t!]
\centerline{\psfig{figure=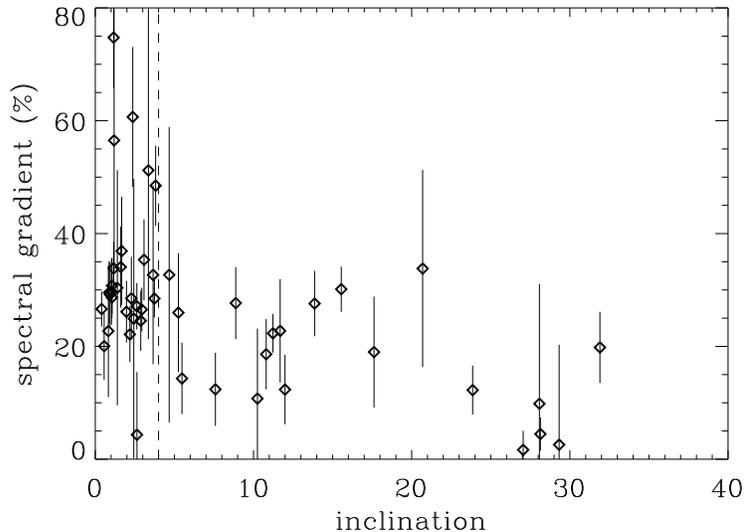,height=8.cm
%,bbllx=10cm,bblly=-20cm,bburx=20cm,bbury=0cm
}}
\caption{Color gradient versus inclination in the classical Kuiper belt (from
  \cite{MB}, using the database in \cite{Hainaut}). Color 
gradient is the slope of the spectrum, in \% per 100nm, 
with 0\% being neutral and large
numbers being red. The hot and cold classical objects have 
significantly different
distributions of color.
} 
\label{colors} 
\end{figure}

\paragraph{Physical evidence for two populations in the classical belt.\,}

The co-existence of a hot and a cold population in the classical belt could be
caused in one of two general manners. Either a subset of an initially
dynamically cold population was excited, leading to the creation of the hot
classical population, or the populations are truly distinct and formed
separately. One manner in which one can attempt to determine which of these
scenarios is more likely is to examine the physical properties of the two
classical populations. If the objects in the hot and cold populations are
physically different it is less likely that they were initially part of the
same population.

The first suggestion of a physical difference between the hot and the cold
classical objects came from \cite{LS01} who noted that the intrinsically
brightest classical belt objects (those with lowest absolute magnitudes) are
preferentially found on high inclination orbits.  This conclusion has been
recently verified in a bias-independent manner in \cite{TB03}, with a survey
for bright objects which covered $\sim$70\% of the ecliptic and found many hot
classical objects but few cold classical objects.

The second possible physical difference between hot and cold classical Kuiper
belt objects is their colors, which relates in an unknown manner to surface
composition and physical properties. Several possible correlations between
orbital parameters and color were suggested by \cite{TeglRom} and further
investigated by \cite{Doressoundiram}. The issue was clarified by \cite{TB02}
who quantitatively showed that for the classical belt, the inclination is
correlated with color.  In essence, the low inclination classical objects tend
to be redder than higher inclination objects (see Fig.~\ref{colors}).  This
correlation has then been confirmed by several other authors
\cite{Doressoundiram05} \cite{elliot}.  Whether or not there is also a
correlation between colors and perihelion distances is still a matter of
debate \cite{Doressoundiram05}.

More interestingly, we see that the colors naturally divide into distinct low
inclination and high inclination populations at precisely the location of the
divide between the hot and cold classical objects. These populations differ at
a 99.9\% confidence level. Interestingly, the cold classical population also
differs in color from the Plutinos and the scattered objects at the 99.8\% and
99.9\% confidence level, respectively, while the hot classical population
appears identical in color to these other populations \cite{TB02}.  The
possibility remains, however, that the colors of the objects, rather than
being markers of different populations, are actually {\it caused} by the
different inclinations.  For example \cite{Stern02} has suggested that the
higher average impact velocities of the high inclination objects will cause
large scale resurfacing by fresh water ice which could be blue to neutral in
color. However, a careful analysis shows that there is clearly no correlation
between average impact velocity and color (\cite{ThebDor03}).

In summary, the significant color and size differences
between the hot and cold classical objects imply that
these two populations are physically different, in addition
to being dynamically distinct.

\paragraph{The radial extent of the Kuiper belt.\,}

Another important property of interest for understanding the primordial
evolution of the Kuiper belt is its radial extent. While initial expectations 
were that the mass of the Kuiper belt should smoothly decrease with
heliocentric distance --or perhaps even increase in number density by a factor
of $\sim$100 \cite{stern96}, back to the level given by the extrapolation of
the minimum mass solar nebula \cite{hayashi} beyond the region of Neptune's
influence-- the lack of detection of objects beyond about 50 AU soon began to
suggest a drop off in number density (\cite{Jewitt98}, \cite{ChiangBrown},
\cite{Trujillo}).  It was often argued that this lack of detections was the
consequence of a simple observational bias caused by the extreme faintness of
objects at greater distances from the sun, but \cite{Allen01}, \cite{Allen02}
showed convincingly that for a fixed absolute magnitude distribution, the
number of objects with semi-major axis less than 50 AU was larger than the
number greater than 50 AU, and thus some density decrease is present.

\begin{figure}[t!]
\centerline{\psfig{figure=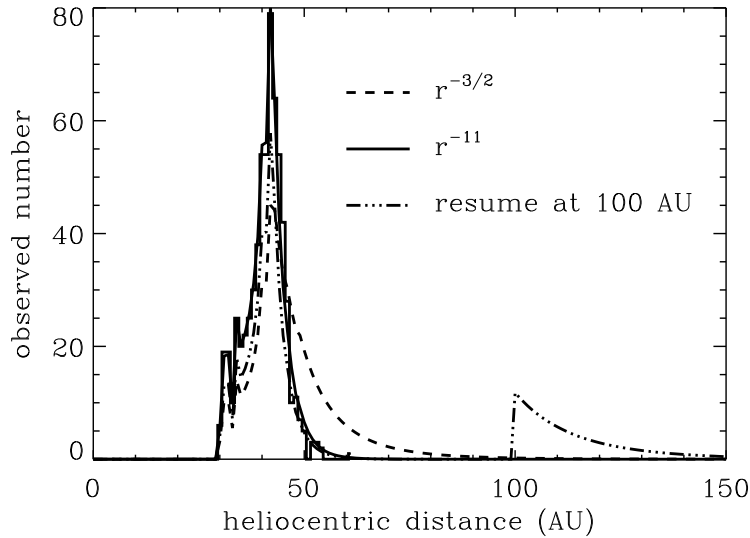,height=8.cm
%,bbllx=10cm,bblly=-20cm,bburx=20cm,bbury=0cm
}}
\caption{The observed radial distribution of Kuiper belt objects 
(solid histogram) compared to
observed radial distributions expected for models where the surface
density
of Kuiper belt objects decreases by $r^{-3/2}$ beyond 42 AU (dashed
curve), 
where the
surface density decreases by $r^{-11}$ beyond 42 AU (solid curve), and
where the
surface density at 100 AU increases by a factor of 100 to the value 
expected from an extrapolation of the minimum mass solar nebula
(dashed-dotted
curve). From \cite{MB}.
} 
\label{limits} 
\end{figure}

The characterization of the density drop beyond 50 AU was hampered by the
small numbers of objects and thus weak statistics in individual surveys. A
method to use all detected objects to estimate a radial distribution of the
Kuiper belt and to test hypothetical radial distributions against the known
observations was developed in \cite{TB01}. The analysis reported in that work
is reproduced in Fig.~\ref{limits}. The drop off beyond 42~AU of the
heliocentric distance distribution of Kuiper belt objects at discovery is
clearly inconsistent with a smooth decline of the surface density distribution
proportional to $r^{-3/2}$. Instead, it can be fitted with a surface density
distribution with a much sharper decay, as $r^{-11\pm 4}$ (error bars are
$3\sigma$), i.e.  assuming the existence of an effective edge in the radial
Kuiper belt distribution.  This steep radial decay should presumably hold up
to $\sim 60$~AU, beyond which a much flatter distribution due to the scattered-disk objects should be found.

It has been conjectured \cite{stern96} that, beyond some range of Neptune's
influence, the number density of Kuiper belt objects could increase back up to
the level expected for the minimum mass solar nebula \cite{hayashi}.  Such an
increase can be ruled out at the 3$\sigma$ level within 115 AU from the Sun.
Beyond this distance the biases due to the slow motions of the objects also
become important, so few conclusion can be drawn from the current data about
objects beyond this threshold.  If the model is slightly modified to make the
maximum object mass proportional to the surface density at a particular
distance, a 100 times resumption of the Kuiper belt can still be ruled out
inside 94~AU.

Although the drop off in the heliocentric distance distributon starts at
42~AU, a visual inspection of Fig.~\ref{aeai} shows that the edge of the
Kuiper belt in semi-major axis space is precisely at the location of the 1:2
mean-motion resonance with Neptune. This is a very important feature, which
points to a role of Neptune in the final positioning of the edge. I will come
back to this in sect.~\ref{sculpting}.

\paragraph{The missing mass of the Kuiper belt.\,}

The absolute magnitude\footnote{The absolute magnitude $H$ is a measure of the
  intrinsic brightness of an object. It corresponds to the visual magnitude
  that an object would have in the paradoxical situation of being
  simultaneously at 1 AU from the Sun and the Earth, at opposition!}
distribution of the Kuiper belt objects can be determined from the so-called
{\it cumulative luminosity function}, which is given by the number of
detections that surveys reported as a function of their limiting magnitude,
weighted by the inverse area of sky that the surveys covered. If one assumes
that the albedo distribution of Kuiper belt objects is size independent, the
slope of the absolute magnitude distribution can be readily converted into the
slope of the cumulative size distribution.

\begin{figure}[t!]
\centerline{\psfig{figure=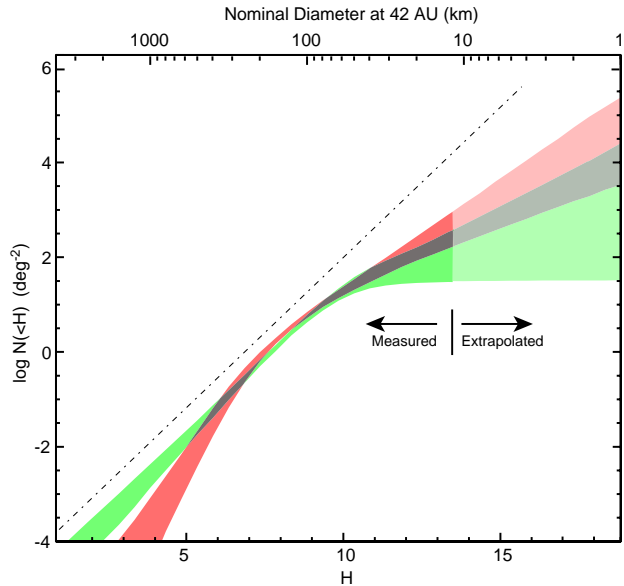,height=8.cm
%,bbllx=10cm,bblly=-20cm,bburx=20cm,bbury=0cm
}}
\caption{The $H$- or size-distribution in the Kuiper belt (adapted from
  \cite{bernstein} with the courtesy of Bernstein. The red and green bands
  show the uncertainties for the cold and the hot population, respectively
  (although the definition for hot and cold used in that work do not exactly
  match those adopted in this paper).  Absolute magnitudes have been computed
  assuming that all detections occurred at 42~AU (the maximum of the radial
  surface density distribution of the Kuiper belt), and the conversion to
  diameters uses the assumption that the mean albedo is 4\%.  }
\label{bernst} 
\end{figure}

The size distribution turns out to be very steep, with exponent of the
cumulative power law exponent between $-3.5$ and $-3$ for bodies larger in
diameter than $\sim$200~km \cite{pencilbeam}. Actually, the
size distribution is slightly shallower for the hot population than for the
cold population, as shown in a recent analysis \cite{bernstein} (see
Fig.~\ref{bernst}). This is not surprising, given that --as we have seen
above--  the hot and the cold
populations contain roughly the same total number of objects, but the former
hosts the largest members of the Kuiper belt.

The HST survey in \cite{bernstein} also reported the detection of a change in
the size distribution for objects fainter than about 100~km in diameter.  The
slopes of the size distribution below this limit, however, remain very
uncertain because of small number statistics. Some researchers still dispute
the validity of the detection of any turnover of the size distribution
(Kaavelars, private communication). Given these uncertainties, as well as
uncertainties on the mean albedo of the Kuiper belt objects (required to
convert a given absolute magnitude into a size) and their bulk density, the
total mass of the Kuiper belt is uncertain up to an order of magnitude at
least, its estimates ranging from 0.01~$M_\oplus$ (\cite{bernstein}) to
0.1~$M_\oplus$ (\cite{pencilbeam}).

Whatever the real value in this range (or slightly beyond), it appears
nevertheless secure that the total mass of the Kuiper belt is now very small,
in particular compared to the mass of the disk of solids from which the Kuiper
belt objects had to form. There are two lines of argument to estimate the
primordial mass.

A first argument follows the reasoning which led Kuiper to conjecture 
the existence of a band of small planetesimals beyond Neptune (\cite{kuiper}).
The minimum mass solar nebula 
inferred from the total planetary mass
(plus lost volatiles; \cite{hayashi}) smoothly declines from the orbit of
Jupiter until the orbit of Neptune; 
why should it abruptly drop beyond the last planet? 
The extrapolation and the integration of this surface density distribution 
predicts that the original total mass of solids in the 30--50~AU range 
should have been $\sim 30 M_\oplus$.

The second argument for a massive primordial Kuiper belt 
was brought to attention by Stern (\cite{stern95}) 
who found that the objects currently in the Kuiper belt
could not have formed in the present environment:
collisions are sufficiently
infrequent that 100 km objects
cannot be built by pairwise accretion
of the current population over the age of the solar system. Moreover,
owing to the large eccentricities and inclinations of Kuiper belt objects 
--and consequently to their high encounter velocities-- 
collisions that do occur tend to be 
erosive rather than accretional, making bodies smaller 
rather than larger. Stern suggested that
the resolution of this dilemma is that the primordial Kuiper belt
was both more massive and dynamically colder, so that more collisions occurred,
and they were gentler and therefore generally accretional. 

Following this idea, detailed modeling of accretion in a massive primordial
Kuiper belt was performed \cite{stern96},
\cite{SternColwell97a},\cite{SternColwell97b} \cite{KLuu98}, \cite{KLuu99a},
\cite{KLuu99b}. While each model includes different aspects of the relevant
physics of accretion, fragmentation, and velocity evolution, the basic results
are in approximate agreement.  First, with $\sim$10 M$_\oplus$ or more of
solid material in an annulus from about 35 to 50 AU on very low eccentricity
orbits ($e \le 0.001$), all models naturally produce of order a few objects
the size of Pluto and approximately the right number of $\sim$ 100km objects,
on a timescale ranging from several $10^7$ to several $10^8$~y.  The models
suggest that the majority of mass in the disk was in bodies approximately 10km
and smaller.  The accretion stopped when the formation of Neptune or other
dynamical phenomena (see section~\ref{sculpting}) began to induce
eccentricities and inclinations in the population high enough to move the
collisional evolution from the accretional to the erosive regime.

A massive and dynamically cold primordial Kuiper belt is also required by the
models that attempt to explain the formation of the observed numerous binary
Kuiper belt objects (\cite{goldreich}, \cite{weidensh-sats}, \cite{funato},
\cite{astakhov}).

Therefore, the general formation picture of an initial massive Kuiper belt
appears secure, and understanding the ultimate fate of the 99\% of the initial
Kuiper belt mass that appears to be no longer in the Kuiper belt is a crucial
step in reconstructing the history of the outer Solar System.

\subsection{Dynamics in the Kuiper belt}
\label{KBdyn}

I now come to overview the dynamical properties in the Kuiper belt.  Without
pretension of being exhaustive, the goal is to understand which properties of
the Kuiper belt orbital structure can be explained from the evolution of the
objects in the framework of the current architecture of the Solar System and
which, conversely, require an explanation built on a scenario of primordial
sculpting (as in section~\ref{sculpting}).

\begin{figure}[t!]
\centerline{\psfig{figure=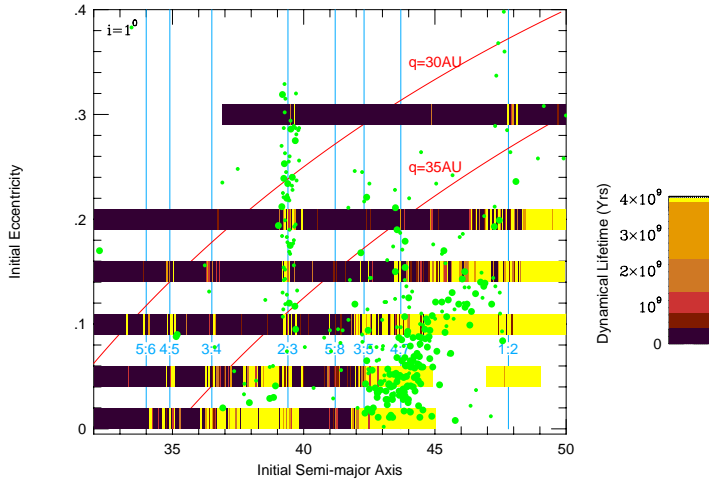,height=8.cm,angle=90
%,bbllx=10cm,bblly=-20cm,bburx=20cm,bbury=0cm
}}
\caption{ The dynamical lifetime for
  small particles in the Kuiper belt derived from 4 billion year integrations
  \cite{DLB}.  Each particle is represented by a narrow vertical strip of
  color, the center of which is located at the particle's initial eccentricity
  and semi-major axis (initial orbital inclination for all objects was 1
  degree).  The color of each strip represents the dynamical lifetime of the
  particle.  Strips colored yellow represent objects that survive for the
  length of the integration, $4\times 10^9$ years.  Dark regions are
  particularly unstable on short timescales.  For reference, the locations of
  the important Neptune mean-motion resonances are shown in blue and two
  curves of constant perihelion distance, $q$, are shown in red.  The $(a,e)$
  elements of the Kuiper belt objects with well determined orbits are also
  shown as green dots. Large dots are for $i<4^\circ$, small dots otherwise. }
\label{Hal} 
\end{figure}

Figure~\ref{Hal} shows a map of the dynamical lifetime of trans-Neptunian
bodies as a function of their initial semi-major axis and eccentricity, for an
inclination of $1^\circ$ and a random choice of the orbital angles $\lambda,
\varpi$ and $\Omega$ (\cite{DLB}). Similar maps, referring to different
choices of the initial inclination or different projections on the orbital
element space can be found in \cite{kuchner} and \cite{DLB}.  These maps have
been computed numerically, by simulating the evolution of massless particles
from their initial conditions, under the gravitational perturbations of the
giants planets. The latter have been assumed to be initially on their current
orbits. Each particle was followed until it suffered a close encounter with
Neptune.  Objects encountering Neptune, would then evolve in the scattered
disk for a typical time of order $\sim$$10^8$ years (but much longer residence
times in the scattered disk occur for a minority of objects), until they are
transported by planetary encounters into the inner planets region or to the
Oort cloud, or are ejected to the interstellar space.  This issue is described
in more detail in section~\ref{comets}.

In Figure~\ref{Hal} the colored strips indicate the timespan required for a
particle to encounter Neptune, as a function of its initial semi-major axis
and eccentricity.  Strips that are colored yellow represent objects that
survive for the length of the simulation, $4\times 10^9$ years (the
approximate age of the Solar System) without encountering the planet.  The
figure also reports the orbital elements of the known Kuiper belt objects. Big
dots refer to bodies with $i<4^\circ$, consistent with the low inclination at
which the stability map has been computed. Small dots refer to objects with
larger inclination and are plotted only for completeness.

As can be seen in the figure, the Kuiper
belt has a complex dynamical structure, although some general trends
can be easily explained. 

\paragraph{Stability limits imposed by close encounters with Neptune.\,}

Most objects with perihelion distances less than $\sim 35$~AU are unstable.
This is due to the fact that they pass sufficiently close to Neptune so that
they are destabilized during the encounters. In fact, in these cases Neptune's
gravity is no longer a `small perturbation' relative to that of the Sun. The
regularity of the oscillation of the orbital elements is broken. The
semi-major axis suffers jumps at each encounter with the planet, and the
eccentricity has correlated jumps in order to keep the perihelion distance
roughly constant (more precisely, to conserve the {\it Tisserand parameter},
see sect.~\ref{comets}). One encounter with Neptune after another, the object
wanders over the $(a,e)$ plane: the object is effectively a member of the
scattered disk. Consequently, the $q=35$~AU curve can be considered as the
approximate border between the Kuiper belt and the scattered disk, in the
30--50 AU semi-major axis range. The real border, however, has a more
complicated, fractal structure, illustrated by the boundary between the black
and the yellow regions in Fig.~\ref{Hal}.

Not all bodies with $q<35$~AU are unstable. The exception is represented by
objects in mean-motion resonances with Neptune.  These objects, despite
approaching (or even intersecting) the orbit of Neptune at perihelion, never 
approach the planet at short distance. This happens because the resonance
plays a protection role against close encounters. 

\begin{figure}[t!]
\centerline{\psfig{figure=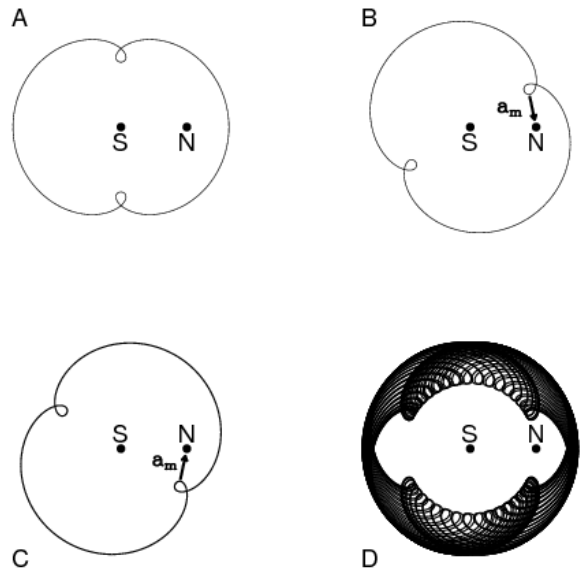,height=10.cm,angle=0
%,bbllx=10cm,bblly=-20cm,bburx=20cm,bbury=0cm
}}
\caption{ The dynamics of an object in the 2:3 mean-motion resonance with Neptune.  The double-lobed curve represents the orbit
  of an object with the eccentricity of Pluto. The coordinate frame rotates
  counterclockwise at the average speed of Neptune.  Thus, Neptune (dot
  labeled `N') is stationary in this figure. The location of the Sun is
  labeled `S'.  A) The orbit of an object whose semi-major axis is equal to
  that characterizing the exact location of the resonance.  The gravitational
  perturbations of Neptune cancel out due to the symmetry in the geometry.
  Thus, this orbit does not precess in the rotating frame. B) If the symmetry
  is broken, there is a net acceleration due to Neptune.  Here, the strongest
  perturbation ($a_m$) is at the upper lobe.  The object is leading Neptune at
  this lobe, so the net acceleration will decrease its semi-major axis. C) The
  strongest perturbation is in the lower lobe.  Consequently, the object's
  semi-major axis has to increase.  D) The orbit of an object that {\it
    librates} in the resonance. Courtesy of H. Levison.}
\label{pluto} 
\end{figure}

The stabilizing role of a mean-motion resonance can be understood in simple,
qualitative terms. For instance, Figure~\ref{pluto} illustrates the mechanism
for the case of Pluto (2:3 mean-motion resonance).  The trajectory of Pluto is
shown in the figure in a frame that rotates with Neptune.    Pluto moves
in a clockwise direction when further from the Sun than Neptune and moves in a
counter-clockwise direction when closer to the Sun.  In the figure, an object
with Pluto's eccentricity and exactly at Neptune 2:3 mean-motion resonance
would have a trajectory that is a double-lobed structure oriented as in
Fig.~\ref{pluto}a.  The configuration shown in the figure will remain fixed
only if the object's semi-major axis is {\it exactly} equal to that
characterizing the location of the resonance.  For an
object with semi-major axis slightly displaced,
the double-lobed structure will slowly precess in the rotating frame.

If the semi-major axis of the object is slightly larger than that
corresponding to the exact location of the resonance the double-lobed
trajectory will slowly precess towards that shown in Fig.~\ref{pluto}b.  If
the precession continued indefinitely, eventually the trajectory would pass
over the location of Neptune and a close encounter or a physical collision
would occur. However, because the new trajectory is no longer symmetric with
respect to Neptune, the object receives its largest acceleration ($a_m$) from
Neptune when in or near the upper lobe.  At this point, the object is leading
Neptune in its orbit and thus it is slowed down in its heliocentric motion.
Consequently its semi-major axis decreases.

When the semi-major axis of the object becomes smaller than that corresponding
to the exact location of the resonance the situation reverses. Now the the
double-lobed trajectory slowly precesses in the opposite direction. The
configuration of Fig.~\ref{pluto}a is restored, and then the trajectory
continues to precess towards the configuration of Fig.~\ref{pluto}c.  In this
case, the object gets its largest acceleration when it is near perihelion and
is trailing Neptune in their orbits (near the lower lobe of the trajectory).
Thus, the object's orbital velocity is increased, increasing its semi-major
axis.

When the semi-major axis of the object becomes again larger than the exact
resonant value, the precession of the double-lobed
trajectory reverses again. The trajectory goes back to the configuration 
of Fig.~\ref{pluto}a and then to that of Fig.~\ref{pluto}b, and the cycle
repeats indefinitely. Each cycle is called a {\it libration}.  Over a full
libration cycle the pattern drawn by the object's dynamics in the frame
co-rotating with Neptune is that illustrated in Fig.~\ref{pluto}d. 

Therefore, the mean-motion resonance exerts on the object
a {\it restoring torque} which reverses the precession of its 
double-lobed trajectory before a close encounter can occur. This of
course happens only if the object is not too far from the exact resonance
location, otherwise the precession is too fast and the magnitude of the
restoring torque is not sufficient. The limiting distance from the exact 
resonance location within which the restoring torque is effective defines the
{\it resonance width}. 

The analytic computation of resonance widths is detailed in \cite{morbybook}.
This calculation, however, overestimates the width of the 
region where resonant objects are stable over the Solar System's age. 
In fact, the situation is complicated by the interaction between the libration
motion inside the resonance and the precession motion of the orbits of the
object and of the perturbing planet. A detailed exploration of the stability
region inside the two main mean-motion resonances of the Kuiper belt, the 2:3
and 1:2 resonances with Neptune, has been done in \cite{NesRoig1}
\cite{NesRoig2}. Its description goes beyond the purposes of this chapter. 

\paragraph{Secular resonance instabilities.\,}

In Fig.~\ref{Hal} one can see that the dark region extends significantly
below the $q=35$~AU line for $40<a<42$~AU (and also for $35<a<36$~AU). 
The instability in these regions is due to the presence of a secular
resonance, due to the fact that ${\rm d}\varpi/{\rm d}t\sim 
{\rm d}\varpi_N/{\rm d}t$, where $\varpi$ is 
the perihelion longitude the object and $\varpi_N$ that of Neptune.
\begin{figure}[t!]
\centerline{\psfig{figure=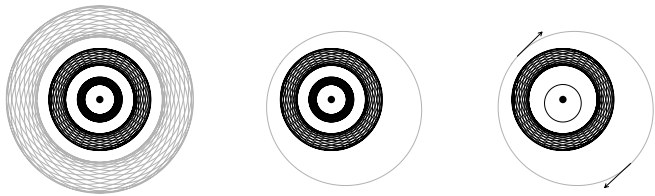,height=3.cm,angle=90}
%,
%\psfig{figure=secres_b.ps,height=3.cm},
%\psfig{figure=secres_c.ps,height=3.cm}
%,bbllx=10cm,bblly=-20cm,bburx=20cm,bbury=0cm
}
\caption{ The dynamics of a secular
  resonance.  Three orbits are shown in each panel.  The inner two are
  planets, which are shown as black lines.  The outer orbit (gray line) is for
  a small object.  The orbits of each object are ellipses, and the ellipses
  are precessing due to the mutual gravitational effects of the planets. Left:
  The orbits of the objects over a period of time that is long compared to the
  precession time of the orbits.  Here, we are looking in a fixed,
  non-rotating reference frame.  Each orbit sweeps out a torus of possible
  positions.  Center: The same as in the left plot, except that we are looking
  in a frame that rotates at the precession rate of the small outer body.
  Thus, its orbit is again an ellipse.  This panel shows the geometry if no
  secular resonance exists.  Note that the trajectories of the planets look
  axisymmetric.  Therefore, there is no net torque on the outer small object.
  Right: Same as the middle plot, except that the outer object is in a secular
  resonance with the inner planet, i.e. both orbits precess at the same rate.
  As a result, the outer object no longer sees an axisymmetric gravitational
  perturbation from the inner planet.  Indeed, it feels a significant torque.
  Courtesy of H.  Levison.}
\label{secres} 
\end{figure}

This resonance forces large variation of the eccentricity of the
trans-Neptunian object, so that --even if the initial eccentricity is null--
the perihelion distance eventually decreases below 35~AU, and the object
enters the scattered disk \cite{HolmanWisdom} \cite{MTM}. 

The destabilizing effect of a secular resonance between the longitude of
perihelia can be understood in simple qualitative terms.  Consider a simple
case where the orbits of the object and of {\it two} planets are in the same
plane. The presence of two planets is necessary, otherwise the planetary orbit
would be a fixed, non-precessing ellipse.  The orbit of the small body also
precesses under the planets' perturbations.  The left plot in
Figure~\ref{secres} shows the long-term trajectories of these objects in a
fixed frame.  The middle plot shows the same system in a frame that rotates
with the precession rate of the small body.  Note that the orbit of the small
body (the outermost orbit) is, in this frame, a fixed ellipse.  If the
precession rates of the planetary orbits are different from that of the small
body, the trajectories of the two planets in the rotating frame are still, on
average, axisymmetric and thus the small body experiences no long-term
torques.  However, if one of the planets precesses at the same rate as the
small body, as in the right plot in Figure~\ref{secres}, its long-term
trajectory is also a fixed ellipse in the rotating frame, and it is no longer
axisymmetric.  Thus the small body feels a significant long-term torque, which
can lead to a significant change in its eccentricity (which is related to the
angular momentum).

The location of secular resonances in the Kuiper belt has been computed in
\cite{knezevic}. This work showed that the secular resonance is present only at
small inclination. Large inclination orbits with $q>35$~AU and $40<a<42$~AU
are therefore stable. Indeed, Figure~\ref{Hal} shows that many objects with
$i>4^\circ$ (small dots) are present in this region. Only large dots,
representing low-inclination objects, are absent.

\paragraph{Chaotic diffusion in the Kuiper belt.\,}

Fig.~\ref{Hal} also shows the presence of narrow bands with brown colors,
crossing the yellow stability domain. These bands correspond to orbits which
become Neptune-crossing only after billions of years of evolution. What is the
nature of these weakly unstable orbits?

\begin{figure}[t!]
\centerline{\psfig{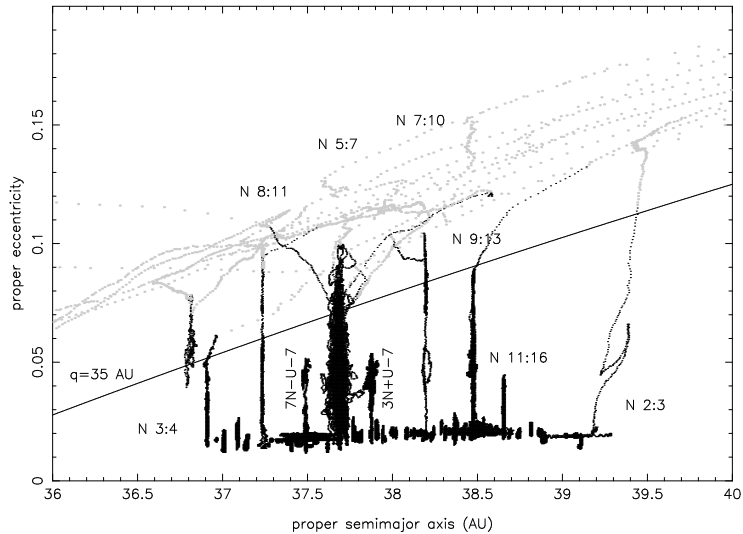}}
\caption{The evolution of objects initially at $e=0.015$ and semi-major axis
  distributed in the 36.5--39.5 AU range. The dots represent the proper semi-major axis and the eccentricity of the objects --computed by averaging their
  $a$ and $e$ over 10~My time intervals-- over the age of the Solar System.
  They are plotted in grey after that the perihelion has decreased below 32 AU
  for the first time. Labels $Nk:k_N$ denote the $k:k_N$ two-body resonances
  with Neptune.  Labels $k_N$N$+k_U$U$+k$ denote the three-body
  resonances with Uranus and Neptune, that correspond to the equality
  $k_N\dot\lambda_{\rm N}+k_U\dot\lambda_{\rm U}+k\dot\lambda=0$.  Reprinted
  from \cite{NesRoig2}.}
\label{diffusion} 
\end{figure}

It has been found \cite{NesRoig2} that these orbits are in general associated
either with high order mean-motion resonances with Neptune (i.e. resonances
for which the equivalence $k{\rm d}\lambda/{\rm d}t=k_N{\rm d}\lambda_N/{\rm
  d}t$ holds only for large values of the integer coefficients $k, k_N$) or
three--body resonances with Uranus and Neptune (which occur when 
$ k{\rm d}\lambda/{\rm d}t+k_N{\rm d}\lambda_N/{\rm d}t+
k_U{\rm d}\lambda_U/{\rm d}t=0$ occurs for some integers $k, k_N$ and $k_U$). 

The dynamics of objects in these resonances is chaotic.  The semi-major axis of
the objects remain locked at the corresponding resonant value, while the
eccentricity of objects is slowly modified. In an $(a,e)$-diagram like
Fig.~\ref{diffusion}, each object's evolution leaves a vertical trace.  This
phenomenon is called {\it chaotic diffusion}.  Eventually the growth of the
eccentricity can bring the diffusing object above the $q=35$~AU curve. These
resonances are too weak to offer an effective protection against close
encounters with Neptune, unlike the low order resonances considered above.
Thus, once above this critical curve, the encounters with Neptune start to
change the semi-major axis of the objects, which leave their original
resonance and evolve --from that moment on-- in the scattered disk.

Notice from Fig.~\ref{diffusion} that some resonances are so weak that,
despite the resonant objects diffuse chaotically, they cannot reach the
$q=35$~AU curve within the age of the Solar System. Therefore, these objects
are `stable' from the astronomical point of view.

Notice also that chaotic diffusion is effective only for selected resonances.
The vast majority of the simulated objects are not affected by any macroscopic
diffusion. They preserve their initial small eccentricity for the entire age
of the Solar System. Thus, the current moderate/large eccentricities and
inclinations of most of the Kuiper belt objects cannot be obtained from
primordial circular and coplanar orbits by dynamical evolution in the
framework of the current orbital configuration of the planetary system.
Likewise, the region beyond the 1:2 mean-motion resonance with Neptune is
totally stable.  Thus, the absence of bodies beyond 48 AU cannot be explained
by current dynamical instabilities. Also, the overall mass deficit of the
Kuiper belt cannot be due objects escaping through resonances, because most of
the inhabited Kuiper belt is stable for the current planetary architecture.
Therefore, all these intriguing properties of the Kuiper belt's structure must
find an explanation in the framework of the formation and primordial evolution
of the Solar System. This will be the topic of sect.~\ref{sculpting}.

\subsection{Note on the scattered disk}
\label{SD}

We have seen above that the bodies that escape from the Kuiper belt and
decrease their perihelion distance below 35~AU, without being protected by a
low-order mean-motion resonance, enter the scattered disk. 

Their subsequent evolution has been studied in detail in \cite{LD97}.  It was
found that the median dynamical lifetime is $\sim 50$~My, the typical
end-states being the transport towards the inner Solar System (and the eventual
ejection from the Solar System due to an encounter with Jupiter or Saturn; see
sect,~\ref{comets}), a collision with a planet or the outward transport
towards the Oort cloud (see sect.~\ref{Oort}).  This result suggests that the
scattered disk could be a population of transient objects, which is sustained
in steady state by a continuous flux of objects escaping from the Kuiper
belt. In this case, the scattered disk would be, relative to the Kuiper belt,
what the population of Near Earth Asteroids is, relative to the main asteroid
belt.

However, \cite{DL97} showed that about 1\%
of the scattered-disk objects can survive on trans-Neptunian orbits for the age
of the Solar System. This opens the possibility that the current scattered
disk is the remnant of a $\sim 100\times$ more massive primordial structure,
which presumably formed when the planets chased the left-over
planetesimals from their neighborhoods. In this case, the scattered disk would
not be in steady state, and it would have be no direct relationship with the
Kuiper belt. 

How can we discriminate between these two hypotheses on the origin of the
scattered disk? In the first case, if the scattered disk is sustained in
steady state by the objects leaking out of the Kuiper belt, the number ratio
between the Kuiper belt and scattered-disk populations must be large. 
Indeed: 
$$
N_{SD}=N_{KB}\times f_{esc}\times L_{SD}
$$
where $N_{SD}$ is the number of scattered-disk objects (larger than a
given size), $N_{KB}$ is the number of Kuiper belt objects (down to the same
size), $f_{esc}$ is the fraction of the Kuiper belt population that escapes
into the scattered disk in the unit time (due to chaotic diffusion or
collisional ejection) and $L_{SD}$ is the mean lifetime in the scattered disk.
Both $L_{SD}$ and $f_{esc}$ are small, so that $N_{KB}>>N_{SD}$.  In fact, in
the case of the main asteroid belt and the NEA population, the number ratio is
about 1,000. 

In the second case, if the current scattered disk is the remnant of a much
more massive primordial scattered population, there is no casual relationship
between $N_{KB}$ and $N_{SD}$. The current population of the scattered disk
depends only on its primordial population, and not on the current Kuiper belt
population. 

Discovery statistics \cite{Trujillo} suggest that the scattered disk and the
Kuiper belt now contain roughly equal populations. This rules out (by orders
of magnitude) the possibility that the scattered disk is sustained in steady
state by the Kuiper belt. Only the scenario of \cite{DL97} remains  
valid for the origin of the scattered disk.
 
\section{The dynamics of comets}
\label{comets}

\begin{figure}[t!]
\centerline{\psfig{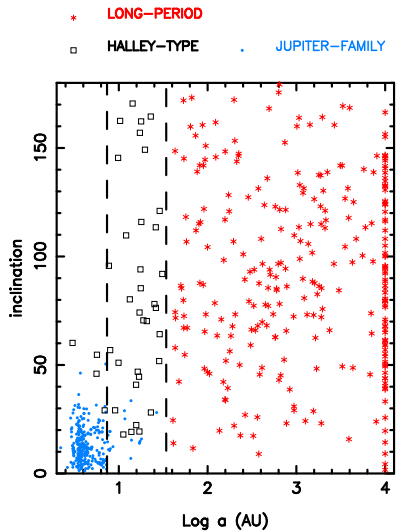}}
\caption{The distribution of comets according to their orbital semi-major axis
  and inclination. Here the orbital elements are defined at the moment of the
  comet's last aphelion passage. Long period, Halley-type and Jupiter family
  comets are plotted as red stars, black squares and blue dots,
  respectively. The separation between Halley-types and Jupiter family comets
  has been made according to the value of their Tisserand parameter, following
  \cite{hal_classes}. The vertical dashed lines correspond to orbital periods  
  $P=20$~y and $P=200$~y, respectively. All long period comets with $a>$10,000
  AU have been represented on the $\log a=$4 line.}
\label{fig-comets} 
\end{figure}

Comets are usually classified in categories according to their orbital period
(Figure~\ref{fig-comets}). Comets with orbital period $P>200$~y are called
{\it long period comets} (LPCs); those with shorter period are called {\it
  short period comets} (SPCs).  The threshold of 200~y is arbitrary, and has
been chosen mostly for historical reasons: modern instrumental astronomy is
about two centuries old, so that the long period comets that we see now are
unlikely to have been observed in the past.

If the orbital distribution of the comets is plotted, like in
Fig.~\ref{fig-comets}, using the orbital elements that the comets had when
they last passed at {\it aphelion} --which can be computed through a backward
numerical integration-- one sees a clustering of long period comets with
$a\sim 10^4$~AU.  These comets are called {\it new comets} because they must
pass through the giant planets system for the first time.  In fact, after a
passage through the inner Solar System, it is unlikely that the semi-major
axis remains of order $10^4$~AU. It either decreases to $\sim 10^3$~AU or the
orbit becomes hyperbolic. The reason is that the binding energy of a new comet
is $E=-{\cal G} M_\odot/2a\sim 10^{-4}$, but typically, during a close
perihelion passage, the energy suffers a change of order of the mass of
Jupiter relative to the Sun: $10^{-3}$.  This change is not due to close
encounters with the planet (which might not occur). It is due to the fact that
the comet has a barycentric motion when it is far away, an heliocentric motion
when it is close, and the distance of the barycentre from the Sun is of the
order of the relative mass of Jupiter.

The short period comets are in turn subdivided in {\it Halley-type} (HTCs) and
{\it Jupiter family} (JFCs). Historically, the partition between the two
classes is done according to the orbital period being respectively longer or
shorter than 20~y. This threshold has been chosen because there is an evident
change in the inclination distribution at the corresponding value of
semi-major axis (see Fig.~\ref{fig-comets}). However, comets continuously
change semi-major axis as a consequence of their encounters with the planets.
In particular, all short period comets had to have a larger semi-major axis in
the past, given that they come from the trans-planetary region. Thus, by
adopting a partition between Halley-type and Jupiter family comets based on
orbital period, one is confronted with the unpleasant situation of objects
changing their classification during their lifetime.

This has motivated Levison \cite{hal_classes} to classify short period comets
according to their {\it Tisserand parameter} relative to Jupiter 
\begin{equation}
T_J={{a_J}\over{a}}+2\sqrt{{{a}\over{a_J}}(1-e^2)}\cos i\ .
\label{Tiss}
\end{equation}
This new classification makes sense, because the Tisserand parameter
is quite well preserved during the comet's evolution.  In Levison's
classification, Halley-type and Jupiter family comets have $T_J$ respectively
smaller and larger than 2. Fig.~\ref{fig-comets} adopts this classification and
shows that for most of the objects the classifications based on orbital period
and on Tisserand parameter are in agreement, but a few objects (those with
$P<20$~y and large inclination or those with $P>20$~y and low inclination)  
do change their classification depending on the adopted criterion. 

\paragraph{Tisserand parameter.\,}

Given the importance of the Tisserand parameter in comet dynamics, it is
useful to derive its expression (which outlines the limitations of its use) 
and discuss its properties. 

The Tisserand parameter is an approximation of the Jacobi constant, which is
an invariant of the dynamics of a small body in the framework of the
restricted, circular, three-body problem.

The expression of the Jacobi constant is:
\begin{equation}
C_J=-(\dot{x}^2+\dot{y}^2+\dot{z}^2)+2\left({{1}\over{r}}+{{m_p}\over{\Delta}}\right)+2H_z\
,
\label{Jacobi}
\end{equation}
where ${\cal G}M_\oplus=a_p=1$ are assumed, and $a_p, m_p$ are the semi-major
axis and mass of the perturbing planet and $H_z$ is the $z$--component of the
small body's angular momentum. The quantity $\Delta$ is the distance between
the small body and the planet. 

We write the kinetic energy of the small body as a function of its semi-major
axis and heliocentric distance:
\begin{equation}
{{1}\over{2}}(\dot{x}^2+\dot{y}^2+\dot{z}^2)=-{{1}\over{2a}}+{{1}\over{r}}\ ,
\label{T}
\end{equation}
while the $z$-component of the angular momentum can be written:
\begin{equation}
H_z=\sqrt{a(1-e^2)}\cos i\ .
\label{Hz}
\end{equation}
Substituting (\ref{T}) and (\ref{Hz}) into (\ref{Jacobi}) and neglecting the
term ${{m_p}/{\Delta}}$ one gets 
\begin{equation}
C_J\sim T\equiv {{1}\over{a}}+2\sqrt{a(1-e^2)}\cos i\ ,
\label{Jacobi-Tiss}
\end{equation}
where the right hand side coincides with (\ref{Tiss}), given that $a$ is
expressed in units of the planet's semi-major axis. 

This derivation of the Tisserand formula shows that the Tisserand parameter is
constant as long as the Jacobi constant is preserved, and $m_p/\Delta$ is
small. This last condition requires that the comet is not in a close encounter
with the planet. During a close encounter, the Tisserand parameter has
large and abrupt changes, but it returns to the value that it had before the
encounter, once the distance to the planet increases back to large values. 
The conservation of the Jacobi constant, conversely, requires that the
conditions of the restricted three-body problem are fulfilled, namely one
planet must dominate the comet's evolution, and the effects of the 
planet's eccentricity must be negligible. This requires that the comet is not
in a region where it can have encounters with {\it two} planets, otherwise the
one-planet approximation does not hold. Also, it requires that the comet is
not in a secular resonance with the planet, otherwise the effects of the
planet's small eccentricity are enhanced. 

One can demonstrate that, if a comet intersects the orbit of a planet, the
Tisserand parameter $T$ is related to the unperturbed relative 
velocity $U$ at which it encounters the planet:
$$
U=\sqrt{3-T}\ ,
$$
where $U$ is expressed in units of the planet's orbital velocity.
The formula is not defined for $T>3$, which implies that comets with such
values of Tisserand parameter cannot intersect the orbit of the planet. Note
however that comets non-intersecting the orbit of the planet can have $T<3$.
Only objects with $T<2\sqrt{2}\sim 2.83$ (the value for a parabolic trajectory
with $i=0$ and $q=a_p$) can be ejected on hyperbolic orbit 
in a single encounter with a planet. 

\subsection{Origin and evolution of Jupiter family comets}
\label{JFC}

The fact that the JFCs have (by definition) a Tisserand parameter with respect
to Jupiter that is distinct from that of HTCs and LPCs suggests that the
former are not the small semi-major axis end of the distribution of the
latter. The average low inclination of the JFCs, and the absence of retrograde
comets in the JFC population (whichever of the two definitions for JFCs is
adopted, see Fig.~\ref{fig-comets}) argues that the source of JFCs must be a
disk-like structure. In 1980 \cite{fernandez1980b} proposed that the source of
JFCs was the --at the time still putative-- Kuiper belt, an hypothesis
supported later in \cite{DQT-jfcs}.

However, today we know that there are two distinct disk-like structures in the
trans-Neptunian region: the Kuiper belt and the scattered disk.  Which of the
two is the source of JFCs?  We have seen in sect.~\ref{SD} that the scattered
disk is too populated to be sustained in steady state by the objects leaking
out of the Kuiper belt. If the scattered disk is not sustained in steady
state, it means that the number of objects that leave the scattered disk
--mostly evolving towards the inner solar system-- is larger than the number of
objects entering the scattered disk from the Kuiper belt. Thus, the scattered
disk must dominate the JFC production, over the Kuiper belt. 

\begin{figure}[t!]
\centerline{\psfig{figure=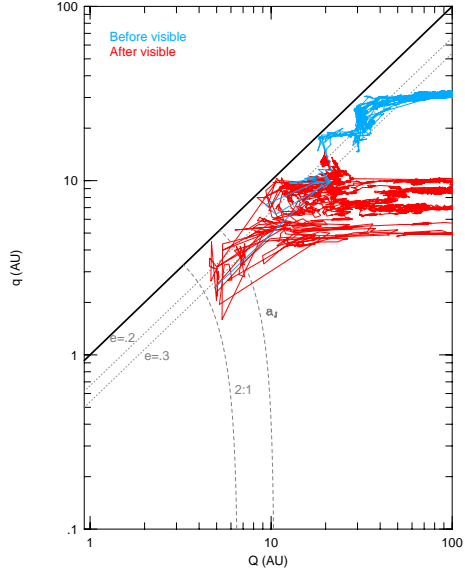,height=9.cm}}
\caption{The evolution of an object from the scattered disk up to its ultimate ejection, projected over the plane representing
  perihelion vs. aphelion distance. The horizontal structure at $q\sim 30$~AU
  represents the scattered disk. When the object evolves along a line
  $q=$constant or $Q=$constant its dynamics are essentially dominated by one
  single planet. This happens at least down to 10~AU, and during the final
  ejection phase. Blue lines denote the evolution before that the object
  becomes a visible JFC, red lines after. The criterion for first visibility
  is that $q$ has decreased below 2.5~AU for the first time. From \cite{LD97}}
\label{qQevolve} 
\end{figure}

The dynamical evolution of objects from the scattered disk to the JFC region
has been studied in detail in \cite{LD97}, with statistics made on a large
number of numerical simulations. The results illustrated in that paper
essentially supersede all the results from the previous
literature. Thus, most of what I report below is taken from that source. The
origin and dynamics of JFCs has also been exhaustively reviewed in
\cite{duncan_cometII}.  

To evolve from the scattered disk to the JFC region, a comet has to pass from
a Neptune-dominated dynamics to a Jupiter-dominated dynamics (see
Fig.~\ref{qQevolve}). The transfer process involving multiple planets, in
principle the Tisserand parameter is not preserved. However, the planetary
system is structured in such a way that the transfer chain from Neptune to
Jupiter is piece-wise dominated by one single planet (see
Fig.~\ref{qQevolve}), and the values of the Tisserand parameters relative to
the dominating planets are almost the same.  For instance, consider a
scattered-disk body with Tisserand parameter relative to Neptune $T_N=2.98$.
The conservation of the Tisserand parameter implies that the smallest
perihelion distance to which Neptune can scatter this object is $q=17.7$~AU,
just enough to become Uranus-crosser. In this orbit, the body has $T_U=2.96$.
If Uranus takes the control of this body, it can scatter it inwards down to
$q=9.0$~AU, barely a Saturn crosser. The body has now $T_S=2.94$ and thus
Saturn can lower its $q$ to only $3.8$~AU.  With such a perihelion, the comet
has a Tisserand parameter $T_J=2.82$. Thus, the body never spends much time in
a region where it can encounter two planets.  The Tisserand parameter is
therefore piece-wise conserved, and the final Tisserand parameter (with
respect to Jupiter) is very close to the initial one (with respect to
Neptune). Now, the bulk of the observed population in the scattered disk has
$2<T_N<3$.  Thus, at the end of the transfer chain, the bodies coming from the
scattered disk will have $2<T_J<3$, namely they will be JFCs.

\begin{figure}[t!]
\centerline{\psfig{figure=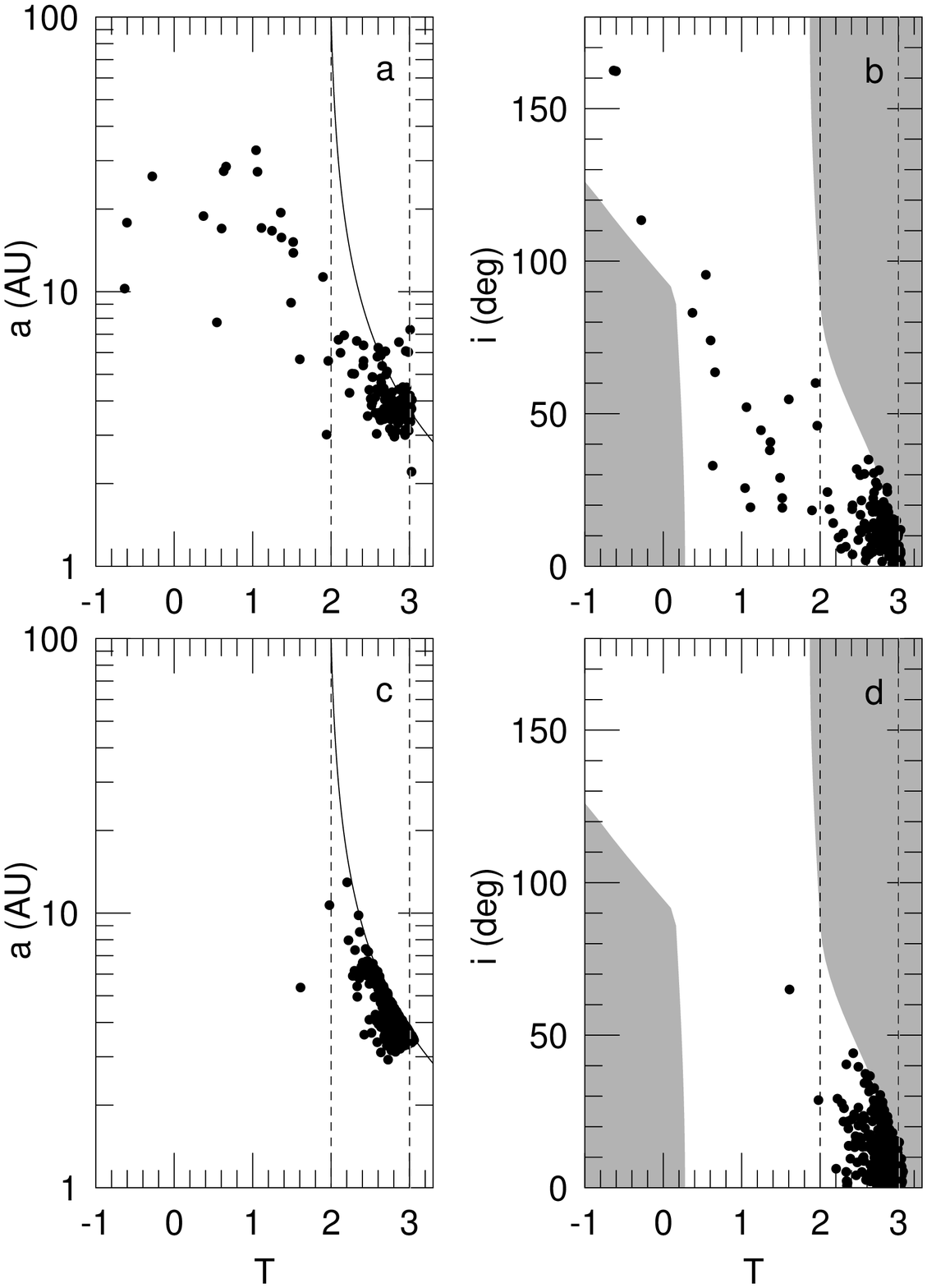,height=9.cm}}
\caption{The distribution of short period comets projected over the $(T_j,a)$
  and $(T_j,i)$ planes. Top panels: the observed distribution. Bottom panels:
  the distribution of the objects coming from the scattered disk, when they
  are visible ($q<2.5$~AU) for the first time. From \cite{LD97}.}
\label{aTi} 
\end{figure}

Because the Tisserand parameter remains close to 3, the inclination cannot
grow to large values (the growth of $i$ would decrease $T$, see (\ref{Tiss})).
So, the final inclination distribution is comparable to the inclination
distribution in the scattered disk, i.e. mostly confined within 30 degrees.
Figure~\ref{aTi} compares the $(a,i,T_J)$ distribution of the observed short
period comets (top panels) with the one obtained in the numerical simulation 
for the objects coming from the
scattered disk, when their perihelion distance first decreases below
2.5~AU (a criterion for visibility as an active comet). As one sees, the
objects with $T_J<2$ (HTCs) are not reproduced, while the observed and
simulated distributions of the JFCs agree with each other in a remarkable way.

Nevertheless, a quantitative comparison would show that the inclination
distribution of the simulated comets when they first become visible is
slightly skewed towards low values relative to the observed distribution.
Similarly, the distribution of the distances of the comets' nodes from
Jupiter's orbit is also skewed towards small values. However, the dynamical
lifetime of comets after they first become visible is of order $10^5$~y. As
time passes, the conservation of the Tisserand parameter degrades, as a result
of the combined effects of Jupiter and Saturn and of secular resonances. Thus,
the inclination is puffed up, and the distribution of $\omega$ (initially
strongly peaked around $0^\circ$ and $180^\circ$) is randomized. As a
consequence, the nodal distance distribution is also puffed up\footnote{Some
  comets eventually evolve towards the $T_J<2$ region, although they never
  manage to reproduce the $(a,i,T_J)$ distribution illustrated in the top
  panels of Fig.~\ref{aTi}}. Consequently, the agreement between the observed
and simulated distributions first improves with the age of the comets, and
then eventually degrades. Thus, \cite{DL97} considered the distribution of all
simulated objects, from the time they first become visible up to time $\tau$.
Using a Kolmogorov-Smirnov test to measure quantitatively the statistical
agreement between simulated and observed distributions, \cite{DL97} concluded
that the best match is achieved --simultaneously for the inclination and the
nodal distance distributions-- for $\tau\sim$~12,000~y.  The interpretation of
this result is that this value of $\tau$ corresponds to the typical physical
lifetime of JFCs, after which the comets loose their activity and are no
longer observed. Comparing the physical lifetime with the dynamical lifetime,
\cite{DL97} concluded that, if all faded JFCs are dormant objects with
asteroidal appearance, the ratio between the number of dormant vs. active JFCs
should be $\sim 4$.

The comparison between the $q$ distribution of the simulated and observed JFCs
suggests that the population of comets is observationally complete up to
$q\sim 2$~AU. There are $\sim$40 known JFCs with total absolute magnitude
$H_{10}<9$\footnote{The total absolute magnitude is computed from the apparent
  magnitude $V$ (of nucleus plus coma), the heliocentric and geocentric
  distances $r$ and $\Delta$ by the formula $H_{10}=V+5\log\Delta+10\log{r}$,
  instead of the usual formula for dormant bodies $H=V+5\log\Delta+5\log{r}$.
  The coefficient 10, instead of 5, accounts for the fact that the intensity
  of the activity of the comet is proportional to $r^{-2}$.} and $q<2$~AU. The
simulated $q$ distribution indicates that there should be about 100 comets
with $q<2.5$~AU, with the same total magnitude. If all faded JFCs are dormant,
then we should expect an additional 400 bodies of asteroidal appearance on
similar orbits. About 100 of them should have $q<1.3$~AU and belong to the NEO
population. The size of these putative bodies is badly constraint, because
the conversion from total magnitude to nuclear magnitude (i.e. the absolute
magnitude of the nucleus, in absence of cometary activity) is poorly known.
Published estimates for the nucleus size for $H_{10}=9$ comets range from
$D=0.8$~km \cite{bailey} to $D=4.5$~km \cite{fernandez99}, with a mean of
about $2$~km \cite{fernandez99}. I will return to the
nature of faded comets in sect.~\ref{fade-fate}.

With this estimate of the total number of JFCs, from the rate at which
scattered-disk bodies become JFCs and the mean lifetime of JFCs measured in
the simulations, \cite{DL97} computed that there should be $4\times 10^8$ of
such objects (i.e.  big enough to have total magnitude $H_{10}<9$ when active)
in the scattered disk. The extrapolation of the size distribution observed in
the scattered disk \cite{bernstein} is roughly consistent with this estimate. 

\paragraph{The orbit of comet P/Encke.\,}

Despite the overall good agreement between the observed and the simulated
distribution of JFCs shown in Fig.~\ref{aTi} there is one important difference
that should not be overlooked: the orbit of comet P/Encke is not re-produced
in the simulation of \cite{LD97}. P/Encke is particular. It is the only active
comet with an orbit totally interior to the orbit of Jupiter and $T_J>3$.  The
aphelion distance of P/Encke is currently 4.1~AU, so that it is not scattered
by Jupiter's encounters.  This implies that encounters with Jupiter cannot
have emplaced the comet onto its current orbit.

It has been proposed that P/Encke reached its orbit from the $T_J<3$ region
due to close encounters with the terrestrial planets, to the effect of
non-gravitational forces\footnote{for a recent review on non-gravitational
  forces acting on comet dynamics see \cite{Yeomans}}, or both
\cite{Valsecchi}\cite{bailey-harris}\cite{FTB02}\cite{pittich}. Neither of
these aspects have been included in the simulations of \cite{LD97}.

A quantitative model of the orbital distribution of JFCs accounting for
terrestrial planets encounters and/or non-gravitational forces has never been
done, so that we do not know if either of these effects naturally explains the
existence of {\it one} active comet on orbits decoupled from Jupiter 
like that of P/Encke (as opposed
to several comets or none).  From the observational point of view, it seems
that only very few objects should have followed Encke's dynamical path. In
fact, a search for objects with albedo typical of dormant
cometary nuclei among the NEOs with $T_J>3$ (\cite{yan}) has showed
that these objects, if they exist, are rare. 
  
\subsection{Origin and evolution of Long period comets}
\label{LPC}

\begin{figure}[t!]
\centerline{\psfig{figure=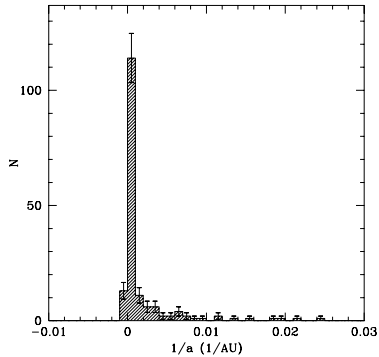,height=5.cm}}
\caption{The differential distribution of LPCs as a function of the inverse
  semi-major axis. The big spike at $1/a<10^{-4}$ is due to the new comets,
  and is usually called the {\it Oort spike}. From \cite{WT99}.}
\label{spike} 
\end{figure}

In an historical paper, Oort (\cite{Oort}) pointed out that the presence of
numerous {\it new comets} with $a>10^4$~AU --which appears as a spike in the
distribution of $1/a$ of the LPCs (see Fig.~\ref{spike})-- argues for the
existence of a reservoir of objects in that distant region. The fact that the
inclination distribution of new comets is essentially isotropic, not only in
$\cos i$ (from -1 to 1, i.e. including also retrograde orbits), but also in
$\omega$ and $\Omega$, indicates that this reservoir must have a
quasi-spherical symmetry, namely it has the shape of a cloud surrounding the
Solar System. This cloud is now generally called the {\it Oort cloud}. In
Oort's view, {\it all } long period comets come from this cloud. The LPCs with
$a<10^4$ AU are returning comets, which originally belonged to the new comet
group when they first entered into the inner Solar System, but subsequently
they had their orbit perturbed and acquired a more negative binding energy
(smaller semi-major axis). This view remains essentially valid even today.

At such large distances from the Sun, the evolution of the comets in the Oort
cloud is strongly affected by the overall gravitational field due to the mass
distribution in the galaxy (the so--called {\it galactic tide}), and by
sporadic passing stars and giant molecular clouds (GMCs).

Assuming that the galaxy has a disk-like structure and considering that the
Sun is not at the center, the galactic tide has both ``disk'' and ``radial''
force components. In a coordinate system centered on the Sun, with $x$-axis
pointing away from the galactic center, $y$-axis in the direction of the
galactic rotation and $z$-axis towards the south galactic pole, the radial
component of the tide can be expressed with forces along the $x$ and $y$
directions, respectively:
\begin{equation}
F_x=\Omega_0^2 x\ ;\quad F_y=-\Omega_0^2 y\ ,
\label{radialtide}
\end{equation}
where $\Omega_0$ is the frequency of revolution of the Sun around the galaxy.
The disk component of the tide can be represented with a force along the $z$
direction:
\begin{equation}
F_z=-4\pi{\cal G}\rho_0 z\ ,
\label{disktide}
\end{equation}
where $\rho_0$ is the mass density in the solar neighborhood
\cite{HeislerTremaine}. The disk component dominates over the radial component
by a factor 8--10, so that typically only the disk component (\ref{disktide}) 
is considered. 

The effect of the disk tide is analogous to the Kozai effect for the dynamics
of asteroids with high inclination relative to Jupiter's orbit \cite{kozai}.
In the following, I denote the inclination of the comet relative to the
galactic plane by $\tilde{\iota}$ and the argument of perihelion by
$\tilde\omega$ (not to be confused with the inclination $i$ and the argument
of perihelion $\omega$ relative to the Solar System plane; the two planes are
inclined at 120 degrees relative to each other).  The disk tide preserves $a$
and the $z$-component of the angular momentum
$H_z=\sqrt{1-e^2}\cos\tilde\iota$ of the comet, while its $e$ and
$\tilde\iota$ change with the precession of $\tilde\omega$. The evolution is
periodic; $e$ has a maximum and $\tilde\iota$ a minimum when
$\tilde\omega=90^\circ, 270^\circ$, while $e$ has a minimum and $\tilde\iota$
a maximum when $\tilde\omega=0^\circ, 180^\circ$\footnote{Here I assume that
  $\tilde\iota$ is defined in the range between $-90^\circ$ and $90^\circ$
  (negative $\tilde\iota$ corresponding to retrograde orbits relative to the
  galactic plane),
  and by `maximum' and `minimum' I mean the maximum and minimum of
  $|\tilde\iota|$}.  The difference between the maximum and minimum values of
$e$ and $\tilde\iota$ increases when $a$ increases or $H_z$ decreases. There
is no variation of $e$ and $\tilde\iota$ if $\tilde\iota=0$.

Thus, Oort cloud comets with high inclination relative to the galactic plane,
under the effect of the tide, increase their orbital eccentricity; their
perihelion distance decreases and the object becomes planet-crosser. If this
evolution is fast enough that $q$ decreases from beyond 10~AU to less than
$\sim 3$~AU within half an orbital period the comet becomes active during its
first dive into the inner solar system (i.e. without having interacted with
Jupiter or Saturn during its previous orbits), namely it appears as a `new
comet'.  The perturbations from the planets remove the planet-crossing comets
from the Oort cloud, by either decreasing their semi-major axis or ejecting
them from the Solar System on hyperbolic orbits.  Thus, the high inclination
portion of the Oort cloud is progressively depleted. The role of passing stars
and GMCs is to reshuffle the comet distribution in the Oort cloud, and to
refill the high inclination region where comets are pushed into the planetary
region by the disk's tide. Of course, stars and GMCs can also directly deflect
the cometary trajectories, injecting the comets into the inner solar system
without the help of the galactic tide.  This happens paticularly during comet
showers caused by close encounters between the Sun and these external
perturbers \cite{Hut}\cite{Heisler}. These directly injected comets do not
need to have a large inclination relative to the galactic plane.

\begin{figure}[t!]
\centerline{\psfig{figure=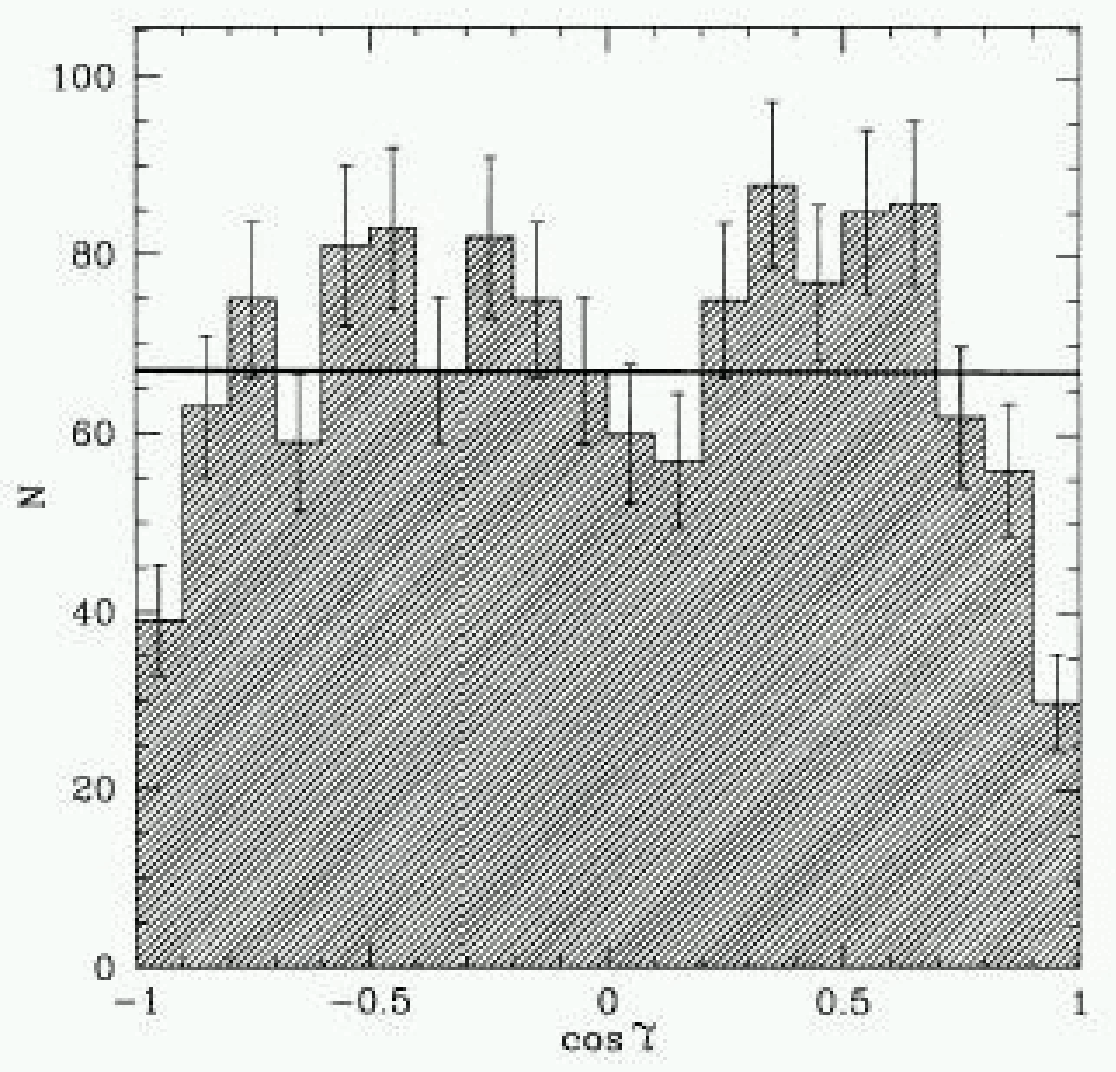,height=5.cm}\psfig{figure=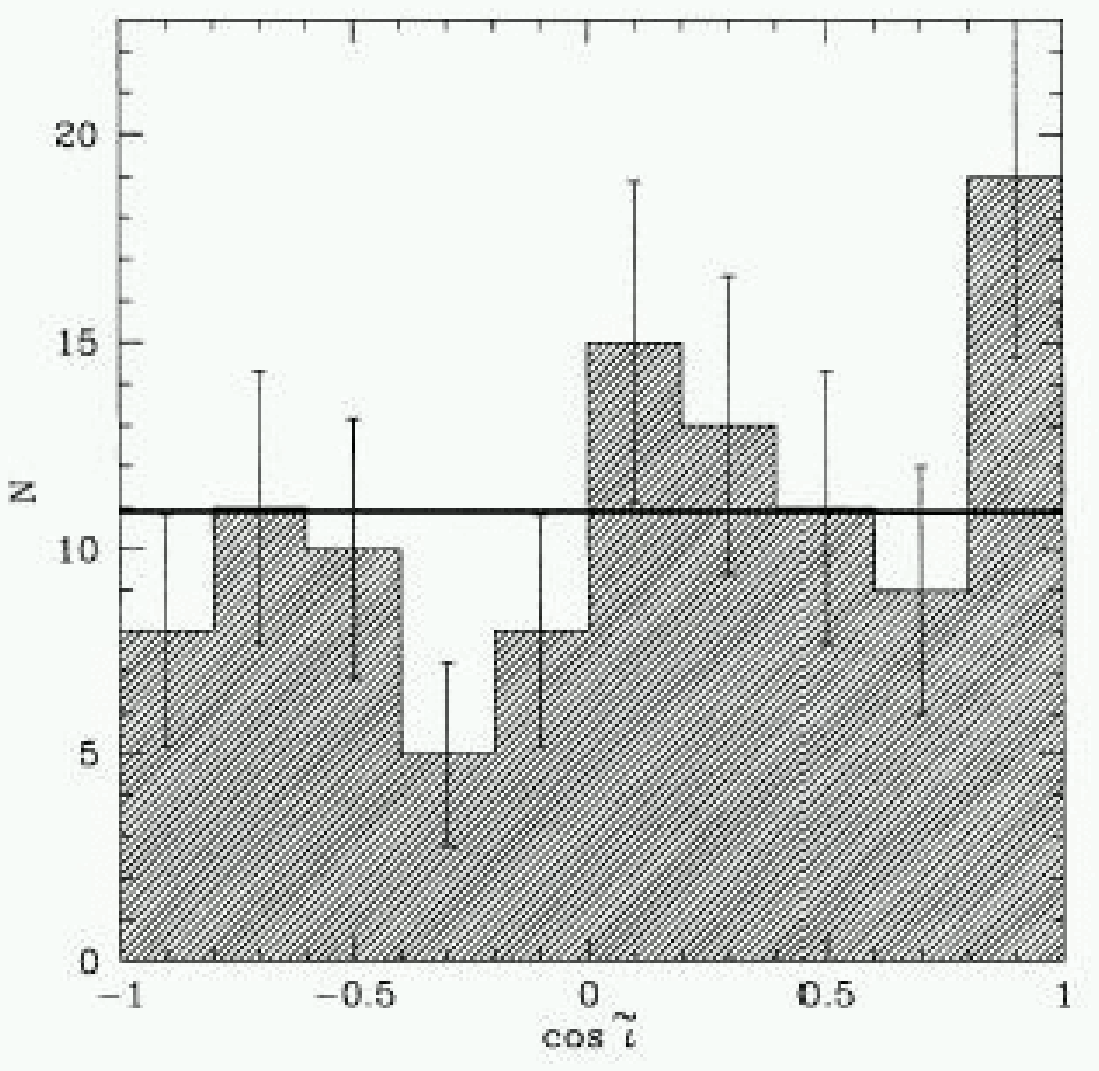,height=5.cm}}
\caption{The inclination distribution relative to the galactic plane for new
  comets. (a) (left): result of a numerical simulation. (b) (right): the
  observed distribution. Here $\tilde\iota$ is defined in the range between
  $0^\circ$ and $180^\circ$; values of $\tilde\iota$ larger than $90^\circ$
  correspond to retrograde orbits relative to the galactic plane. From
  \cite{WT99}.}
\label{WT-i} 
\end{figure}

The transfer of comets from the Oort cloud to the inner Solar System has been
simulated by many authors, in particular by \cite{Heisler}, \cite{weissman}
and, more recently \cite{WT99}. In what follows I will mostly refer to this
latter, most modern work. 

\begin{figure}[t!]
\centerline{\psfig{figure=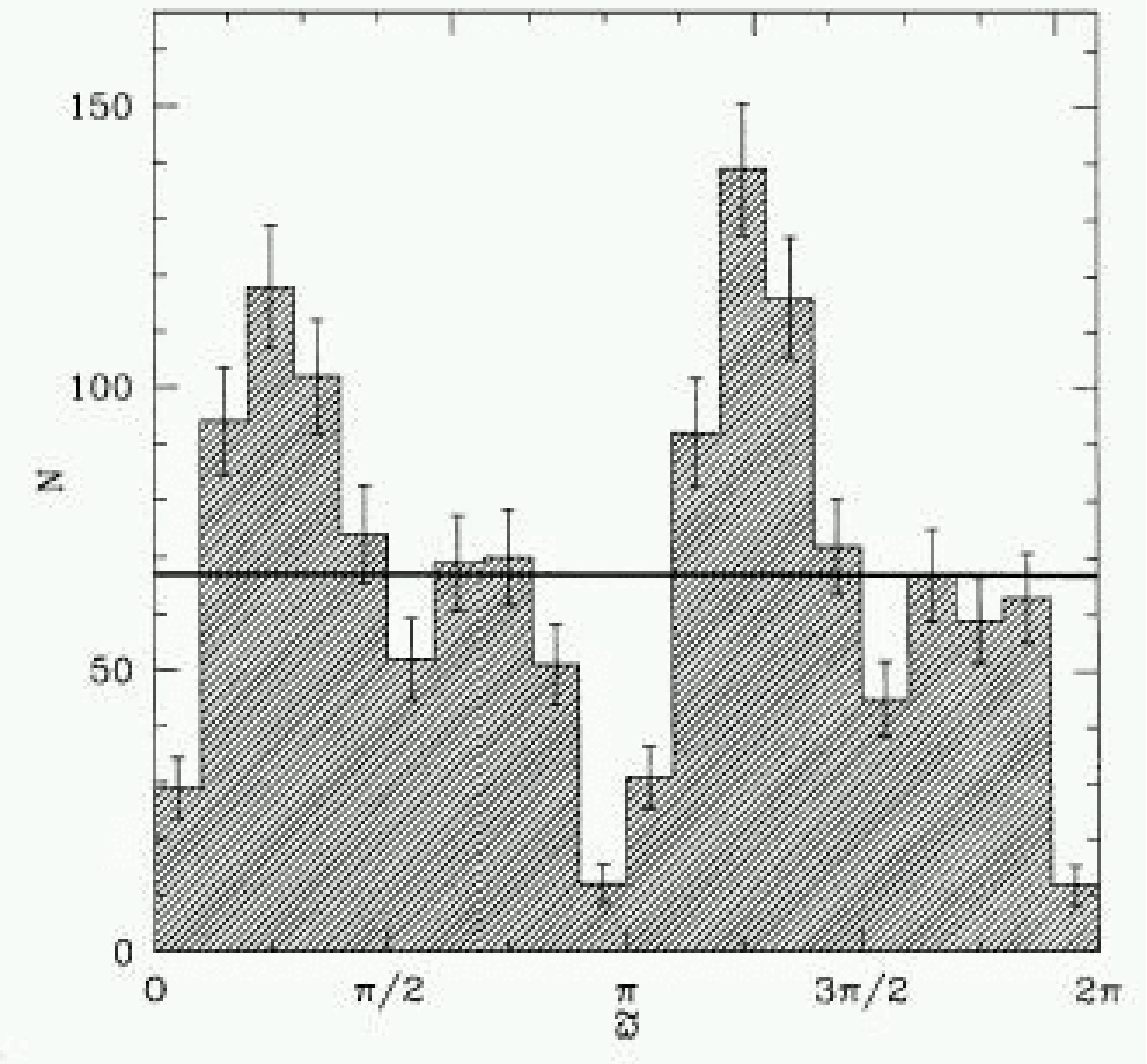,height=5.cm}\psfig{figure=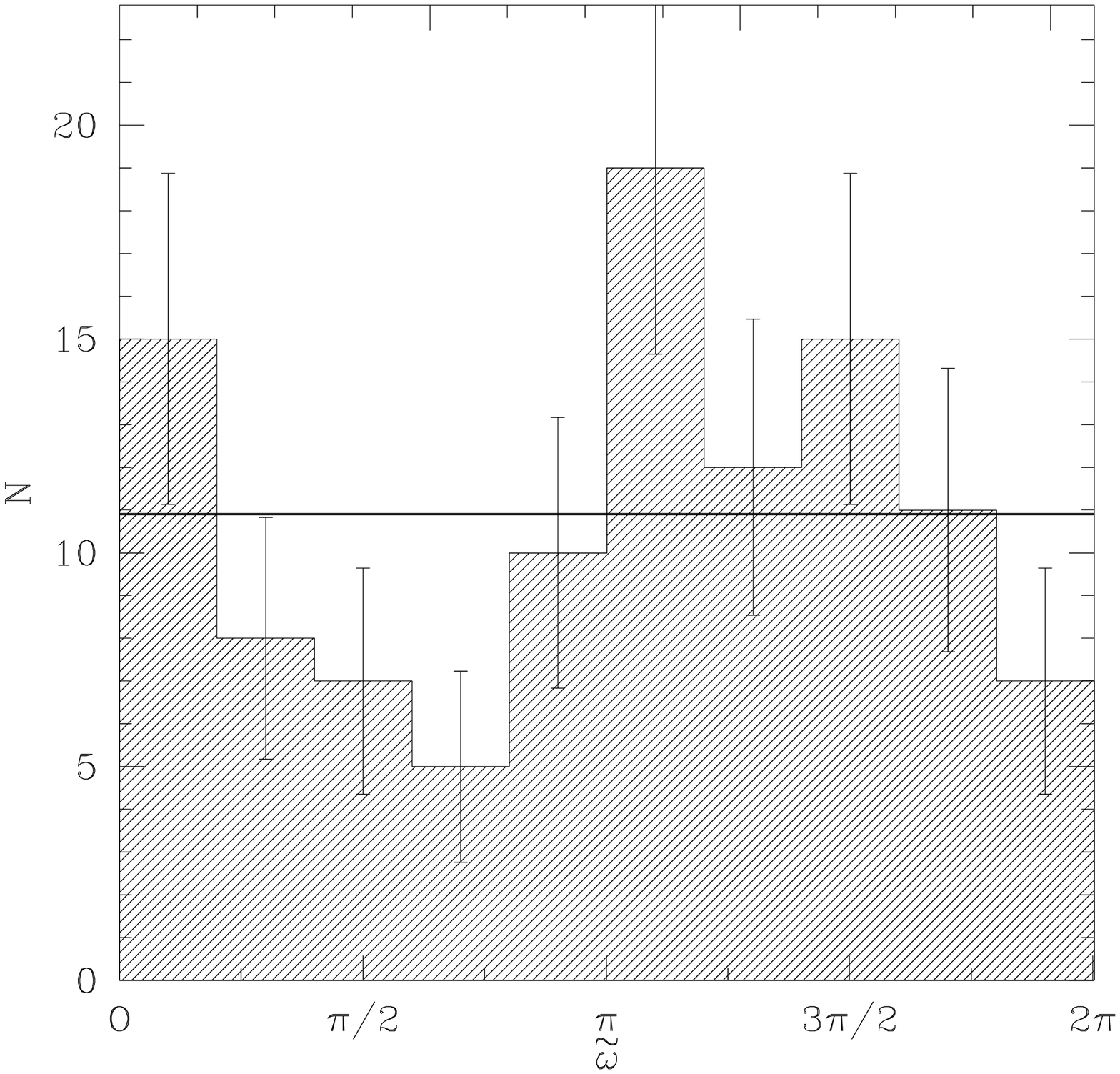,height=5.cm}}
\caption{The same as Fig.~\ref{WT-i}, but for the distribution of the argument
  of perihelion relative to the galactic plane. From \cite{WT99}.}
\label{WT-omega} 
\end{figure}

In \cite{WT99} the Oort cloud was modeled as a collection of objects with
10,000$<a<$50,000~AU, differential distribution $N(a){\rm d}a\propto a^{-1.5}$
and uniform distribution on each energy hyper-surface, consistent with an
earlier model of Oort cloud formation\cite{DQT87}. The evolution of the
comets was followed numerically, under the influence of the galactic disk's
tide and of the 4 giant planets, the latter assumed to be on coplanar circular
orbits. Stellar and GMCs passages, as well as the radial component of the
galactic tide, were neglected. Figure~\ref{WT-i}a shows the $\cos\tilde\iota$
distribution of the simulated comets at their first passage within 3~AU from
the Sun (the limit assumed for comet physical activity and visibility). The
distribution peaks at $\cos\tilde\iota=\pm 0.5$ and is relatively depleted at
$\cos\tilde\iota=\pm 1$ and 0. This is the signature of the galactic tide.
Comets with $\tilde\iota\sim 0^\circ$ (or equivalently, $\tilde\iota\sim
180^\circ$) have an oscillation of the perihelion distance which is too small
to bring them from the trans-planet region into the visibility region. Comets
with initial $\tilde\iota\sim 90^\circ$ have their inclination decreased to
lower values by the time that the perihelion distance is decreased below 3~AU.
Similarly, Figure~\ref{WT-omega}a shows the $\tilde\omega$ distribution. The
peaks at $\tilde\omega\sim 1/4\pi$ and $3/4\pi$ are, again, a signature of the
galactic tide. In fact, the precession of $\tilde\omega$ is counter-clockwise
and the minimal $q$ is achieved when $\tilde\omega=\pi/2, 3/2\pi$. Thus the
perihelion distance decreases below the imposed threshold $q=3$~AU when
$\tilde\omega$ is {\it en route} from 0 to $\pi/2$ or from $\pi$ to $3/2\pi$.
Figures~\ref{WT-i}b and~\ref{WT-omega}b show the same distributions for the
observed new comets. The observed and simulated distributions are quite
similar, which confirms the dominant role of the galactic tide. However, the
peak and valleys observed in the simulated distributions are not nearly as
pronounced. This suggests that the direct injection of comets from the Oort
cloud due to passing stars and/or GMCs (neglected in the simulation) has
non-negligible importance.

\begin{figure}[t!]
\centerline{\psfig{figure=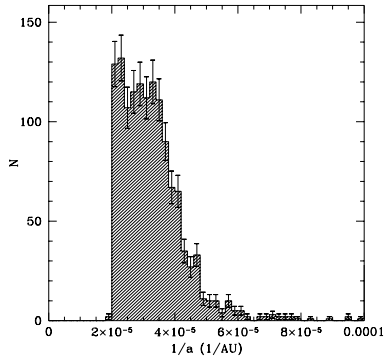,height=5.cm}}
\caption{The distribution of $1/a$ of the comets at their first appearance
  ($q<3$~AU) from the Oort cloud, according to \cite{WT99}. The sharp fall-off
  at $1/a=2\times 10^{-5}$~AU$^{-1}$ is due to the choice of the initial
  conditions ($a<50,000$~AU).}
\label{WT-a} 
\end{figure}

Figure~\ref{WT-a} shows the distribution of $1/a$ for the comets at their
first apparition, still according to the simulation in \cite{WT99}. Notice the
sharp fall off at $a\lesssim 20,000$~AU 
($1/a\gtrsim 5\times
10^{-5}$~AU$^{-1}$), that reproduces well the one observed in
the $1/a$ distribution of LPCs (see Fig.~\ref{spike}).  Thus, essentially all
comets at their first apparition have semi-major axis beyond 20,000~AU and
therefore would be classified as `new comets' by an observer. This sharp fall
off is due to the so-called {\it Jupiter barrier}. It is due to the fact that
new comets must have decreased their $q$ from $>10$~AU to $<3$~AU in less than
one orbital period, otherwise, they would have encountered Jupiter and Saturn
during an earlier evolution, and most likely they would have been ejected from
the Solar System. This condition is fulfilled only if the semi-major axis is
larger than $\sim$20,000~AU. The implication of this result is that LPCs do
not probe the Oort cloud inside this semi-major axis threshold, except during
rare showers due to a very close encounter of a passing star with the Solar
System (which allows a rapid decrease of $q$ even for $a<$20,000~AU; see
\cite{Heisler-shower}). Therefore, our information on the inner Oort cloud does
not come from the observations of comets, but solely from models of Oort cloud
formation (see sect.~\ref{Oort}).

From the fraction of the Oort cloud population that enters the visibility
region per unit time, and the flux of new comets with $H_{10}<11$ and $q<3$~AU
estimated from observations, \cite{WT99} concluded that the Oort cloud
population with $a>$20,000 AU and $H_{10}<11$ is $10^{12}$. This estimate
agrees with \cite{weissman96}, and is 2 times higher than that in
\cite{Heisler}, which gives a measure of its uncertainty. For the reason
explained just above, the estimated population in the Oort cloud with smaller
semi-major axis is totally dependent on its formation model.

\begin{figure}[t!]
\centerline{\psfig{figure=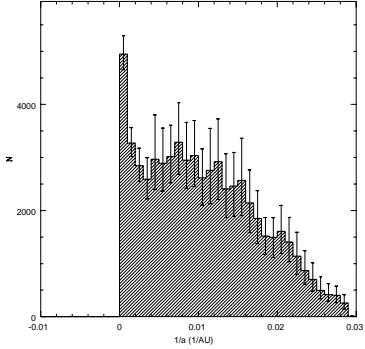,height=5.cm}}
\caption{The distribution of the inverse semi-major axis of all LPCs, 
independent on the number of perihelion passages within 3~AU that they
 already suffered, according to the simulation in \cite{WT99}. This
 distribution is very different from the observed distribution, illustrated on
 the same scale in Fig.~\ref{spike}}
\label{WT-ALLa} 
\end{figure}

The evolution of the comets, from their first apparition to their ultimate
dynamical elimination, has also been followed in \cite{WT99}. If the orbital
elements of all comets at every passage at $q<3$~AU are added up (without
limitation on the number of perihelion passages that they already suffered),
the resulting distribution of $1/a$ (Fig.~\ref{WT-ALLa}) is very different
from the observed distribution (Fig.~\ref{spike}). In particular, the ratio
between the number of comets in the Oort spike and the number of returning
comets is much smaller than observed. This problem was already pointed out in
\cite{Oort}. As suggested by Oort himself, this mismatch indicates that comets
from the Oort cloud have a very limited physical lifetime: after a few
perihelion passages they fade away from visibility, either by becoming inactive
or by disrupting. In \cite{WT99} it was shown that a very good match with the
observed distribution of LPCs can be achieved if one assumes that the
probability $P_m$ that a comet is still active after $m$ perihelion passages
within 3~AU decays as $m^{-0.6}$. This fading law implies that only 10\% of
the comets survive more than 50 passages and only 1\% of them survive more
than 2,000 passages. Other equally drastic fading laws, such as $P_m=1$ for
$m\le 6$ and $p_m=0.04$ for $m>6$ \cite{weissman78}, can also reproduce the
observed distribution of LPCs.

Therefore, the conclusion is that comets from the Oort cloud fade very
quickly, in just a few revolutions. This is a very different behavior with
respect to that of JFCs, which have a physical lifetime of $\sim 10,000$~y,
namely they remain active for about 1,000 revolutions. The fate of faded
comets (disruption versus inactivity) for both LPCs and JFCs is discussed in
sect.~\ref{fade-fate}.

\subsection{Note on Halley-type comets}
\label{HTC}

The Halley-type comets have been traditionally considered as the low semi-major axis end of the returning LPC distribution. Indeed, at a first glance,
the distribution of HTCs and of returning LPCs (apart from the semi-major axis
range that they cover) appear fairly similar. 

Under the effects of close encounters with Jupiter and Saturn, some returning
comets can have their semi-major axis decreased to less than 34.2~AU. At that
point, their orbital period becomes shorter than 200~y, so that, by
convention, they are classified as short period comets. They are predominantly
HTCs, and not JFCs, because their Tisserand
parameter relative to Jupiter is typically smaller than 2. The reason for this
is that new comets from the Oort cloud, having $q<3$,
$a\sim\infty$, $e\sim 1$ must have $T_J< 2.15$, and the Tisserand parameter
remains roughly conserved during the subsequent evolution down to the SPC
region, due to the predominance of Jupiter's scattering action.  The transfer
of comets from the Oort spike to the HTC region typically requires a large
number of revolutions. Thus, the HTCs should belong to the small fraction
($\sim 4$\%) of Oort cloud comets that do not fade away rapidly.

\begin{figure}[t!]
\centerline{\psfig{figure=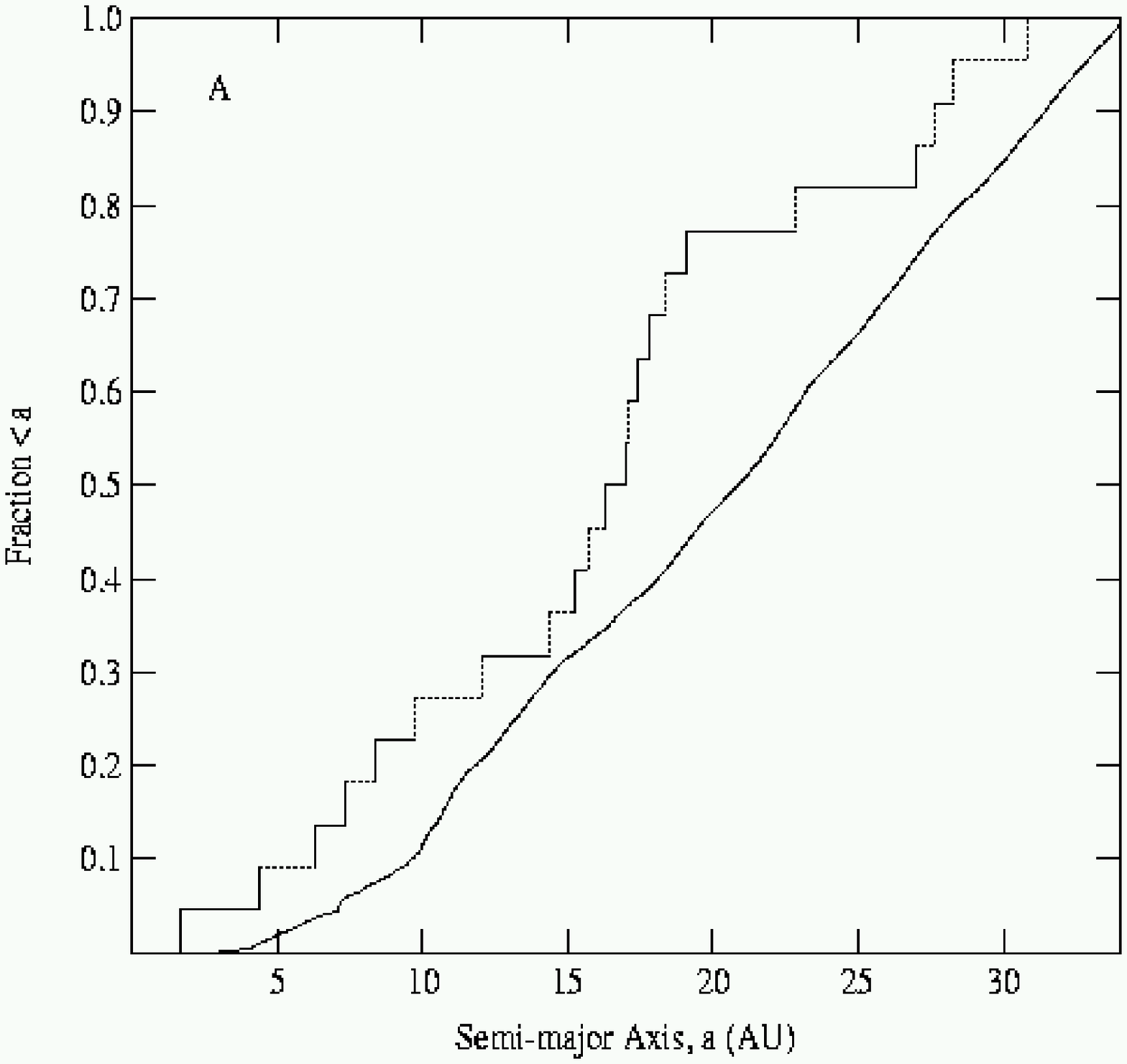,height=5cm}\psfig{figure=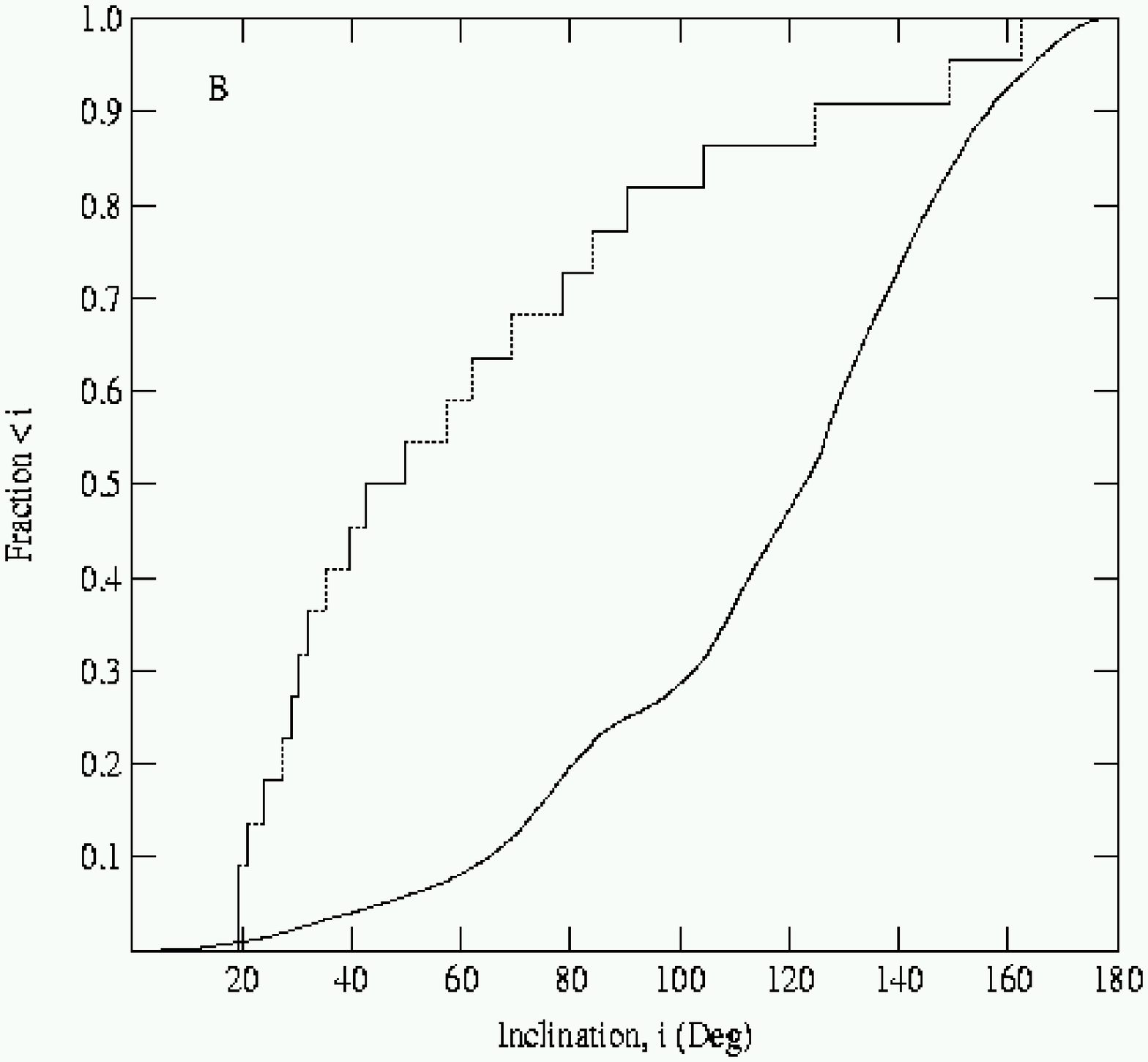,height=5cm}
}
\caption{Comparison between the cumulative orbital element distributions of
  the observed HTCs (dotted line) and those produced in the integrations of
  \cite{Hal-HTC1} (solid line). (a) Semi-major axis distributions ; (b)
  inclination distributions. Note the significant disagreement in the
  inclination distributions. Only comets with $q<1.3$~AU are considered.}
\label{HTCs} 
\end{figure}

This transfer process from the Oort cloud to the HTC region has been recently
revisited in \cite{Hal-HTC1}, using state of the art numerical simulations. It
was found that, if the semi-major axis distribution of the HTCs obtained in
the simulations matches fairly well the observed distribution, the inclination
distributions are profoundly different (Fig.~\ref{HTCs}). In particular, the
median inclination distribution of the observed HTCs is 45 degrees, and 80\%
of them have a prograde orbit, whereas the median inclination of the HTCs
obtained in the simulation is 120 degrees and only 25\% of them have prograde
orbit. The reason for which the simulated distribution is skewed towards 
retrograde objects is that the latter have a longer 
dynamical lifetime  (100,000~y, as opposed to 60,000~y for prograde HTCs). 

In \cite{Hal-HTC1} the solution that the authors proposed to solve the
mismatch between the inclination distributions was that part of the HTCs come
from the inner Oort cloud ($a<$20,000~AU) and that the latter has a disk-like
structure, with inclinations within 50 degrees from the ecliptic. However,
modern formation models of the Oort Cloud (see sect.~\ref{Oort} and
Fig.~\ref{DLDW-4}) show that retrograde orbits in the Oort cloud start to
appear beyond 6,000--7,000~AU, and a flattened region can be found only inside
this boundary in semi-major axis. However, this region is too tightly bound to
the Sun to be an aboundant source of comets.

In \cite{Hal-HTC2} it was proposed that part of the HTCs come from the distant
end of the scattered disk. They would be objects that, pushed outwards by
Neptune, eventually feel the galactic tide and have the perihelion decreased
deeper into the planetary region ($q<25$~AU). However, all these objects would
appear first as prograde long period comets, whereas there is no trace of an
overabundance of prograde comets in the new LPC population (Rickman, private
communication).

In conclusion, the problem of the inclination distribution of HTCs is
currently unsolved.  It is possible that part of the solution is that HTCs,
even if more resistant than new LPCs, cannot be active for more than
$\sim$~10,000~y, as it is the case for JFCs. This would bring the median value
of the inclination distribution of the simulated `active' comets down to $\sim
90^\circ$, or even less (\cite{FernGall}).  Moreover, the median inclination
of the currently observed HTCs might be smaller than in reality, due to
observational biases and/or small number statistics. In fact, an update of the
HTC catalog with respect to that used in \cite{Hal-HTC1}, shows an increase of
the median inclination from 45 to 60 degrees. In addition, the HTCs catalog
might be contaminated by a few prograde objects coming from the JFC population
(see sect.~\ref{JFC}). Finally, I notice that in the simulations of
\cite{WT99}, 65\% of the short period comets were on prograde orbits. Why this
result is different from that in \cite{Hal-HTC1} (25\%) is not clear. The
efficency of transfer of comets from the Oort cloud to the SPC region is very
small, so that it is possible that the results of any model based on numerical
simulations is dominated by small number statistics. Definitely, the issue of
the origin of HTCs needs to be investigated further, and a quantitative model
of their distribution remains to be done.

\subsection{The fate of faded comets}
\label{fade-fate}

We have seen in sections~\ref{JFC} and~\ref{LPC} that there is quite  strong
evidence that comets fade after a limited number of revolutions, and that the
rate at which they do so is different for JFCs and LPCs. What happens to the
faded comets? Do they remain on orbit around the Sun as dormant asteroid-like
objects, or do they disintegrate into smaller, undetectable pieces? 

To answer this question it is necessary to look for asteroid--like objects on
orbits typical of these comets, and compute if their number is consistent with
the one expected assuming that all faded comets are dormant and accounting for
the discovery biases.

Several Near Earth Asteroids (NEA) have been discovered on orbits typical of
JFCs, with $2<T_J<3$. The NEA model developed in \cite{BottkeNEA}, calibrated
on Spacewatch discoveries, argues that the asteroid belt is not a sufficient
source of these objects. This model implies that, among the NEA population,
$60\pm 40$ objects with $H<18$ are dormant JFCs. A similar model
(\cite{whitman}), developed using the more extended dataset provided by the
LINEAR survey, estimates $\sim 70$ dormant JFCs in the NEA population in the
same magnitude range (for comparison, the total number of NEAs with $H<18$ is
estimated to be $\sim 1,200$ \cite{Stuart}). 
Assuming 4\% albedo --typical of cometary nuclei without
activity--, $H=18$ corresponds to $D=1.7$~km. As we have seen in
sect.~\ref{JFC}, \cite{DL97} estimate the existence of $\sim 100$ faded JFCs
in the NEO region with diameter of about 2.0~km.

An independent confirmation that many/most NEAs with $T_J<3$ are dormant
comets come from spectroscopic observations (\cite{yan}, \cite{StuartBinzel}),
which show that the albedo distributions of the NEAs with, respectively,
$T_J>3$ and $T_J<3$ are totally different. The latter have much darker albedos
than the former. In conclusion, there is solid evidence that at least a
significant fraction of JFCs --if not all-- become dormant when they fade.

The situation is totally different for LPCs and HTCs, as shown in
\cite{Hal-faded}. The steep fading law required to explain the observed number
ratio between new and returning comets (see sect.~\ref{LPC}) implies that for
every active returning comet there should be 20 faded comets. Thus, if all
faded comets were dormant, the model in \cite{WT99} would imply the existence
of $4\times 10^6$ objects, with $q<3$~AU and semi-major axis distribution
similar to that of Fig.~\ref{WT-ALLa}. Again, the absolute magnitude $H$ of
these objects, corresponding to comets with $H_{10}<11$ when in activity, is
very uncertain.  Assuming that they have $H<18$ --with cumulative distribution
$N(<H)\propto 10^{0.28 H}$ as in \cite{WeissmanLowry}-- \cite{Hal-faded}
estimated that 1 object out of 20,000 should have been discovered by asteroid
surveys, namely $\sim 200$ objects. However, only 2 `asteroidal' objects have
been discovered on LPC orbits. Similarly, if all faded comets were dormant,
the model in \cite{Hal-HTC1} estimates that there should be 100,000 inactive
HTCs with $H<18$ and $q<2.5$~AU. Of these, \cite{Hal-faded} estimated that
1,000 should have been discovered by asteroid surveys. This estimate is again
100 times larger than the number of actual discoveries of `asteroids' on
corresponding orbits. Thus, the conclusion seems to be that only $\sim 1$\% of
the comets from the Oort cloud become dormant when they fade. The remaining
99\% apparently split in smaller undetectable fragments (like in the case of
comet LINEAR C/2001 A2), if not into dust trails.

In summary, JFCs and LPCs seem to fade away due to different physical
processes.  This may be surprising, given that both are thought to be similar
mixtures of ice and rock.  However, evolutionary processes could affect
comets' susceptibility to disruption.  For example, over long timescales, JFCs
could have lost more volatiles than LPCs because they have been stored in the
scattered disk, at closer heliocentric distances and thus higher temperatures
than in the Oort cloud.  JFCs could be more porous, and thus less susceptible
to disruption resulting from volatile pressure buildup, due to a relatively
violent collisional environment.  Finally, the dynamical pathways that LPCs
and JFCs take on their way into the inner Solar System might lead to very
different thermal histories for the two populations.  To jump over the Jupiter
barrier, in one orbital period LPCs have to evolve from very distant orbits
(with perihelia outside the planetary region) to orbits that closely approach
the Sun.  On the other hand, objects from the scattered disk slowly move
through the planetary region, taking $\sim 10\,$My to evolve onto orbits with
$q<2.5AU$ (\cite{LD97}).  Perhaps LPCs disrupt because of strong thermal
gradients or volatile pressure buildup, while JFCs survive because they are
warmed more slowly.

\section{The formation of the Oort cloud}
\label{Oort}

For the formation of the Oort cloud it is intuitive to invoke the mechanism
described in the previous section for the origin of LPCs, but `played' in
`reverse mode'. Imagine an early time when the Oort cloud was still empty and
the giant planets' neighborhoods were full of icy planetesimals.  The
scattering action of the planets dispersed the planetesimals throughout the
solar system. Some were moved onto eccentric orbits with large semi-major
axis, but with perihelion distance still in the planetary region. Those of
them which reached a semi-major axis of $\sim$10,000~AU started to feel a
galactic tide strong enough to modify their orbit on a timescale of an orbital
period.  During the scattering process, these planetesimals remained
relatively close to the ecliptic plane, so that their inclination relative to
the galactic plane $\tilde\iota$ was $\sim 120^\circ$. Due to their large $e$
and $\tilde\iota$ the effect of the tide on the evolution of $e, \tilde\iota$
was prominent. The planetesimals with $\tilde\omega$ between $90^\circ$ and
$180^\circ$ (or, symmetrically, between $270^\circ$ and $360^\circ$) had their
eccentricity decreased. This lifted their perihelion distances beyond the
planets' reach, so that they could not be scattered any more: they became Oort
cloud objects. The precession of $\tilde\Omega$ and the random passage of
rogue stars randomized the planetesimals' distribution, giving to the Oort
cloud the structure that is inferred from LPCs observations.

This scenario, originally proposed in \cite{kuiper}, was first simulated in
\cite{fernandez1978}, \cite{fernandez1980} using a Monte Carlo method to
represent the effects of repeated, uncorrelated encounters of the
planetesimals with the giant planets and passing stars (the role of the
galactic tide was not yet taken into account). The first simulation of Oort
cloud formation using direct numerical simulations and accounting for the
galactic tide was done in \cite{DQT87}. To save computing time, however, the
simulations were started with comets already on low inclination, high
eccentricity orbits: initial $a=$2,000~AU
and $q$ uniformly distributed between 5 and 35~AU.

The Oort cloud formation has been recently revisited in \cite{DLDW} (see also
\cite{Dones-CII}), using more modern numerical simulation techniques. They
started with more realistic initial conditions, assuming planetesimals
initially distributed in the 4--40~AU zone with small eccentricities and
inclination. The giant planets were assumed to be on their current orbits, and the
migration of planets in response to the dispersion of the planetesimals (see
sect.~\ref{sculpting}), was neglected.  The evolution of the planetesimals was
followed for 4~Gy, under the gravitational influence of the 4 giant planets,
the galactic tide (both radial and disk components - see (\ref{radialtide}),
(\ref{disktide})-), and passing stars. Both the tide and the statistics of
passing stars were calibrated using the current galactic environment of the
Sun. A stellar density of $0.041 M_\odot$/pc$^3$ was assumed, with stellar
masses distributed in the range 0.11--18.24~$M_\odot$ and relative velocities
between 1.7 and 158 km/s (with a median value of 46 km/s).  In total the
simulation in \cite{DLDW} recorded $\sim$50,000 stellar encounters within 1~pc
from the Sun, in 4~Gy. In the following discussion of Oort cloud
formation, I mostly refer to the results of this work.

\begin{figure}[t!]
\centerline{\psfig{figure=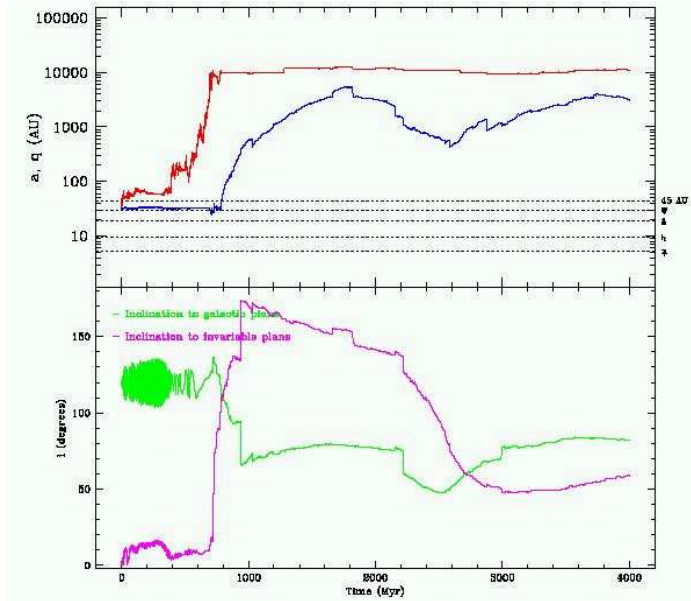,height=8.cm}}
\caption{An example of evolution of a comet from the vicinity of Neptune into
  the Oort cloud, from \cite{DLDW}. The top panel shows the evolution of the
  object's semi-major axis (red) and perihelion distance (blue). The bottom
  panel shows the inclinations relative to the galactic plane (green) and
  Solar System invariable plane (the plane
    orthogonal to the total angular momentum of the planetary system; in
  magenta).
  }
\label{DLDW-14} 
\end{figure}

Fig.~\ref{DLDW-14} shows an example of the evolution of a comet from the
neighborhood of Neptune to the Oort cloud. With a sequence of encounters, the
object is first scattered by Neptune to larger semi-major axis, while keeping
the perihelion distance slightly beyond 30~AU, as typical of scattered-disk
bodies.  After about 700~My, the random walk in semi-major axis brings the
body's semi-major axis to $\sim 10,000$ AU. At this time the galactic tide
starts to be effective, and the perihelion distance is rapidly lifted above
45~AU. Neptune's scattering action ceases and the further changes in
semi-major axis are due to the effects of distant stellar encounters. When the
body starts to feel the galactic tide, its inclination relative to the
galactic plane is 120 degrees. As the perihelion distance is lifted (namely
the eccentricity decreases), the inclination decreases towards 90
degrees\footnote{Notice that, for the dynamical evolution forced by the
  galactic disk tide, the decrease of $\tilde\iota$ from 120 to 90 degrees, is
  equivalent to an increase from 60 to 90 degrees, in agreement with what has
  been said in the previous section on the anti-correlation of the evolutions
  of eccentricity and inclination.}. A stellar passage causes a sudden jump of
$\tilde\iota$ down to $65^\circ$ just before $t=1$~Gy. This allows the tide to
enhance its action, bringing the perihelion distance of the object beyond
1,000~AU and the inclination $\tilde\iota$ up to $80^\circ$. This
configuration is reached at $t=1.7$~Gy, when $\tilde\omega$ is 0 or 180
degrees. From this time onwards the galactic tide reverses it action,
decreasing $q$ and $\tilde\iota$. In principle the action of the galactic tide
is periodic, so that the object's perihelion should be decreased back to
planetary distances. However, the jumps in $a, q, \tilde\iota$ caused by the
stellar encounters break this reversibility. The oscillation of $q$ becomes
more shallow and the object never comes back into the planetary region within
the age of the Solar System.  Notice finally that during this evolution, the
inclination relative to the invariable plane is strongly changed. It is turned
to retrograde, and then back to prograde values, as the longitude of galactic
node $\tilde\Omega$ precesses.

Not all particles follow this evolution, though. Those which come to interact
closely with Jupiter and Saturn are mostly ejected from the Solar System. Those
which have distant encounters with Saturn are transported more rapidly and
further out in semi-major axis with respect to the evolution shown in
Fig.~\ref{DLDW-14}. The strength of the galactic tide increases with $a$;
thus, for the comets that are scattered to $a\sim$20,000y or beyond, the
oscillation period of $q$ and $\tilde\iota$ is shorter than for the particle
in Fig.~\ref{DLDW-14}.

\begin{figure}[t!]
\centerline{\psfig{figure=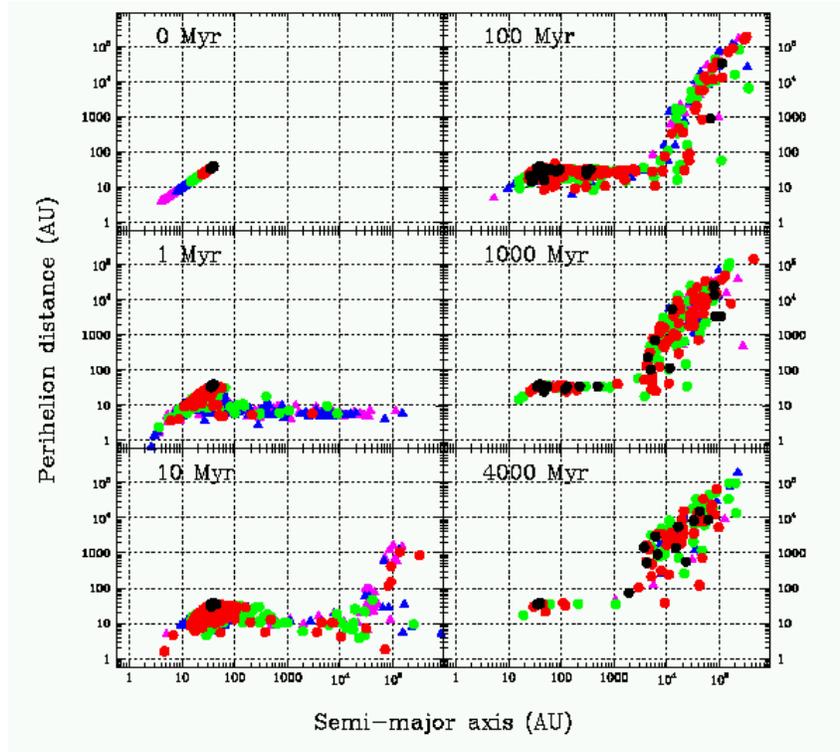,height=10.cm}}
\caption{Scatter plot of osculating barycentric pericenter distance
  vs. osculating barycentric semi-major axis, at various times in the Oort
  cloud formation simulations of \cite{DLDW}. The points are color-coded to
  reflect the region in which the simulated comets formed. 
  Each panel is labeled by the simulation time that it corresponds to.}
\label{DLDW-3} 
\end{figure}

\begin{figure}[t!]
\centerline{\psfig{figure=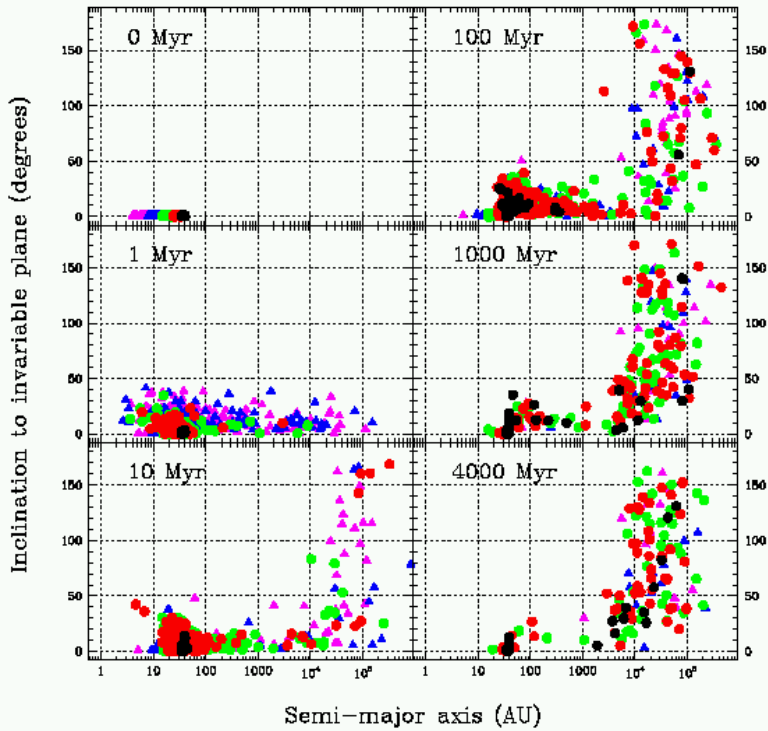,height=10.cm}}
\caption{The same as Fig.~\ref{DLDW-3} but for  
osculating barycentric inclination relative to the Solar System mid-plane 
  vs. osculating barycentric semi-major axis. From \cite{DLDW}. }
\label{DLDW-4} 
\end{figure}

Figures \ref{DLDW-3} and \ref{DLDW-4} give a global illustration of the Oort
cloud formation process, showing snapshots of the $(a,q)$ and $(a,i)$
distributions of all planetesimals at 0 (initial conditions), 1, 10, 100~My
and 1, 4~Gy.  The planetesimals in these plots are color-coded according to
their initial position: Jupiter region objects are magenta; Saturn region
objects are blue; Uranus region objects are green; Neptune region objects are
red and trans-Neptunian objects are black. Figure~\ref{DLDW-3} shows that,
after only 1~My, a scattered disk is formed by Jupiter and Saturn, out of
particles initially in the Jupiter-Uranus region. This scattered disk differs
from the current scattered disk because most of its objects have $q<10$~AU.
Particles originally in Neptune's region or beyond have not been scattered out
yet. At 10~My a signature of the galactic tide starts to be visible. The Oort
cloud begins to form. Particles with $a>30,000$, mostly from Jupiter-Saturn
region, have their perihelia lifted beyond the orbits of the planets. Neptune's
particles start to populate the scattered disk. From 100My to 1Gy, particles
continue to enter the Oort cloud from the scattered disk. The population of
the Oort cloud peaks at 840 My, at which time 7.55\% of the initial particles
occupy the cloud. Objects from the Uranus-Neptune region gradually replace
those from Jupiter-Saturn zone. The latter have been lost during stellar
encounters, as they predominantly occupied the very outer part of the Oort
cloud ($a>30,000$~AU). Due to the longer time over which the galactic tide
has acted and to stellar encounters, the population of bodies with perihelion
distances above 100~AU can have semi-major axes as low as 3,000~AU. The Oort
cloud with $a<20,000$~AU is usually called the inner Oort cloud, or Hills
cloud from \cite{Hills}. The last panel in Fig.~\ref{DLDW-3}, representing the
distribution at 4Gy, should correspond to the current structure of the Oort
cloud.  The distribution remains nearly the same as that at 1 Gy, but the Oort
cloud population has declined slightly in number.

Figure~\ref{DLDW-4} shows the evolution of the particles inclinations. After 1
Myr the planets have scattered the comets into moderately inclined orbits.
After 10 My the particles with $a>$30,000~AU have been perturbed by the
galactic tide and passing stars into a nearly isotropic distribution of
inclinations. As time continues, tides affect the inclinations of particles
closer to the Sun, so that at 4,000 My inclinations are clearly isotropic for
$a>$20,000~AU.

The final Oort cloud contains roughly equal populations in the inner and outer
parts, with radial distribution $N(r){\rm d}r\propto 1/r^3$. About 5\% to 9\%
of the planetesimals initially in the Uranus-Neptune-transneptunian region
remain in the Oort cloud at the end of the simulation. Conversely, only 2\% of
the planetesimals originally in the Jupiter-Saturn region do so. The
scattering action of these planets is too strong to deposit a large fraction
of planetesimals in the Oort cloud.  The reason is the same as invoked to
explain the Jupiter's barrier for the new-LPCs distribution (see
sect.~\ref{LPC}). In energy space, the Oort cloud is $10^{-4}$ wide, whereas
the random walk in energy of particles scattered by Jupiter and Saturn has
steps of width $\sim 10^{-3}$, e.g. proportional to the masses of these
planets relative to that of the Sun. Thus, most of the particles scattered by
these planets go directly from a scattered-disk orbit (Energy$<-10^{-3}$) to
unbound orbit (Energy$>0$), without passing through the Oort cloud
($-10^{-4}<$Energy$<10^{-3}$).

\begin{figure}[t!]
\centerline{\psfig{figure=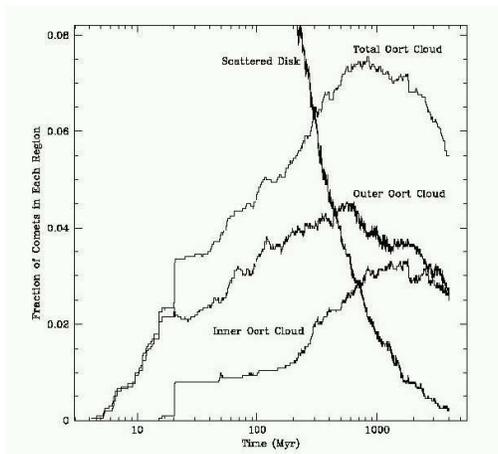,height=6.cm}}
\caption{Fraction of the initial planetesimal population that is
  in the Oort cloud, in its inner and outer parts and in the scattered
  disk, as a function of time. From \cite{DLDW}. }
\label{DLDW-11} 
\end{figure}
Figure~\ref{DLDW-11} shows the evolution of the mass in the Oort cloud as a
function of time. The formation and the erosion of the Oort cloud are not
separate processes. Throughout the Solar System history, in parallel with new
planetesimals entering the Oort cloud from the scattered disk, other comets
left the cloud, because the galactic tide pushed their perihelion back into
the planetary region or passing stars put them on hyperbolic orbits. Thus, on
the one hand, the flux of LPCs started as soon as the first planetesimals
reached $\sim$10,000~AU (10~My). On the other hand the supply of new objects
to the Oort cloud is still ongoing today \cite{fern-SD}. However, as mentioned
above, the mass in the cloud peaks at about 800~My. Before this date the
formation process dominated over the erosion process. Then --because the mass
of the scattered disk dropped-- the erosion process became predominant, and
the total mass in the cloud decayed to $\sim 5.5$\% of the mass originally in
the planetesimals' disk.  The outer Oort cloud formed faster then the inner
cloud --because of the contribution of planetesimals from Jupiter-Saturn
region-- but then it eroded faster, because its objects are less
gravitationally bound to the Sun.

\subsection{Problems with the classical scenario}
\label{Oort-problems}

The classical scenario of Oort cloud formation discussed above meets two
problems, when confronted with the quantitative constraints provided by 
the current Solar System. 

As we have seen in sect.~\ref{LPC}, the outer Oort cloud should currently
contain $10^{12}$ comets with $H_{10}<11$.  The estimates of the nuclear size
of a $H_{10}=11$ comet range from 1~km \cite{BaileyStagg} to
2.3~km\cite{weissman96}.  Assuming, as in \cite{Hal-faded} that a $H_{10}=11$
comet has $D\sim 1.7$km, and assuming also a cumulative size distribution
proportional to $D^{-2}$ and a density of 0.6g/cm$^3$ (as for P/Halley), one
obtains a total mass of $3\times 10^{28}$g, namely $3 M_\oplus$.  Because the
overall efficiency of formation of the outer Oort cloud is small (2.5\%), this
implies that the original planetesimal disk in the Jupiter-Neptune region was
$\sim 100 M_\oplus$. This seems rather high compared to the total mass of
solids associated with the minimal mass solar nebula \cite{hayashi}.  Also,
numerical simulations show that a planetesimal disk more massive than
30--50$M_\oplus$ would have driven Neptune beyond 30~AU and Jupiter and Saturn
would have passed across their mutual 2:5 mean-motion resonance (see
sects.~\ref{sculpting} and~\ref{LHB}). The uncertainty in the conversion
between $H_{10}$ magnitude and size, however, allows enough wiggle room to
make consistent estimates. For instance, it would be enough that the nuclear
size of $H_{10}=11$ comets is 1.3~km (instead of the assumed 1.7) to bring the
required mass of the planetesimal disk to a more reasonable value of $50
M_\oplus$.

A second, more compelling problem concerns the number ratio between the comet
populations in the Oort cloud and in the scattered disk. We have seen in
sect.~\ref{JFC} that the scattered disk, to be a sufficient source of JFCs,
has to contain $4\times 10^8$ comets with $H_{10}<9$. The number of comets
with $H_{10}<11$ depends on the exponent of the $H_{10}$ distribution of
comets, which is highly debated.  Using the largest value available in the
literature (0.7 \cite{fernandez99}) the scattered disk should have $10^{10}$
$H_{10}<11$ comets. Using the exponent for the nuclear magnitude distribution
in \cite{WeissmanLowry}\cite{Hal-faded} (0.28) and assuming a linear scaling
between nuclear magnitude and $H_{10}$, the number of $H_{10}<11$ comets in
the scattered disk reduces to $1.5\times 10^9$. Because the number of comets
in the outer Oort cloud with $H_{10}<11$ is $10^{12}$, the comet number ratio
{\it inferred from observations} between the outer Oort cloud and the
scattered disk is in the range 100--1,000. However, in the simulations in
\cite{DLDW}, the final ratio is $\sim 10$ (see Fig.~\ref{DLDW-11}).

The way out of this problem is much more difficult than for the total mass
problem. The discrepancy does not depend on assumed relationships between
total magnitude and size, nor on density. It cannot be alleviated with any
reasonable assumption of the exponent of the $H_{10}$ distribution. Also,
different assumptions of the initial planetesimal distribution in the disk
would not help. The point is that most of the Oort cloud is made of
planetesimals from the Uranus--Neptune--trans-Neptunian zone, which have to
pass through the scattered disk to reach the cloud. Thus, there is a causal
relationship between the final numbers of comets in the scattered disk and
Oort cloud. To change this relationship, it would be necessary that a much
larger number of planetesimals could reach the Oort cloud without passing
through the scattered disk. This requires that Jupiter and Saturn were more
effective in the real Oort cloud building process than in the simulations of
\cite{DLDW}. A possible scenarioin which this can occur is discussed below.

\subsection{Oort cloud formation in a dense galactic environment}
\label{dense}

It is now known that most stars form in clusters.  In \cite{fernandez97} it
was pointed out that a denser galactic environment would have exerted a
stronger tide on the scattered planetesimals. In addition, stellar encounters
would have been more effective, because of the slower relative velocities and
smaller approach distances typical of a cluster environment. As a consequence,
the threshold semi-major axis value beyond which planetesimals could be
decoupled from the planets would have been $\sim 1,000$~AU, instead of the
current value of $\sim$10,000~AU. In other words, the Oort cloud would have
extended closer to the Sun, covering the region with binding energy down to
$-10^{-3}$ in normalized units. Because this width is of the same order os the
energy change suffered by planetesimals crossing the orbits of Jupiter and
Saturn, the role of these gas giants in building the Oort cloud would 
be greatly enhanced.

Simulations of Oort cloud formation in a dense environment have been done in
\cite{fern-brunini}. Three kinds of environments were considered: (i) a loose
cluster with 10 stars/pc$^3$; (ii) a dense cluster with 25 stars/pc$^3$ and
(iii) a super-dense cluster with 100 stars/pc$^3$. In all cases, all stars
solar were assumed to have a solar mass (compare with the current stellar
density of $0.041 M_\odot$/pc$^3$ \cite{DLDW}). The average relative velocity
among the stars was assumed to be 1~km/s, typical of star
clusters~\cite{BinneyTrem} (instead of the current $\sim 40$~km/s). In
addition, a placental molecular cloud containing $10^5$ molecules of
Hydrogen per cm$^3$ was assumed (the current molecular density is$\sim
3$~g/cm$^3$). The initial conditions of the planetesimals were similar to
those in \cite{DQT87}. Comets were placed on
initial orbits with $100<a<250$~AU and $q$ ranging from 4 to 30 AU.

\begin{figure}[t!]
\centerline{\psfig{figure=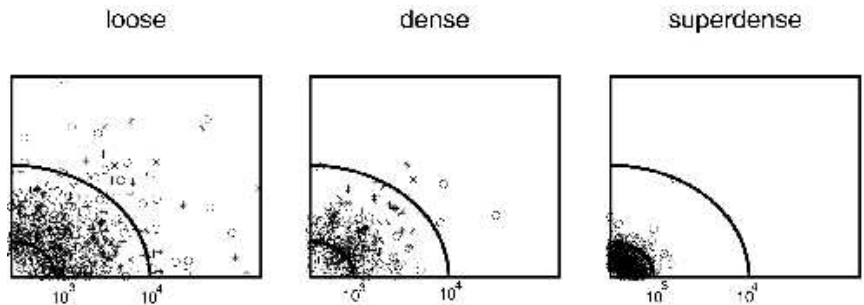,height=4.cm}}
\caption{A sketch showing how comets trapped in the Oort cloud would appear
  distributed in the circumsolar space, for three kinds of star clusters
  surrounding the Sun.  The radii of the circles are expressed in AU. Stars
  denote comets coming from Jupiter-Saturn zone, while open circles denote 
bodies from the Uranus-Neptune zone. From \cite{fern-brunini}. }
\label{FB} 
\end{figure}

Fig.~\ref{FB} shows the result of these simulations. As expected, the denser
the cluster, the more bound to the Sun becomes the resulting Oort cloud. Notice
however that the outer part of the cloud (beyond $10^4$AU) becomes totally
empty, because all comets beyond this limit are stripped off by the passing
stars. Thus, a mechanism would be required to transfer the comets from the
massive inner Oort cloud to the outer cloud, in order to explain the current
flux of LPCs (which come from the outer cloud only). Less effective stellar
encounters, occurring during the dispersal of the cluster and in the current
galactic environment, might be responsible for this process.

In terms of efficiency of Oort cloud formation, \cite{fern-brunini} found that
about 30\% of the initial planetesimals were trapped in the cloud, namely a
factor of 6 higher than in \cite{DLDW}. However, this new efficiency is of the
same order of that found in \cite{DQT87}, which used initial conditions
similar to those in \cite{fern-brunini}, but no star cluster. Thus, it is
unclear if the difference in efficiency between \cite{fern-brunini} and
\cite{DLDW} is due to the different choice of initial conditions (in which
case the efficiency in \cite{DLDW} is more accurate because the initial
conditions are more realistic) or to the presence of the cluster. Moreover,
conversely to what expected, the final contribution of Jupiter and Saturn to
the formation of the Oort cloud (i.e. the fraction of the planetesimal
population with initial $q<10$~AU that ended in the cloud) was minimal. This
happened because the planetesimals scattered by Jupiter and Saturn typically
ended in the outer part of the cloud, and were subsequently stripped away by
the numerous stellar encounters.

Thus, at the current state of the art, the formation of the Oort cloud in a
stellar cluster has not yet been proven to be the solution for the Oort
cloud/scattered disk ratio problem.  More precise simulations --starting from
a disk of planetesimals on low eccentricity orbits as in \cite{DLDW}-- are
required.  Moreover it would be more realistic to do these simulations
accounting for gas-drag, given that the gas-disk was present for most of the
time that the Sun spent in the cluster. Gas drag could protect comets from
ejection (Levison, private communication), thus increasing the fraction of
planetesimals from Jupiter--Saturn zone that are trapped in the cloud.
Moreover, it would be necessary to quantify more precisely which mechanism
could transfer the comets from the massive inner Oort cloud --produced in the
dense environment-- to the outer Oort cloud --where comets must reside
at the current time to produce LPCs.

\paragraph{Sedna: an inner Oort cloud object?\,}

One piece of evidence for a moderate stellar cluster surrounding the early Sun
is provided by Sedna. The distribution of the extended scattered-disk bodies
shows a clear tendency: the perihelion distance is larger for bodies with
larger semi-major axis. The bodies in the 50--100~AU range have $q<41.5$~AU;
2000~CR$_{105}$ ($a=222$~AU) has $q=44.3$~AU and Sedna ($a=495$~AU) has
$q=76$~AU. Although only a few such bodies are known --and one should be
careful about small number statistics-- the lack of objects with perihelion
distances comparable to those of 2000~CR$_{105}$ and Sedna but smaller
semi-major axes seems significant. In fact, observational biases (given an
object's perihelion distance and absolute magnitude, and a survey's limiting
magnitude of detection) sharply favor the discovery of objects with smaller
semi-major axes. So, it would be unlikely that the first two discovered bodies
with $q\!>\!44$~AU have $a\!>\!200$~AU if the real semi-major axis
distribution in the extended scattered disk were skewed toward smaller $a$.

\begin{figure}[t!]
\centerline{\psfig{figure=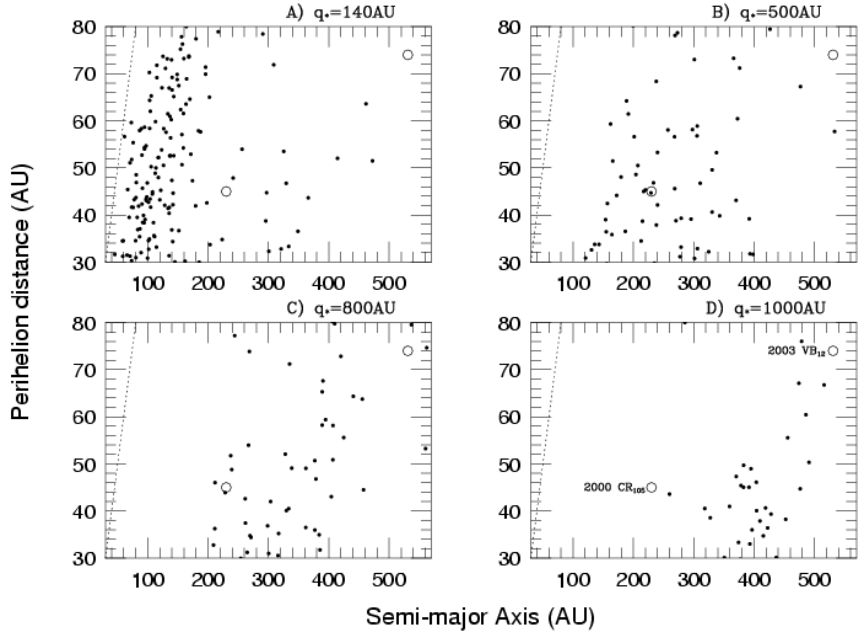,height=10.cm}}
\caption{The extended scattered disk that results from passing stars. 
  In all cases the passing star had 1~$M_\odot$ and was on a hyperbolic orbit
  with relative velocity of 1~km/s. Only the perihelion distance of the
  stellar orbit is varied from panel to panel. The particles were intially in
  the scattered disk created in the simulation of \cite{DLDW}, at $t=10^5$~y.
  The two open circles show the orbits of Sedna and 2000~CR$_{105}$. From
  \cite{morbCR105}. }
\label{cr105} 
\end{figure}
Assuming that the extended scattered-disk bodies belonged to the scattered
disk until a perturbation lifted their perihelion distance beyond Neptune's
reach, the fact that $q$ increases with $a$ is a clear signature that the
perturbation had to grow in magnitude with increasing heliocentric distance.
Passing stars produce this very signature
\cite{fern-brunini}\cite{morbCR105}\cite{rickmanCR105}.  In particular, it was
shown in \cite{morbCR105} that an encounter with a Solar mass star at 800~AU
with an unperturbed relative velocity of 1~km/s (see Fig.~\ref{cr105}) would
have produced a distribution of extended scattered-disk objects that overlaps
the orbits of Sedna and 2000~CR$_{105}$ and does not extend to smaller
semi-major axes. Closer stellar encounters would still produce a distribution
overlapping the orbits of Sedna and 2000~CR$_{105}$, but such distribution
would extend to smaller semi-major axes, inconsistent with the lack of
detections of large-$q$ bodies at small $a$. More distant encounters would not
reproduce the orbits of Sedna and/or 2000~CR$_{105}$ (see Fig.~\ref{cr105}).
The best `fit distance' of the stellar encounter depends on the stellar mass.
A star with $M=1/4M_\odot$ should have passed at $\sim 400$~AU in order to
produce a distribution similar to that in Fig.~\ref{cr105}C.

Stellar encounters at such short distances from the Sun were statistically
possible only if the Sun was embedded in a cluster, supporting the necessity
of building models of Oort cloud formation in the framework of a dense galactic
environment. If this view is correct, then the outer part of the extended
scattered disk smoothly joins the inner Oort cloud. In particular, Sedna
could be considered the first discovered object in the inner Oort cloud! 

\section{The primordial sculpting of the Kuiper belt}
\label{sculpting}

In section~\ref{KB} I have shown that many properties of the
Kuiper belt cannot be explained in the framework of the current Solar System:
\begin{itemize}
\item[i)] the existence of the resonant populations, 
\item[ii)] the excitation of the eccentricities in the classical belt, 
\item[iii)] the co-existence of a cold and a hot population with 
different physical properties, 
\item[iv)]the presence of an outer edge at the
location of the 1:2 mean-motion resonance with Neptune, 
\item[v)]the mass deficit of the Kuiper belt, 
\item[vi)]the existence of the extended scattered-disk population (with the
  exception of 2000~CR$_{105}$ and Sedna, whose orbits can be explained in the
  framework of the Oort cloud formation in a dense galactic environment, as
  just discussed above). 
\end{itemize}

These puzzling aspects of the trans-Neptunian population reveal that the
latter has been sculpted when the Solar System was different, due to
mechanisms that are no longer at work.  Like detectives ot the scene of a
crime, trying to reconstruct what happened from the available clues, 
astronomers have tried to reconstruct how the Solar System formed and evolved
from the traces left in the structure of the Kuiper belt.
A large number of mechanisms have been proposed so far to explain some of the
properties of the Kuiper belt listed above. For space limitation, here I 
debate only those which, in my opinion --in light of our current
observational knowledge of the Kuiper belt-- played a role in in the primordial
sculpting of the trans-Neptunian population. I will try to put the various
scenarios together, in order to build-up a consistent view of the primordial
sculpting of the Kuiper belt. For a 
more exhaustive review see \cite{MB}. 

\subsection{Origin of the resonant populations}
\label{resonant}

\begin{figure}[t!]
\centerline{\psfig{figure=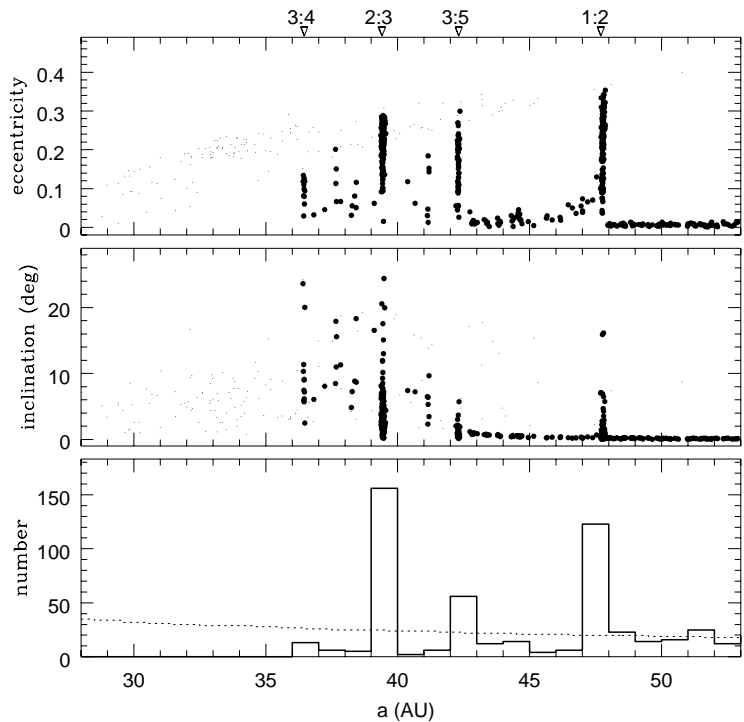,height=11.5cm
%,bbllx=10cm,bblly=-20cm,bburx=20cm,bbury=0cm
}}
\caption{Final distribution of the Kuiper belt bodies according to the
  sweeping resonances scenario (courtesy of R.~Malhotra). The simulation is
  done by numerically integrating, over a 200~My time-span, the evolution of 800
  test particles on initial quasi-circular and coplanar orbits. The planets
  are forced to migrate (Jupiter: -0.2~AU;, Saturn: 0.8~AU; Uranus: 3~AU;
  Neptune: 7 AU) and reach their current orbits on an exponential timescale of
  4~My.  Large solid dots represent `surviving' particles (i.e., those that
  have not suffered any planetary close encounters during the integration
  time); small dots represent the `removed' particles at the time of their
  close encounter with a planet (e.g. bodies that entered in the scattered
  disk and whose evolution was not followed further).  In the lowest panel,
  the solid line is the histogram of semi-major axes of the `surviving'
  particles; the dotted line is the initial distribution. The locations of the
  main mean-motion resonances are indicated above the top panel.}
\label{renu} 
\end{figure}

It was shown in \cite{FernIp} that, while scattering away the primordial
planetesimals from their neighboring regions, the giant planets had to migrate
in semi-major axis as a consequence of angular momentum conservation.  Given
the giant planets' configuration in our Solar System, migration should have had
a general trend. As discussed above concerning Oort cloud formation,
the ice giants have difficulty in ejecting planetesimals on hyperbolic orbits.
Apart from the few percent of planetesimals that they can permanently store in
the Oort cloud or in the scattered disk, the remaining planetesimals (the
large majority) are eventually scattered inwards, towards Saturn and Jupiter.
Thus, the ice giants, by reaction, have to move outwards. Jupiter, on the
other hand, eventually ejects from the Solar System almost all of the
planetesimals that it encounters: thus it has to move inwards. The fate of
Saturn is more difficult to predict, a priori. However, modern numerical
simulations show that this planet also moves outwards, although only by a few
tenths of AU for reasonable disk's masses \cite{HM}\cite{GomesMigr}.

In \cite{malhotra93}\cite{malhotra95} it was realized that, following
Neptune's migration, the mean-motion resonances with Neptune also migrated
outwards, sweeping the primordial Kuiper belt until they reached their present
positions.  From adiabatic theory \cite{henrard82}, some of the Kuiper belt
objects swept by a mean-motion resonance were captured into resonance; they
subsequently followed the resonance in its migration, while increasing their
eccentricities.  This model accounts for the existence of the large number of
Kuiper belt objects in the 2:3 mean-motion resonance with Neptune (and also in
other resonances) and explains their large eccentricities (see
Fig.~\ref{renu}).  Neptune had to migrate $\sim 7$~AU in order to reproduce
quantitatively the observed range of eccentricities of the resonant bodies.
In \cite{malhotra95}, it was also showed that the bodies captured in the 2:3
resonance can acquire large inclinations, comparable to those of Pluto and
other objects.  The mechanisms that excite the inclination during the capture
process have been investigated in detail in \cite{gomes00}, who concluded
that, although large inclinations can be achieved, the resulting proportion of
high inclination vs. low inclination bodies, as well as their distribution in
the eccentricity vs. inclination plane, do not reproduce well the
observations.  According to \cite{gomes03} (see sect,~\ref{sec.gomes}) most
high inclination Plutinos were captured from the scattered-disk population
during Neptune's migration, rather than from an originally cold Kuiper belt as
in \cite{malhotra95}.

The mechanism of adiabatic capture into resonance requires that Neptune's
migration happened very smoothly. If Neptune had encountered a 
significant number of large bodies (Lunar mass or more), its 
jerky migration 
would have jeopardized capture into resonances.
For instance, direct simulations of Neptune's migration in \cite{HM} 
--which modeled the disk with Lunar to Martian-mass
planetesimals-- did not obtain any permanent capture.
Adiabatic captures into resonance can be seen in numerical simulations only if
the disk is modeled using many more, smaller mass planetesimals 
\cite{gomes03}\cite{GomesMigr}\cite{HM05}.
The constraint set by the capture process on the maximum size 
of the planetesimals comprising the bulk of the mass in the disk  
has been recently estimated in \cite{MC}.

\subsection{Origin of the hot population}
\label{sec.gomes}

An appealing mechanism for the origin of the hot population, also in the framework of the planet migration scenario, has been proposed
in \cite{gomes03}.
Like in \cite{HM}, \cite{gomes03} simulated Neptune's migration by the
interaction with a massive planetesimal disk, extending from beyond Neptune's
initial position.  But, taking advantage of improved computer technology,
10,000 particles were used to simulate the disk population, with individual
masses roughly equal to twice Pluto's mass. For comparison, \cite{HM} used
only 1,000 particles, with Lunar to Martian masses.  Moreover, Neptune was
started at $\sim 15$~AU, instead of 23~AU as in \cite{HM}.

\begin{figure}[t]
\centerline{\psfig{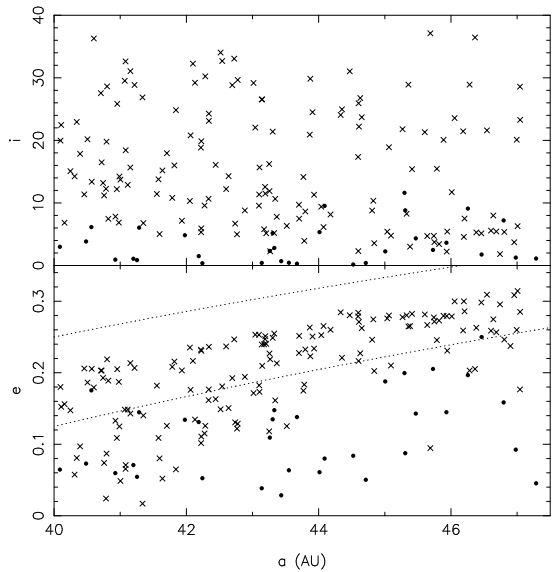}}
%\centerline{\psfig{figure=RIO/fig10tris.ps,height=7cm,bbllx=3cm,bblly=12cm,bbur%x=18cm,bbury=22cm}}
\vspace*{-.3cm}
\caption{The orbital distribution in the classical belt according to the 
  simulations in \cite{gomes03}. The dots denote the local population, which
  is only moderately dynamically excited. The crosses denote the bodies that
  were originally inside 30~AU. Therefore, the resulting Kuiper belt
  population is the superposition of a dynamically cold population and a
  dynamically hot population, which gives a bi-modal inclination distribution
  comparable to that observed.  The dotted curves in the eccentricity vs.
  semi-major axis plot correspond to $q=30$~AU and $q=35$~AU. Courtesy of R.
  Gomes.  } \vspace{0.3cm}
\label{gomes}
\end{figure}

In the simulations of \cite{gomes03}, during its migration Neptune scattered
the planetesimals and formed a massive scattered disk.  Some of the scattered
bodies decoupled from the planet, by decreasing their eccentricities through
the interaction with some secular or mean-motion resonance. If Neptune had not
been migrating, the decoupled phases would have been transient --as often
observed in the integrations of \cite{DL97}. In fact, the dynamics are
reversible, so that the eccentricity would have eventually increased back to
Neptune-crossing values.  But Neptune's migration broke the reversibility, and
some of the decoupled bodies managed to escape from the resonances and
remained permanently trapped in the Kuiper belt.  As shown in
Fig.~\ref{gomes}, the current Kuiper belt would therefore be the result of the
superposition in $(a,e)$-space of these bodies with the local population,
originally formed beyond 30~AU, which stays dynamically cold because they were
only moderately excited (by the resonance sweeping mechanism, as in
Fig.~\ref{renu}).

The migration mechanism is sufficiently slow (several $10^7$~y) that the
scattered particles have the time to acquire very large inclinations,
consistent with the observed hot population.  The resulting inclination
distribution of the bodies in the classical belt is bimodal, because it
results from the superposition of two different populations, each having its
own inclination distribution.  If the number of objects in the cold population
is properly scaled\footnote{The cold population is not depleted in \cite{gomes03},
while only a fraction of a percent of the scattered disk remains trapped in
the hot population. So the former
would outnumber the latter by orders of magnitudes unless some
other mechanism trimmed it down; see sect.~\ref{mass.deficit}.}, the resulting
distribution can quantitatively reproduce the de-biased inclination
distribution computed in \cite{Brown} from the observations.

Assuming that the bodies' color varied
in the primordial disk with heliocentric distance, the scenario proposed in
\cite{gomes03} qualitatively explains
why the scattered objects and hot classical belt objects --which
mostly come from regions inside $\sim$30 AU-- appear to have similar
color distributions, while the cold classical objects --the only
ones that actually formed in the trans-Neptunian region-- have a different
distribution. Similarly, assuming that the maximum size of the objects 
was a decreasing function of the heliocentric distance at which they formed, 
the scenario also explains why the biggest Kuiper belt objects are all in the
hot population. 

The mechanism uncovered in \cite{gomes03} would also have important
implications for two other trans-Neptunian sub-populations: the Plutinos and
the extended scattered disk. In the simulations, some scattered objects also
reached stable Plutino orbits, with orbital properties remarkably similar to
those of the observed objects. Because, on the contrary, the final $(e,i)$
distribution of the Plutinos captured by resonance sweeping from the cold
population is not consistent with observations \cite{gomes00}, this suggests
that the Plutinos have been predominantly captured from the scattered disk.
The fact that the Plutinos have a color distribution similar to that of the
hot population (see \cite{TB03}), without a predominant red component typical
of the cold population, also supports this scenario.

An extended scattered disk is also formed in \cite{gomes03} (see also
\cite{gomes03b}\cite{gomesESD}), beyond 50~AU.  However, orbits similar to
that of Sedna are not achieved in these simulations.  Orbits like that of
2000~CR$_{105}$ are obtained in \cite{gomes03b}, but the resulting population
with $q\sim 45$~AU is skewed towards small semi-major axis, which --as
discussed before-- is probably inconsistent with observations. It is probable
that this large perihelion distance population simulated in \cite{gomes03b}
really exists, but it is very small in number so that none of these objects
has yet been discovered. In this case, 2000~CR$_{105}$ (and Sedna of course)
would be representative of a more conspicuous population with $a>200$~AU,
decoupled from the planets by a stellar encounter during the Oort cloud
formation time \cite{morbCR105}. Conversely, the observed extended
scattered-disk bodies with $q\sim 39$--40~AU and $a\sim$50--100~AU most likely
achieved their current orbits as shown in \cite{gomes03}\cite{gomes03b}.

\subsection{Origin of the outer edge of the Kuiper belt}
\label{edge}

The existence of an outer edge of the Kuiper belt is a very intriguing
property. Several mechanisms for its origin have been proposed, none
of which has raised the general consensus of the community of the experts. 
These mechanisms can be grouped in three classes.

\paragraph{Destroying the distant planetesimal disk.\,}

It has been shown with numerical simulations in \cite{BruniniMelita} that a
Martian mass body residing for 1~Gy on an orbit with $a\sim 60$~AU and $e\sim
0.15$--0.2 could have scattered into Neptune-crossing orbits most of the
Kuiper belt bodies originally in the 50--70~AU range, leaving this region
strongly depleted and dynamically excited.  As shown in Fig.~\ref{limits} the
apparent edge at 50~AU might simply be the inner edge of a similar gap in the
distribution of Kuiper belt bodies.  A problem with this scenario is that
there are no evident dynamical mechanisms that would ensure the later removal
of the massive body from the system. In other words, the massive body should
still be present, somewhere in the $\sim 50-70$~AU region.  A Mars-size body
with 4\% albedo at 70~AU would have apparent magnitude brighter than 20.  In
addition its inclination should be small, both in the scenario where the it
was originally a scattered-disk object whose eccentricity (and inclination)
were damped by dynamical friction (\cite{BruniniMelita}) and in that where the
body reached its required heliocentric distance by migrating through the
primordially massive Kuiper belt (\cite{GomesMigr}). Thus, in view of its
brightness and small inclination, it is unlikely that the putative Mars-size
body could escape detection in the numerous wide field ecliptic surveys that
have been performed up to now, and in particular in that led by Trujillo and
Brown (\cite{TB03}).

A second possibility is that the planetesimal disk was truncated by the
passage of a star in the vicinity of the Sun.  The eccentricities and
inclinations of the planetesimals resulting from a stellar encounter depend
critically on $a/D$, where $a$ is their semi-major axis and $D$ is the
heliocentric distance of the stellar encounter \cite{Ida}\cite{Kobay}.  A
stellar encounter at $\sim 200$~AU would make most of the bodies beyond 50~AU
so eccentric that they intersect the orbit of Neptune, which eventually would
produce the observed edge \cite{Melita}.  An interesting constraint on the
time at which such an encounter occurred is set by the existence of the Oort
cloud. It was shown in \cite{LMD} that the encounter had to occur much earlier
than $\sim 10$~My after the formation of Uranus and Neptune, otherwise most of
the existing Oort cloud would have been ejected to interstellar space.
Moreover, many of the planetesimals at that time in the scattered disk would
have had their perihelion distance lifted beyond Neptune, decoupling them from
the planet.  As a consequence, the extended scattered-disk population, with
$a>50$~AU and $40<q<50$~AU, would have had a mass comparable or larger than
that of the resulting Oort cloud, hardly compatible with the few detections of
extended scattered-disk objects achieved up to now. As discussed in
sect.~\ref{dense}, a close encounter with a star during the first million years
of planetary formation is a possible event if the Sun formed in a stellar
cluster.  However, at such an early time, presumably the Kuiper belt objects
were not yet fully formed \cite{stern96}\cite{KLuu98} (unless they accreted
very rapidly by gravitational instability). In this case the edge of the belt
would be at a heliocentric distance corresponding to a post-encounter
eccentricity excitation of $\sim 0.05$, a threshold value below which
collisional damping is efficient and accretion can recover, and beyond which
the objects rapidly grind down to dust \cite{KB02}.
  
An edge-forming stellar encounter could not be the responsible for the origin
of the peculiar orbit of Sedna, unlike what has been proposed in \cite{KB04a}.
In fact, such a close encounter would also produce a relative overabundance of
bodies with perihelion distance similar to that of Sedna but with semi-major
axis in the 50--200~AU range \cite{morbCR105}. These bodies have never been
discovered, despite of their more favorable observational biases.

\paragraph{Forming a bounded planetesimal disk from an extended gas-dust disk.\,}

In \cite{weidenshilling}, it was suggested that the outer edge of the Kuiper
belt is the result of two facts: {\it i)} accretion takes longer with
increasing heliocentric distance and {\it ii)} small planetesimals drift
inwards due to gas drag.  This leads to a steepening of the radial surface
density gradient of solids.  The edge effect is augmented because, at whatever
distance large bodies can form, they capture the $\sim$m-sized bodies
spiraling inwards from farther out.  The net result of the process, as shown
by numerical modeling in \cite{weidenshilling}, is the production of an
effective edge, where both the surface density of solid matter and the mean
size of planetesimals decrease sharply with distance.

A variant of this scenario has been proposed in \cite{YoudinShu}. In their
model, planetesimals could form by gravitational instability in the regions
where the local ratio solid/gas was 2-10 times that corresponding to cosmic
abundances. According to the authors, this large ratio could be achieved
because of a radial variations of orbital drift speeds of millimeter-sized
particles induced by gas drag.  However, this mechanism would have worked only
within some threshold distance from the Sun, so that the resulting planetesimal
disk would have had a natural edge.

A third possibility is that planetesimals formed only within a limited
heliocentric distance, because of the effect of turbulence. If turbulence in
proto-planetary disks is driven by magneto-rotational instability (MRI), one
can expect that it was particularly strong in the vicinity of the Sun and at
large distances (where solar and stellar radiation could more easily ionize
the gas), while it was weaker in the central, optically thick region of the
nebula, known as the `dead zone' \cite{stone}.  The accretion of planetesimals
should have been inhibited by strong turbulence, because the latter enhanced
the relative velocities of the grains. Consequently, the planetesimals could
have formed only in the dead zone, with well defined outer (and inner)
edge(s).

\paragraph{Truncating the original gas disk.\,}

The detailed observational investigation of star formation regions has
revealed the existence of many {\it proplyds}, i.e. anomalously small
proto-planetary disks. It is believed that these disks were originally much
larger, but in their distant regions the gas was 
photo-evaporated by the very energetic radiation emitted by the massive stars
of the cluster \cite{adams}. Thus, it has been proposed that the outer edge of
the Kuiper belt reflects the size of the original solar system proplyd
\cite{hollenbach}.  

\paragraph{Remark on the location of the Kuiper belt edge.\,}

In all the scenarios discussed above, the location of the edge can be adjusted
by tuning the relevant parameters of the corresponding model. In all cases,
however, Neptune plays no direct role in the edge formation. In this context,
it is particularly important to remark from Fig.~\ref{aeai} that the edge of
the Kuiper belt appears to coincide precisely with the location of the 1:2
mean-motion resonance with Neptune. This strongly suggests that, whatever
mechanism formed the edge, the planet was able to adjust the final location of
the edge through gravitational interactions. I will return to this in  
sect.~\ref{pushout}.

\subsection{The mass deficit of the cold population}
\label{mass.deficit}

The scenario proposed in \cite{gomes03} (see sect.~\ref{sec.gomes}) confines
the problem of the mass depletion of the Kuiper belt to just the cold
population. In fact, in \cite{gomes03} only $\sim$0.2\% of the bodies
initially in the disk swept by Neptune remained in the Kuiper belt on stable
high-$i$ orbits at the end of Neptune's migration. This naturally explains the
current low mass of the hot population. However, the population originally in
the 40-50~AU range --which would constitute the cold population in the
scenario of \cite{gomes03}-- should have been only moderately excited and not
dynamically depleted, so that it should have preserved most of its primordial
mass.

Two general mechanisms have been proposed for the mass depletion: the
dynamical ejection of most of the bodies from the Kuiper belt to the
Neptune-crossing region and the collisional comminution of most of the mass of
the Kuiper belt into dust.

\begin{figure}[t]
\centerline{\psfig{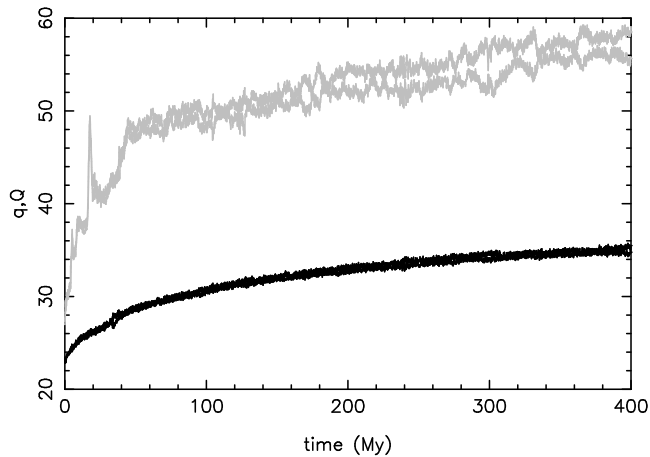}}
%\centerline{\psfig{figure=RIO/fig10tris.ps,height=7cm,bbllx=3cm,bblly=12cm,bbur%x=18cm,bbury=22cm}}
\vspace*{-.3cm}
\caption{A self-consistent simulation of the 
  scenario proposed in \cite{Petit} for the excitation and dynamical depletion
  of the Kuiper belt (from \cite{GomesMigr}).  Neptune is originally placed at
  $\sim 23$~AU and an Earth-mass embryo at $\sim 27$~AU. Both planets are
  embedded in a 30~$M_\oplus$ disk, extending from 10 to 50~AU with a $r^{-1}$
  surface density profile (7.5~$M_\oplus$ between 40 and 50 AU). The black
  curve shows the evolution of Neptune's semi-major axis (its eccentricity
  remains negligible), while the grey curves refer to perihelion and aphelion
  distances of the embryo.  Notice that
  the embryo is never scattered by Neptune, unlike in \cite{Petit}. It
  migrates through the disk faster than Neptune, up to the disk's outer edge.
  Neptune interacts with the entire mass of the disk, thanks to the dynamical
  excitation of the disk due to the presence of the embryo. Therefore, it
  migrates much further that it would if the embryo were not present, and
  reaches a final position well beyond 30~AU (40~AU after 1~Gy).}
\vspace{0.3cm}
\label{m1}
\end{figure}

The dynamical depletion mechanism was proposed in \cite{MV} and later
revisited in \cite{Petit}. In this scenario, a planetary embryo,
with mass comparable to that of Mars or of the Earth, was scattered by
Neptune onto a high-eccentricity orbit that crossed the Kuiper belt for $\sim
10^8$~y.  The repeated passage of the embryo through the Kuiper belt excited
the eccentricities of the Kuiper belt bodies, the vast majority of which
became Neptune crossers and were subsequently dynamically eliminated by the
planets' scattering action. The integrations in \cite{Petit}, however, treated
the Kuiper belt bodies as test particles, and therefore their encounters with
Neptune did not alter the position of the planet. Thus, similar simulations
have been re-run in \cite{GomesMigr}, in the framework of a more
self-consistent model accounting for planetary migration in response to
planetesimal scattering. As expected, the dynamical depletion of the Kuiper
belt greatly enhanced Neptune's migration. The reason for this is that, thanks
to the dynamical excitation of the distant disk provided by the embryo,
Neptune interacted not only with the portion of the disk in its local
neighborhood, but with the entire mass of the disk at the same time. As shown
in Fig.~\ref{m1} even a low mass disk of 30~$M_\oplus$ between 10 and 50~AU
(7.5~$M_\oplus$ only in the Kuiper belt) could drive Neptune well beyond
30~AU.  Halting Neptune's migration at $\sim 30$~AU requires a disk mass of
$\sim 15 M_\oplus$ or less (depending on Neptune's  initial location).
Such a mass and density profile would imply only 3.75~$M_\oplus$ of material
originally in the Kuiper belt between 40 and 50~AU, which is less than the
mass required (10--30~$M_\oplus$) by models of the accretion of Kuiper belt
bodies \cite{SternColwell97a}\cite{KLuu99b}.

A priori, for what concerns Neptune's migration, there is no evident
difference between the case where the Kuiper belt is excited to
Neptune-crossing orbits by a planetary embryo or by some other mechanism, such
as the primordial secular resonance sweeping proposed in \cite{makiko}.
Therefore, we conclude that Neptune never `saw' the missing mass of the Kuiper
belt. The remaining possibility for a dynamical depletion of the Kuiper belt
is that the Kuiper belt objects were kicked directly to hyperbolic or 
Jupiter-crossing orbits and consequently were eliminated without interacting
with Neptune.  Only the passage of a star through the Kuiper belt seems to be
capable of such an extreme excitation \cite{Kobay}.

The collisional grinding scenario was proposed in \cite{SternColwell97b}
\cite{DF97} \cite{DF98} and then pursued in \cite{KLuu99a} \cite{KB02}
\cite{KB04}. In essence, a massive Kuiper belt with large
eccentricities and inclinations would undergo a very intense collisional
activity. Consequently, most of the mass originally in bodies
smaller than 50--100~km in size could be comminuted into dust, and then
evacuated by radiation pressure and Poynting-Robertson drag, causing  
a substantial mass depletion.

To work, the collisional erosion scenario requires that two essential
conditions are fulfilled. First, it requires a peculiar primordial size
distribution, such that all of the missing mass was contained in small, easy
to break, objects, while the number of large object was essentially identical
to that in the current population. Some models support the existence of such a
size distribution at the end of the accretion phase \cite{KLuu98}
\cite{KLuu99b}.  However, the collisional formation of the Pluto--Charon
binary \cite{robin}, the capture of Triton onto a satellite orbit around
Neptune \cite{stern1000} and the discovery of 2003~UB$_{313}$ in the extended
scattered disk \cite{planet}, suggest that the number of big bodies was much
larger in the past, consisting of about 1,000 Pluto-sized objects
\cite{stern1000}.  In principle, it is possible that all of these large bodies
were in the planetesimal disk inside 30~AU, swept by Neptune's migration,
while the primordial Kuiper belt contained only the number of large bodies
inferred from the current discovery statistics, but this would require that
the size distribution in the planetesimal disk had a very sensitive dependence
on heliocentric distance.

The second essential condition for substantial collisional grinding is that
the massive primordial Kuiper belt had a large eccentricity and inclination
excitation, comparable to the current one ($e\sim 0.25$ and/or $i\sim
7^\circ$). However, as reported at the beginning of this section, in light of
\cite{gomes03}, the mass depletion problem concerns only the cold Kuiper belt,
and the dynamical excitation of the cold population is significantly smaller
than that required by the collisional grinding models.

I remark, moreover, that even assuming that the two conditions above are
fulfilled, the collisional grinding models still have problems in reducing the
total mass of the belt down to the current values of a few percent of an Earth
mass.  As the mass decreases, the collisional grinding process progressively
slows down, and eventually stops when the total mass is still about an Earth
mass. The most advanced of the collisional models \cite{KB04} can reduce the
total mass to few $0.01 M_\oplus$ only if a very low specific disruption
energy $Q_*$ is assumed; if more reasonable values of $Q_*$ (similar to those
obtained in hydro-code experiments \cite{BenzAsph}) are adopted, the final
mass achieved by collisional grinding is at least one tenth of the initial
mass, namely about $1~M_\oplus$ or more.

It is very difficult to reach a firm conclusion on the possibility of
collisional grinding of the Kuiper belt from the collisional models alone,
because of the sensitivity of the latter on the assumed parameters.
Perhaps the best strategy is to assume that the collisional grinding was
effective, explore its general consequences and compare them with the
available constraints.  This work is mostly in progress, but I briefly outline
below its preliminary results.   

First, most of the binaries in the cold population would not have survived the
collisional grinding phase \cite{PetitMousis}. In fact,  
the observed Kuiper belt binaries have large separations, so that it can be
easily computed that the impact on the satellite of a 100 times less
massive projectile with a speed of 1km/s would give the satellite
an impulse velocity sufficient to escape to an unbound orbit.  
If the collisional activity was strong enough to cause an effective reduction
of the overall mass of the Kuiper belt, these kind of collisions had to be
extremely common, so that we would not expect a significant fraction of widely
separated binary objects in the current remaining population.

Second, if the conditions favorable for collisional grinding in the Kuiper
belt are assumed for the entire planetesimal disk (5-50 AU), the Oort cloud
would not have formed: the planetesimals would have been destroyed before
being ejected as in \cite{SternW} (Charnoz private communication).

Third, as the Kuiper belt mass decreased during the grinding process, the
precession frequencies of Neptune and the planetesimals had to change.
Consequently, secular resonances had to move, potentially sweeping the belt.
Assuming that, when Neptune reached 30~AU, the disk was already depleted
inside 35~AU but was still massive in the 35--50~AU region, \cite{GomesMigr}
showed that the $\nu_8$ secular resonance would have started sweeping through
the disk as soon as the mass decreased below 10~$M_\oplus$.  The $\nu_8$
resonance sweeping would have excited the eccentricity of the bodies to
Neptune-crossing values and --given the large mass that the Kuiper belt would
have still had when this phenomenon started-- Neptune would have continued its
radial migration well beyond its current location.

\subsection{Pushing out the Kuiper belt}
\label{pushout}

\begin{figure}[t]
\centerline{\psfig{figure=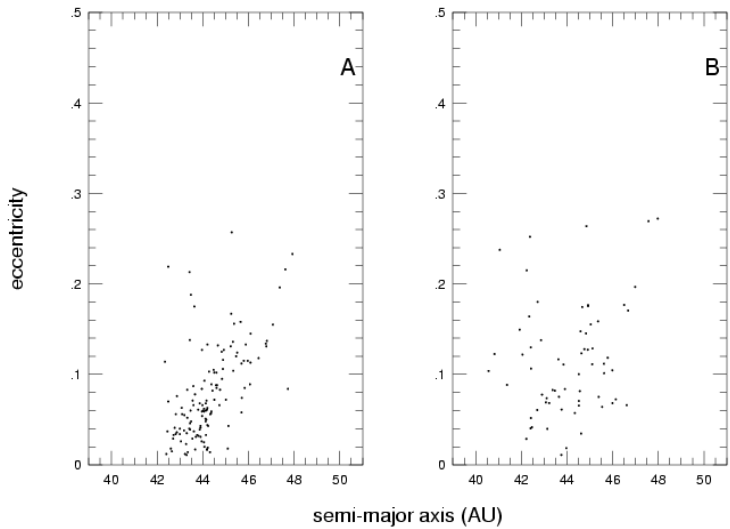,height=8.5cm}}
%\centerline{\psfig{figure=RIO/fig10tris.ps,height=7cm,bbllx=3cm,bblly=12cm,bbur%x=18cm,bbury=22cm}}
\vspace*{-.3cm}
\caption{Left: the observed semi-major axis vs. eccentricity distribution of
  the cold population. Only bodies with multi-opposition orbits and
  $i<4^\circ$ are taken into
  account. Right: the resulting orbital distribution in the scenario proposed
  in \cite{LevM03}. }
\vspace{0.3cm}
\label{LM03}
\end{figure}

Given the difficulties of the collisional grinding scenario for the cold
Kuiper belt, a dynamical way to solve the mass depletion problem has been
proposed in \cite{LevM03}.  In this scenario, the primordial edge of the
massive proto-planetary disk was somewhere around 30--35~AU and the {\it
  entire} Kuiper belt population --not only the hot component as in
\cite{gomes03}-- formed within this limit and was transported to its current
location during Neptune's migration. The transport process for the cold
population had to be different from the one found in \cite{gomes03} for the
hot population (but work in parallel with it), because the inclinations of the
hot population were excited, while those of the cold population were not.

In the framework of the classical migration scenario \cite{malhotra95}
\cite{GomesMigr}, the mechanism proposed in \cite{LevM03} was the following:
the cold population bodies were initially trapped in the 1:2 resonance with
Neptune; then, as they were transported outwards by the resonance, they were
progressively released due to the non-smoothness of the planet migration.
In the standard adiabatic migration scenario \cite{malhotra95} there would be
a resulting correlation between the eccentricity and the semi-major axis of
the released bodies.  However this correlation was broken by a secular
resonance embedded in the 1:2 mean-motion resonance. This secular resonance
was generated because the precession rate of Neptune's orbit was modified by
the torque exerted by the massive proto-planetary disk that drove the
migration.

Simulations of this process can match the observed $(a,e)$ distribution
of the cold population fairly well (see Fig.~\ref{LM03}), while the initially
small inclinations are only very moderately perturbed.  In this scenario, the
small mass of the current cold population is simply due to the fact
that only a small fraction of the massive disk population was
initially trapped in the 1:2 resonance and released on stable non-resonant
orbits.  The preservation of the binary objects would not be a problem because
these objects were moved out of the massive disk in which they formed by a
gentle dynamical process. The final position of Neptune would simply reflect
the primitive truncation of the proto-planetary disk, as in \cite{GomesMigr}. 
Most important, this model explains why the current edge of the Kuiper belt is
at the 1:2 mean-motion resonance with Neptune, despite that
none of the mechanisms proposed for the truncation of the
planetesimal disk involves Neptune in a direct way (see sect.~\ref{edge}). The
location of the edge was modified by the migration of Neptune via its
resonance.  

On the flip side, the model in \cite{LevM03}
re-opens the problem of the origin of different physical properties of the
cold and hot populations, because both would have originated within
35~AU, although in somewhat different parts of the disk.  

I stress, however, that the strength of \cite{LevM03} is in the idea that
pushing out the cold Kuiper belt could solve both the problems related to
mass deficit and edge location. The specific mechanism for pushing out the
cold belt depends on the specific model of giant planet
evolution that is adopted.  The classical planet migration scenario used in
\cite{LevM03} might not reflect the real evolution of the system (see
sect.~\ref{LHB}). In this case, alternative push--out mechanisms should be
investigated. Whatever the preferred mechanism, it will have to give a
predominant role to the 1:2 mean-motion resonance with Neptune in order to
explain the current location of the Kuiper belt edge.

\section{Origin of the Late Heavy Bombardment of the Terrestrial planets}
\label{LHB}
 
The models proposed in the previous sections for the formation of the Oort
cloud and the sculpting of the Kuiper belt seem to offer a quite complete view
of the formation and evolution of the Solar System. But they are not 
entirely satisfactory, because they ignore an important fact in the
history of the Solar System: the late heavy bombardment (LHB) of the
terrestrial planets.
 
Below, I review the observational constraints on the LHB, then I describe 
the models proposed in the past to explain a spike in the bombardment rate
and, finally, I will focus on an emerging view of what happened $\sim 650$~My
after the formation of the planets.  In section~\ref{perspectives} I will
discuss how our understanding of Oort cloud and Kuiper belt formation needs to
be modified in light of the LHB evidence, pointing also to open problems
and prospects for future research.

\paragraph{Evidence for a late cataclysmic bombardment.\,}

The crust of the Moon crystallized around 4.44~Gy ago, and the morphology of
its highlands records a dense concentration of impact craters, excavated prior
to the emplacement --around 3.8~Gy ago-- of the first volcanic flows in the
mare plains \cite{Wilhems1987}. Thus, a period of intense bombardment --the
LHB-- occurred in the first 600-700~My of the Moon's history. However, the
magnitude and the chronology of the collisions between 4.5 and 4~Gy remains a
topic of controversy.

Two explanations have been proposed.  According to \cite{Hartmann1975}
\cite{Wilhems1987}, the frequency of impacts declined
slowly and progressively since the end of the accretion period, up to 3.9~Gy
ago.  In this view, the LBH is not an exceptional event.  Rather it is a
600~My tail of the collisional process that built the terrestrial planets.

Another view advocates a rapid decline in the frequency of impacts after the
formation of the Moon, down to a value comparable to the current one. This was
followed by a cataclysmic period between $\sim 4.0$ and $\sim 3.8$~Gy ago,
marked by an extraordinarily high rate of collisions \cite{Tera1974}
\cite{Ryder1990} \cite{Cohen2000} \cite{Ryder2000} \cite{Ryder2002}.

Today, the majority of authors favors the cataclysmic scenario of the LHB. 
The latter is supported by a series of arguments:

\noindent{\it i)} 600 million years of continual impacts should have left an
obvious trace on the Moon.  So far, no such trace has not been found. The
isotopic dating of both the samples returned by the various Apollo and Luna
missions revealed no impact melt-rock older than 3.92~Gy \cite{Ryder1990}
\cite{Ryder2000}. The lunar meteorites confirm this age limit.  The meteorites
provide a particularly strong argument because they likely originated from
random locations on the Moon \cite{Cohen2000}, unlike the lunar samples
collected directly on the Moon. A complete resetting of all ages all over
the Moon is possible \cite{grinspoon} but highly unlikely, considering the
difficulties of completely resetting isotopic ages at the scale of a full
planet \cite{DeutschScharer1994}.  The U-PB and Rb-Sr isochrones of lunar
highland samples indicate a single metamorphic event at 3.9~Gy
ago and between 3.85 and 4~Gy ago respectively \cite{Tera1974}.  There is no
evidence for a resetting of these isotopic systems by intense collisions
between 4.4 and 3.9~Gy.

\noindent{\it ii)} The old upper crustal lithologies of the Moon do not 
show the expected enrichment in siderophile elements (in particular the
Platinum Group Elements) implied by an extended period of intense collisions
\cite{Ryder2000}. 

\noindent{\it iii)} If the elevated mass accretion documented in the
period around 3.9~Gy is considered to be the tail end of an extended period of
collisions, the whole Moon should have accreted at about 4.1~Gy ago instead of
4.5~Gy \cite{Ryder2002} \cite{Koeberl2004}.  

\noindent{\it iv)} The 15 largest impact structures on the Moon, the so-called
basins, with diameters between 300 and 1200 Km, have been dated to have formed
between 4.0 and 3.9~Gy ago. If the bombardment had declined monotonically
since 4.5~Gy ago, it would be strange that the largest impacts all occurred at
the end of the bombardment period. 

\noindent{\it v)} On Earth, the oxygen isotopic
signature of the oldest known zircons (age: 4.4~Gy) indicates formation
temperatures compatible with the existence of liquid water \cite{Valley2002}.
This argument seems contradictory with an extended period of intense
collisions, which would have brought the Earth's temperature to exceed the
water evaporation threshold. 

Therefore, it can be concluded that there is strong evidence for a cataclysmic
Late Heavy Bombardment event around 3.9~Gy ago.  This cataclysm did not
concern just the Moon, but has now been clearly established throughout the
inner solar system \cite{KringCohen2002}. The exact duration of the cataclysm
is difficult to estimate, though.  Based on the cratering record of the Moon,
it varies between 20 and 200~My, depending on the mass flux estimate used in
the calculation.

\paragraph{Early models of LHB origin.\,}

The occurrence of a cataclysmic LHB challenges our naive view of a Solar
System gradually evolving from chaos to order.
Several ideas have been proposed to explain what could have abruptly changed
the evolution of the system, causing a spike in the bombardment rate.  

The possibility of a stochastic break-up of
an asteroid close to a resonance in the main belt has been investigated in
\cite{zap}. The flux of projectiles inferred from the crater density would
require the break-up of an object larger than Ceres. This event is very
implausible and would have left a huge asteroid family in the main belt, 
of which we see no trace. 

If a stochastic break-up is ruled out, then the remaining possibility is that
a reservoir of small bodies, which remained stable up to the time of the LHB,
suddenly became unstable, with most of its objects achieving planet crossing
orbits.

A comet shower from the Oort cloud, possibly triggered by a stellar encounter,
is a first possibility. However, a new LPC has a probability to collide with
the Earth of about $10^{-9}$. Because the mass hitting the Earth during the
LHB is estimated to be $\sim 10^{-5} M_\oplus$ (\cite{Hartmann}), this would
require an Oort cloud initially containing $10^4$ Earth masses, which --as
discussed in sect.~\ref{Oort}-- is impossible.

In \cite{ChambersLiss} it was proposed that a fifth terrestrial planet, with a
mass comparable to that of Mars, became unstable after $\sim 600$~My of
evolution, and crossed the asteroid belt before being dynamically removed.
Invaded by this new perturber, the asteroid belt became unstable and most of
its objects acquired planet crossing eccentricities. The simulations presented
in   \cite{ChambersLiss} show that a late instability of a 5 terrestrial 
planet system is indeed possible, but it requires that the rogue planet was
initially at about 1.9~AU, with an inclination of $\sim 15^\circ$. Whether
this initial configuration was consistent with terrestrial planets formation
models was not discussed. Similarly, the resulting orbital distribution in the
asteroid belt, after the removal of the rogue planet, was not
investigated. Moreover, in most simulations the rogue planet was removed by a
collision with Mars, whereas the red planet does not show any sign of such a
gigantic strike.  

In \cite{HalLHB} it was proposed that the LHB was associated with a `late
appearance' of Uranus and Neptune in the planetesimal disk. That paper showed
that the planetesimals scattered away from the neighborhoods of the ice giants
would have been sufficient to cause a bombardment on the Moon with a magnitude
comparable to that of the LHB. Moreover, the dynamical removal of these
planetesimals would have caused a radial migration of Jupiter and Saturn,
which in turn would have forced the $\nu_6$ secular resonance to sweep across
the main asteroid belt \cite{gomes_secres}. Their eccentricities being excited
by the resonance passage, most asteroids would have acquired planet crossing
orbits.  Consequently, they would have contributed to --or even dominated-- the
terrestrial planets cratering process. The problem in this work was that the
`late appearance' of Uranus and Neptune was postulated, rather than explained.
The authors argued that these planets might have formed very slowly, although
this seems implausible given that they accreted hydrogen atmospheres of 1--2
Earth masses from the proto-solar nebula \cite{guillot}, which should have
dissipated within $\sim 10$~My (\cite{nebula}).  Later, in
\cite{HalFairy} it was proposed that Uranus and Neptune formed in between
Jupiter and Saturn. The system remained stable for 600~My, until an
instability was produced. Consequently, Uranus and Neptune were scattered
outwards by Jupiter and Saturn; the interaction with the disk eventually
damped the planets' eccentricities and parked them on stable orbits. In this
process, the planetesimal disk was destroyed as in \cite{HalLHB}. The
simulations in \cite{HalFairy} showed that a late instability of a
Jupiter-Uranus-Neptune-Saturn system is indeed possible.  However, the
instability time depends critically on the initial conditions, and it is
unclear if those adopted in the successful simulations could be consistent
with giant planet formation models. More importantly, the scattering of Uranus
and Neptune by Jupiter and Saturn would have destabilized the regular
satellites systems of all the planets. Finally, the massive planetesimal disk
required to stabilize the orbits of Uranus and Neptune would have forced the
latter to migrate well beyond its current position. Thus, as admitted by the
authors themselves, this scenario had to be considered as a `fairy tale'.

\paragraph{The great comet-asteroid alliance: an emerging view on the LHB
  origin.\,} 

Starting from two key considerations:
\begin{itemize}
\item[{\it i)}] planet migration
through the planetesimal disk induces a bombardment of the terrestrial planets
of sufficient magnitude to explain the LHB (from \cite{HalLHB}),
\item[{\it i)}] at the end of the migration phase, the Solar
System is essentially identical to the current one (namely there
are no more reservoirs of planetesimals to destabilize), 
\end{itemize}
it was realized in \cite{Morby-science} that solving the problem of the LHB
origin required to find a plausible mechanism to trigger planet migration at a
late time.

Pursuing this goal, in \cite{gomesLHB} the authors remarked that, in all
previous simulations, planet migration started immediately because
planetesimals were placed close enough to the planets to be violently
unstable.  While this type of initial condition was reasonable for the goals
of those works, it is unlikely.  Planetesimal driven migration is probably not
important for planet dynamics as long as the gaseous massive nebula exists
(the nebula accounts for about 100 times more mass than the planetesimals).
The initial conditions in simulations of the planetesimal driven migration
should therefore represent the system that existed at the time the nebula
dissipated.  Thus, the planetesimal disk should contain only those particles
that had dynamical lifetimes longer than the lifetime of the solar nebula (a
few million years), because the planetesimals initially on orbits with shorter
dynamical lifetimes should have been eliminated earlier, during the nebula
era.  If this constraint on the initial conditions is fulfilled, then the
resulting migration is necessarily slow, because it depends on the rate at
which disk particles evolve onto planet-crossing orbits, which is long by
definition.  If the planetary system, in absence of planetesimals, is stable,
this slow migration can continue for a long time, slightly accelerating or
damping depending on the disk's surface density \cite{GomesMigr}. Conversely,
if the planet system is --or becomes-- unstable, then the planets tend to
increase their orbital separation. The outermost planet penetrates into the
disk and this starts a fast migration, similar to that obtained in previous
simulations, where the planets are embedded in the disk from the very
beginning.  Thus the problem of triggering the LHB is reduced to the problem
of understanding how the giant planets, during their slow migration, could
pass from a stable configuration to an unstable one.

\begin{figure}[t!]
\centerline{\psfig{figure=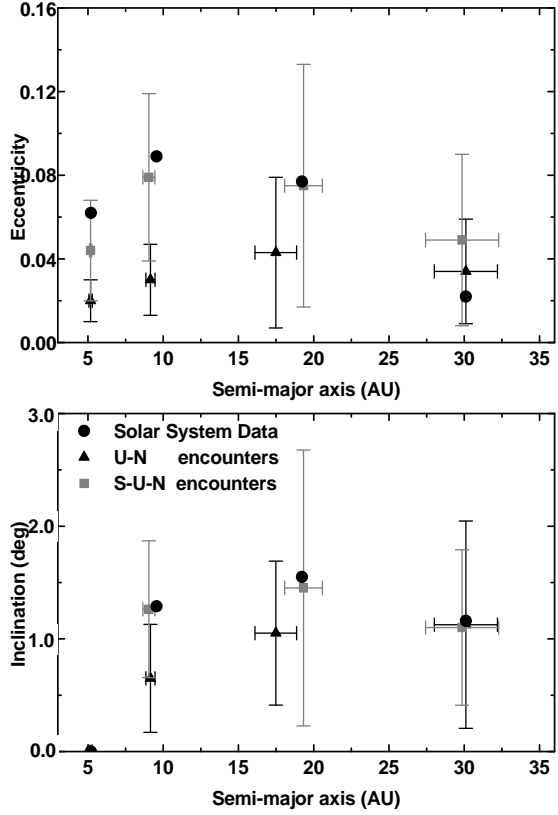,height=11.cm}}
%\centerline{\psfig{figure=RIO/fig10tris.ps,height=7cm,bbllx=3cm,bblly=12cm,bbur%x=18cm,bbury=22cm}}
\vspace*{-.3cm}
\caption{Comparison of the synthetic final planetary 
  systems obtained in \cite{menios} with the real outer Solar System. Top:
  Proper eccentricity vs.  semi-major axis. Bottom: Proper inclination vs.
  semi-major axis.  Here, proper eccentricities and inclinations are defined
  as the maximum values acquired over a 2~My time-span and were computed from
  numerical integrations. The inclinations are measured relative to Jupiter's
  orbital plane.  The values for the real planets are presented with filled
  black dots.  The gray squares mark the mean of the proper values for the
  runs with no planetary encounters involving Saturn, while the black
  triangles mark the same quantities for the runs where at least one ice giant
  encountered the ringed planet (about 15 runs in each case). The error bars
  represent one standard deviation of the measurements.  From \cite{menios}. }
\vspace{0.3cm}
\label{menios_fig}
\end{figure}

A solution of this problem has been proposed in \cite{menios}.  This work
postulated that, at the time of the dissipation of the gas disk, the four giant
planets were in a compact configuration, with quasi-circular, quasi-coplanar
orbits with radii ranging from 5.5 to 13--17~AU. Saturn and Jupiter were close
enough to have a ratio of orbital periods less than 2.  This choice of the
initial conditions for the two giant planets is supported by simulations of
their evolution during the gas-disk phase \cite{MassSnell} \cite{MorbDPS05}.
The assumption of initial small eccentricities and inclination is consistent
with planet formation models.  The small eccentricities ensure the stability
of such a compact planet configuration. In the scenario of \cite{menios},
during their migration in divergent directions, Jupiter and Saturn eventually
crossed their mutual 1:2 mean-motion resonance. This resonance crossing
excited their eccentricities to values comparable to those currently observed
(for eccentricity excitation due to resonance crossing see also
\cite{Chiang2003}). The acquired eccentricities of Jupiter and Saturn
destabilized the planetary system as a whole.  The planetary orbits became
chaotic and started to approach each other.  Thus, a short phase of encounters
followed the resonance-crossing event. Consequently, both ice giants were
scattered outward, deeply into the disk. As discussed above, this abruptly
increased the migration rates of the planets.  During this fast migration
phase, the eccentricities and inclinations of the planets decreased by the
dynamical friction exerted by the planetesimals and the planetary system was
finally stabilized.

With a planetesimal disk of about $35 M_\oplus$, the simulations in
\cite{menios} remarkably reproduced the current architecture of the giant
planets orbits, in terms of semi-major axes, eccentricities and
inclinations. In particular, this happened in the simulations where at least
one of the ice giants encountered Saturn (see Fig.~\ref{menios_fig}).
Conversely, in the simulations where encounters with Saturn never occurred,
Uranus typically ended its evolution on an orbit too close to the Sun and the
final eccentricities and inclinations of all planets were too small.

\begin{figure}[t!]
\centerline{\psfig{figure=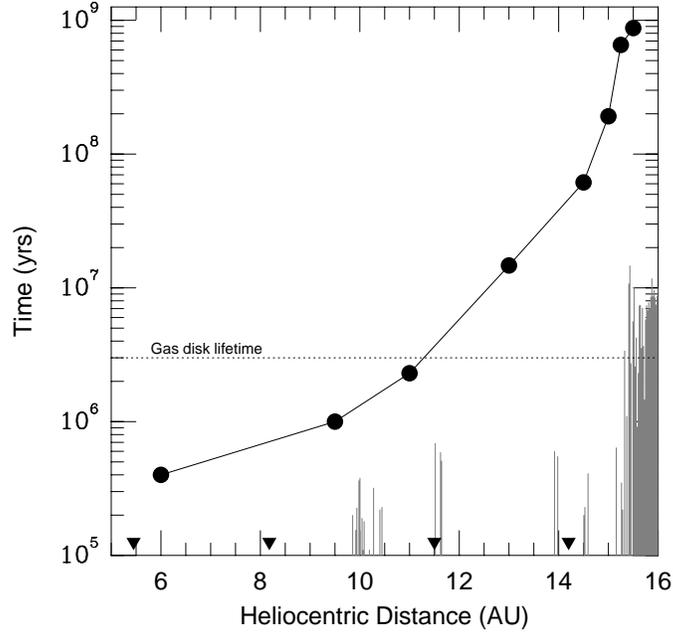,height=10.cm}}
%\centerline{\psfig{figure=RIO/fig10tris.ps,height=7cm,bbllx=3cm,bblly=12cm,bbur%x=18cm,bbury=22cm}}
\vspace*{-.3cm}
\caption{Disk location and LHB timing. The
  histogram reports the average dynamical lifetime of massless test particles
  placed in a planetary system with Jupiter, Saturn and
  the ice giants on nearly-circular, coplanar orbits at $5.45$, $8.18$, 11.5,
  $14.2$~AU, respectively (marked as black triangles on the plot).  The
  dynamical lifetime was computed by placing 10 particles with $e\!=\!i\!=\!0$
  and random mean anomaly at each semi-major axis.  Each vertical bar in the
  plot represents the average lifetime for those 10 particles, after having
  removed stable Trojan cases. The `lifetime' is defined as the time required
  for a particle to encounter a planet within a Hill radius.  A comparison
  between the histogram and the putative lifetime of the gaseous nebula
  \cite{nebula} argues that, when the latter dissipated, the inner edge of the
  planetesimal disk had to be about 1--1.5~AU beyond the outermost ice giant.
  The time at which Jupiter and Saturn crossed their 1:2 mean-motion resonance,
  as a function of the location of the planetesimal disk's inner edge, is
  shown with filled dots.  From \cite{gomesLHB}. } \vspace{0.3cm}
\label{LHB1}
\end{figure}

With this result in hand \cite{gomesLHB} could put all the elements together
in a coherent scenario for the LHB origin. Assuming an initial planetary
system as in \cite{menios}, the planetesimal disk fulfilled the lifetime
constraint discussed above only if its inner edge was located about 1~AU
beyond the position of the last planet. With this kind of disk, the 1:2
resonance crossing event that destabilized the planetary system occurred at a
time ranging from $192$~My to $875$~My (see Fig.~\ref{LHB1}). Modifying the
planetary orbits also led to changes in the resonance crossing time, pushing
it up to 1.1~Gy after the beginning of the simulation. This range of
instability times well brackets the estimated date of the LHB from lunar data.

\begin{figure}[t!]
\centerline{\psfig{figure=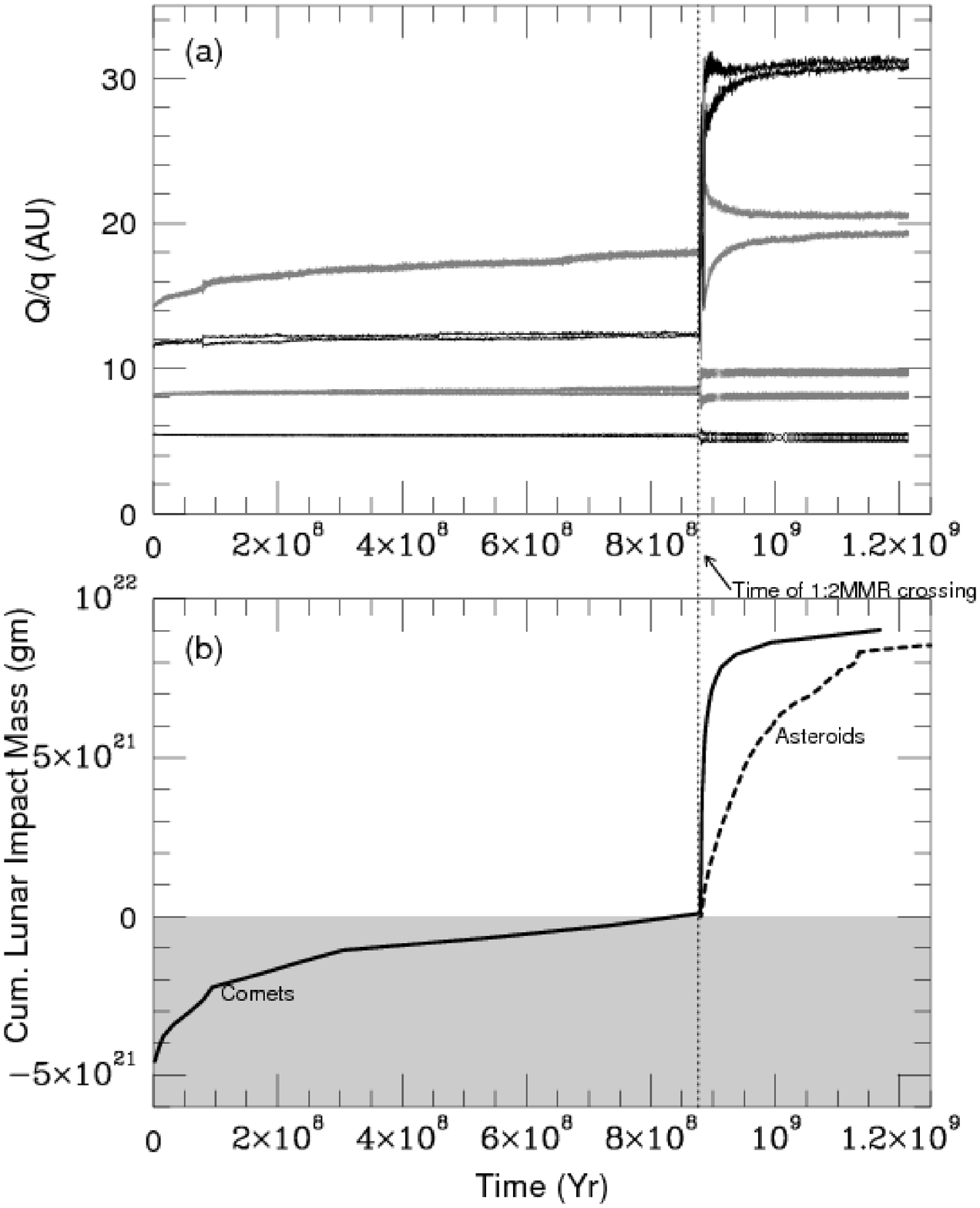,height=12.cm}}
%\centerline{\psfig{figure=RIO/fig10tris.ps,height=7cm,bbllx=3cm,bblly=12cm,bbur%x=18cm,bbury=22cm}}
\vspace*{-.3cm}
\caption{Planetary migration and the associated mass
  flux towards the inner Solar System from a representative simulation of
  \cite{gomesLHB}.  Top: the evolution of the 4 giant planets.  Each planet is
  represented by a pair of curves -- the top and bottom curves are the
  aphelion and perihelion distances, respectively.  Jupiter and Saturn cross
  their 1:2 mean-motion resonance at 880~My.  Bottom: the cumulative mass of
  comets (solid curve) and asteroids (dashed curve) accreted by the Moon.  The
  comet curve is offset so that the value is zero at the time of 1:2 resonance
  crossing.  The estimate of the total asteroidal contribution is very
  uncertain, but should be roughly of the same order of magnitude as the
  cometary contribution, and occur over a longer time-span.  From
  \cite{gomesLHB}. } \vspace{0.3cm}
\label{LHB3}
\end{figure}

The top panel of Fig.~\ref{LHB3} shows the giant planets' evolution in a
representative simulation of \cite{gomesLHB}. Initially, the giant planets
migrated slowly due to the leakage of particles from the disk (Figure~3a).
This phase lasted $875$~My, at which point Jupiter and Saturn crossed their 1:2
resonance.  At the resonance crossing event, as in \cite{menios}, the orbits
of the ice giants became unstable and they were scattered into the disk by
Saturn.  They disrupted the disk and scattered objects all over the Solar
System, including the inner regions. Eventually they stabilized on orbits very
similar to the current ones, at $\sim$20 and $\sim$30 AU respectively. The
solid curve in the bottom panel shows the amount of material that struck the
Moon as a function of time.  As predicted in \cite{HalLHB}, the amount of
material hitting the Moon after resonance crossing is consistent with the mass
($6 \times 10^{21}$ g) estimated from the number and size distribution of
lunar basins that formed around the time of the LHB epoch \cite{Hartmann}.

As discussed in \cite{HalLHB} though, the planetesimals from the distant disk
--which can be identified as `comets'-- were not the only ones to hit the
terrestrial planets. The radial migration of Jupiter and Saturn forced the
secular resonances $\nu_6$ and $\nu_{16}$ to sweep across the asteroid belt
\cite{gomes_secres}, exciting eccentricities and inclinations of asteroids.
The fraction of the main belt population that acquired planet-crossing
eccentricities depends quite crucially on the orbital distribution that the
belt had before the LHB, which is not well known.  The asteroid belt could not
be a massive, dynamical cold disk at the time of the LHB. If it did,
essentially all asteroids would have been ejected onto planet-crossing orbits,
the bombardment of the Moon would have been orders of magnitude more intense
than that recorded by the LHB \cite{HalLHB} and the few asteroids surviving in
the belt after the secular resonance sweeping would have an orbital
distribution inconsistent with that currently observed \cite{gomes_secres}.
Presumably, the asteroid belt underwent a first phase of dynamical depletion
and excitation at the time of terrestrial planet formation
\cite{wetherill_ast} \cite{petit_ast} and then a second dynamical depletion at
the time of the LHB. If, at the end of the first phase, the orbital
distribution in the belt was comparable to the current one, then the secular
resonance sweeping at the time of the LHB would have left $\sim$10\% of the
objects in the asteroid belt \cite{gomesLHB}. Assuming this figure, the
pre-LHB main belt contained roughly $5\times 10^{-3} M_\oplus$ (10 times the
current mass) and the total mass of the asteroids hitting the Moon was
comparable to that of the comets (see Fig.~\ref{LHB3}). However, slight
changes in the pre-LHB asteroid distribution and in the migration rate of
Jupiter and Saturn (also highly variable from simulation to simulation,
depending on the chaotic evolution of Neptune), can change this result for the
asteroidal contribution to the Lunar cratering rate by a factor of several. In
conclusion, the model in \cite{gomesLHB} cannot state whether asteroids or
comets dominated the impact rate on the terrestrial planets. What it can say,
however, is that the asteroidal contribution came later and more slowly than
the cometary contribution (see Fig.~\ref{LHB3}), possibly erasing much of the
signature of the comet bombardment.

The issue of which population dominated the impact rate can be solved by
looking for constraints on the Moon. In \cite{KringCohen2002}, analysis of
Lunar impact melts indicated that at least one of the projectiles that hit the
Moon, and probably more, had a chemistry inconsistent with carbonaceous
chondrites or comets. In \cite{Tagle} it was found that the impact melt at the
landing site of Apollo 17 was caused by a projectile of LL-chondritic
composition.  These results imply that the bombardment was dominated by
asteroids typical of the inner belt.

In \cite{StromNeukum} the comparison of size distributions of the craters
formed at the time of the LHB on Mercury, Mars and the Moon allowed the
calculation of the ratios among the impact velocities on these planets,
leading to the conclusion that most projectiles had a semi-major
axis between 1 and 2 AU. Comets never acquire such a small semi-major axis
during their evolution, so that this argument again favors a predominant
contribution from the inner main belt. More recently, \cite{Strom} found that
the crater size distribution on the lunar highlands is consistent with the
size distribution currently observed in the main belt.

Taken altogether, these results point with little doubt to asteroids being the
dominating (or, possibly, latest-arriving) projectile population for the
terrestrial planets at the time of the LHB. However, they do not imply that
the asteroids {\it triggered} the LHB.  On the contrary, the result in
\cite{Strom} implies that the LHB was triggered by a distant disk
of comets as in \cite{gomesLHB}, for the reasons explained below.

The remarkable match between the size distributions of craters and main belt
asteroids, pointed out in \cite{Strom}, implies that --at the LHB time--
asteroids have been ejected from the main belt onto planet-crossing orbits in
proportions independent of their size\footnote{unlike the current Near Earth
  Asteroids (NEAs) which, escaping from the belt due to non-gravitational
  forces, have a size distribution steeper than that of the main belt
  population}. Only the sweeping of secular resonances can give a
size-independent ejection throughout the main belt. At the time of the LHB,
the gas disk was already totally dissipated. Thus, secular resonance sweeping
could only be caused by the radial displacement of Jupiter and Saturn.  Now,
even assuming that the entire LHB on the terrestrial planets was caused by
asteroids, from the mass hitting the Moon at that time \cite{Hartmann} and the
Moon collision probability typical of NEAs, one can easily compute that the
total asteroid mass on planet crossing orbits was about 0.01~$M_\oplus$. This
mass was too small to cause a significant migration of the giant planets. In
conclusion, a more massive disk --which could only be trans-Neptunian-- had to
trigger and drive planet migration.
Comets mandated the bombardment and asteroids executed it.

\paragraph{Note on Trojans and satellites of the giant planets.\,} 

To validate or reject a model, it is important to look at the largest possible
number of constraints. Two populations immediately come to mind when
considering the LHB scenario proposed in \cite{gomesLHB}: the Trojans and the
satellites. Is their existence consistent with that scenario?

Jupiter has a conspicuous Trojan population. These objects, usually referred
to as `asteroids', follow essentially the same orbit as Ju\-pi\-ter, but lead
or trail that planet by an angular distance of $\sim\!60$ degrees, librating
around the Lagrange triangular equilibrium points. The first Trojan of Neptune
was recently discovered \cite{Chiang}; detection statistics imply that the
Neptune Trojan population could be comparable in number to that of Jupiter,
and possibly even 10 times larger \cite{ChiangLit}.

\begin{figure}[t!]
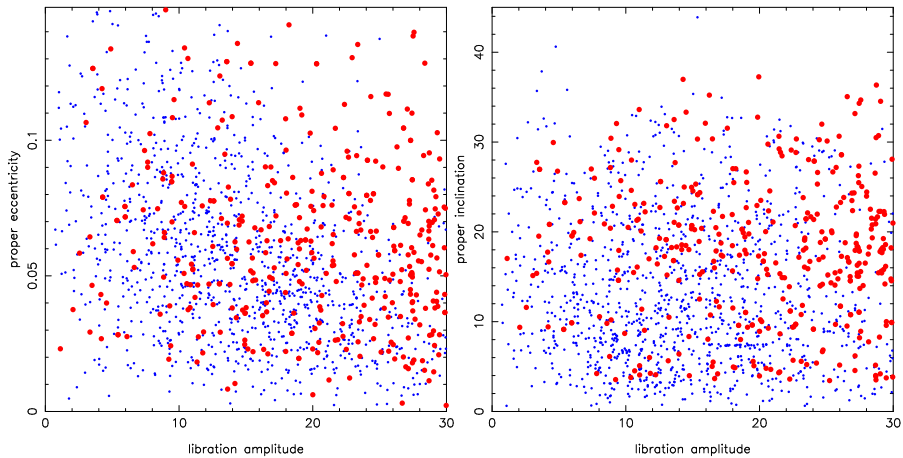

\centerline{\psfig{figure=Troj_De.ps,height=6.cm}\psfig{figure=Troj_Di.ps,height=6.cm}}
%\centerline{\psfig{figure=RIO/fig10tris.ps,height=7cm,bbllx=3cm,bblly=12cm,bbur%x=18cm,bbury=22cm}}
\vspace*{-.3cm}
\caption{Comparison of the orbital distribution of Trojans
  between the simulations in \cite{MorbTroj} and observations.  The simulation
  results are shown as red circles and the observations as blue dots in the
  planes of proper eccentricity vs. libration amplitude (left) and proper
  inclination vs. libration amplitude (right). The distribution of the
  simulated Trojans is somewhat skewed towards large libration amplitudes,
  relative to the observed population. However, this is not a serious problem
  because a fraction of the planetesimals with the largest amplitudes would
  leave the Trojan region during the subsequent 4~Gy of
  evolution\cite{HalTroj}, leading to a better match.  The similarity between
  the two inclination distributions provides strong support for the LHB model
  in \cite{gomesLHB}.}  \vspace{0.3cm}
\label{Trojs}
\end{figure}

The simulations in \cite{menios} \cite{gomesLHB} led to the capture of several
particles on long-lived Neptune Trojan orbits (2 per run, on average, with a
lifetime larger than 80~My). Their eccentricities reached values smaller than
$0.1$.  These particles were eventually removed from the Trojan region, but
this probably is an artifact of the graininess of Neptune's migration in the
simulation, due to the quite large individual mass of the planetesimals
(\cite{HM05}).

Jupiter Trojans are a more subtle issue that is described in detail in
\cite{MorbTroj}. There is a serious argument in the literature against the
idea that Jupiter and Saturn crossed their 1:2 mean-motion resonance: if the
crossing had happened, any pre-existing Jovian Trojans would have become
violently unstable, and Jupiter's co-orbital region would have emptied
\cite{gomes_secres} \cite{Mitch}. However, the dynamical evolution of a
gravitating system of objects is time reversible. Thus, if the original objects
can escape the Trojan region when the latter becomes unstable, other bodies
can enter the same region and be temporarily trapped.  Consequently, a
transient Trojan population can be created if there is an external source of
objects.  In the framework of the scenario in \cite{gomesLHB}, the source
consists of the very bodies that are forcing the planets to migrate, which
must be a very large population given how much the planets must move.  When
Jupiter and Saturn get far enough from the 1:2 resonance, so that the
co-orbital region becomes stable, the population that happens to be there at
that time remains trapped. It becomes the population of permanent Jovian
Trojans still observable today. 

This possibility has been tested with numerical simulations in
\cite{MorbTroj}. Among the particles that were Jupiter or Saturn crossers
during the critical period of Trojan instability, between $2.4\times 10^{-6}$
and $1.8\times 10^{-5}$ remained permanently trapped as Jovian Trojans. 
More importantly, at the end of the simulations, the distribution of the
trapped Trojans in the space of the three fundamental quantities for Trojan
dynamics --the {\it proper} eccentricity, inclination and libration amplitude
\cite{MilaniProTroj}-- was remarkably similar to the current distribution of
the observed Trojans, as illustrated in Fig.~\ref{Trojs}. In particular, this
is the only model proposed thus far which explains the inclination distribution
of Jovian Trojans. The latter was considered to be the hardest problem in the
framework of the classical scenario, according to which the Trojans formed 
locally and were captured at the time of Jupiter's growth \cite{Marzari_rev}.

The capture probabilities reported above allowed \cite{MorbTroj} to conclude
that the total mass of the captured Trojan population was between $\sim
4\times 10^{-6}$ and $\sim 3\times 10^{-5} M_\oplus$. Previous estimates 
from detection statistics \cite{JewittTroj} concluded that
the current mass of the Trojan population is $\sim 10^{-4} M_\oplus$. However,
taking into account modern, more refined knowledge of the Trojans absolute
magnitude distribution (discussed in \cite{MorbTroj}), mean albedo
\cite{FernTroj} and density \cite{marchis}, the estimate of the current mass
of the Trojan population is reduced to $7\times 10^{-6} M_\oplus$, consistent
with the simulations in \cite{MorbTroj}.  The bulk
density of 0.8$^{+0.2}_{-0.1}$g/cm$^3$, measured for the binary Trojan
617~Patroclus \cite{marchis} is a confirmation by itself of the model of
chaotic capture of Trojans from the original trans-Neptunian disk.  In fact,
this density is significantly smaller than any density measured so far in the
asteroid belt, including for the most primitive objects, while it is
essentially identical to the bulk densities inferred for the trans-Neptunian
objects Varuna \cite{JewittVaruna} and 1997~CQ$_{29}$ \cite{Noll}.

In conclusion, the properties of Jovian Trojans are not simply consistent with
the LHB model of \cite{gomesLHB}: they constitute a strong indication --if not
a smoking gun-- in support of the 1:2 mean-motion resonance crossing of
Jupiter and Saturn, which is at the core of the model in \cite{gomesLHB}. 

\vskip 10pt I now briefly come to the satellites of the giant planets. As
discussed above, the non-survivability of the regular satellite systems is one
of the killing arguments against the exotic scenario proposed in
\cite{HalFairy}. Because Saturn, Uranus and Neptune also have encounters with
each other in the model of \cite{menios}\cite{gomesLHB}, it is important to
look at the satellites' fates in this new framework. The issue has been
addressed in \cite{menios}.  The authors recorded all encounters deeper than
one Hill-radius occurring in eight simulations.  Then, they integrated the
evolution of the regular satellite systems of Saturn, Uranus and Neptune
during a re-enactment of these encounters.  They found that, in half of the
simulations, all of the satellite systems survived the entire suite of
encounters with final eccentricities and inclinations smaller than 0.05.  The
difference with respect to the case of \cite{HalFairy} is that, in the latter
model, both ice giants had to have close and strong encounters with Jupiter or
Saturn, whereas in the evolutions of \cite{menios}\cite{gomesLHB} encounters
with Jupiter never occur, and encounters with Saturn are typically distant
ones, with moderate effects.  Thus, the survivability of the regular
satellites is not a problem for the LHB model.  However, the more distant,
irregular satellites would not survive the planetary encounters.  Thus, if the
LHB model is correct they must have been captured at the time of the LHB (see
sect.~\ref{perspectives}).

\section{Building a coherent view of solar system history: perspectives for
  future work.}
\label{perspectives}

From the emerging view of the events that led to the origin of the LHB, it
appears that the evolution of the Solar System was characterized by three main
phases:
\begin{itemize}
\item[i)] {\it the planetary accretion phase}. The giant planets formed in a
  few million years, in a compact orbital configuration embedded in a gas
  disk. The terrestrial planets presumably formed on a timescale of several
  $10^7$~y~\cite{allegre} \cite{Yin} \cite{SasakiAbe}. Planetesimals formed
  out to a threshold distance of $\sim$30--35~AU. The asteroid belt underwent
  a first dynamical depletion and excitation during this phase
  \cite{petit_ast}, while planetesimals in the giant planets region were
  removed, leaving a massive planetesimal disk existing only beyond the orbit
  of the outermost giant planet \cite{gomesLHB}.
\item[ii)] {\it a long quiescent phase} lasting 600~My, during which the
  distant planetesimal disk was gradually eroded at its inner edge by the
  planetary perturbations, leading to a slow migration of the giant planet
  orbits \cite{gomesLHB}. 
\item[iii)] {\it the current phase}, lasting since 3.8~Gy ago, during which
  the Solar System has maintained essentially the same structure \cite{grieve}.
\end{itemize}
The LHB marks the cataclysmic transition between phase ii) and phase iii).

From this template of the Solar System history I will dare to try to 
put in a new context the various scenarios discussed in the previous sections
for the origin of the comet reservoirs, and to suggest new
directions for future research. 

The Oort cloud should have formed in two stages. The first stage occurred as
soon as (or even during the time that) the giant planets formed. This occurred
very early, when the system was still rich in gas, and presumably the Solar
System was still embedded in a stellar cluster. Appropriate simulations should
thus account for a dense galactic environment, close and frequent stellar
encounters as in \cite{fern-brunini}, but using particles on initial
quasi-circular and coplanar orbits in the planetary region and accounting for
gas drag. The decoupling of Sedna and 2000~CR$_{105}$ from the scattering
action of the planets should have occurred in this phase. The second stage
occurred at the LHB time, when the original outer planetesimal disk was
destroyed and a massive scattered disk was formed.  The classical simulations
discussed in sect.~\ref{Oort} are pertinent for this last phase.  The inferred
ratio between the number of comets currently in the Oort cloud and in the
scattered disk (see sect.~\ref{Oort-problems}) argues that the first stage was
more effective than the second one.

The Kuiper belt took shape at the LHB time. As the outer planetesimal disk was
destroyed by the eccentric and migrating ice giants, a fraction of a percent
of the planetesimals managed to be pushed outwards and be implanted in a
region of orbital space that became stable when the planets finally settled on
their current orbits. Thus the principle of the push-out scenario for the
Kuiper belt should remain valid, although the simulations discussed in
sects.~\ref{sec.gomes} and~\ref{pushout} are not really pertinent. In fact,
they assumed a smooth, long-range migration of Neptune, which is not what the
LHB simulations in \cite{gomesLHB} show. Simulations in progress seem to
indicate that the mechanism proposed in \cite{gomes03} for the origin of the
hot population still applies (Gomes, private communication). For the cold
population the mechanism proposed in \cite{LevM03} has to be substituted with
a new one. It turns out that, during the short phase when Neptune is
eccentric, the Kuiper belt is totally unstable up to the 1:2 mean-motion
resonance with Neptune. It can therefore be visited by planetesimals coming
from inside the outer edge of the disk. This builds a sort of steady state
population in the Kuiper belt region, which remains permanently trapped when
Neptune's eccentricity damps by dynamical friction, and the Kuiper belt
becomes stable again (Levison, private communication). This process would
therefore be analogous to that leading to the capture of Jovian Trojans.  If
it the damping of Neptune's eccentricity occurred sufficiently fast, as found
in the LHB model of\cite{gomesLHB} describes, the planetesimals that remained
trapped in the Kuiper belt by this mechanism would not have had enough time to
develop large inclinations, and therefore the population trapped by this
process would produce the cold Kuiper belt. In this scenario, the current size
distribution of the Kuiper belt should be a fossil of that acquired during the
$\sim 600$~My time-span that the objects spent in the massive planetesimal
disk, before being pushed out \cite{ObrienKB}.

The irregular satellites of (at least) Saturn, Uranus and Neptune, if they
existed before the LHB, would have been lost during the phase of encounters
among the planets. Thus, those currently observed had to be captured later,
from the flux of planetesimals coming from the distant disk. At this late
stage, the capture process could not be related to gas drag, nor to a fast
growth of the planetary masses (the so-called pull-down scenario); it is likely
related to three--body interactions (i.e.  interactions between
planetesimals inside the Hill's sphere of a planet), although the exact
mechanism has not been demonstrated yet. This view is consistent with that
proposed in \cite{Jewitt_sats}, from the comparative analysis of the size
distributions of the irregular satellite populations of the 4 giant planets.
Moreover, in this scenario the irregular satellites should have the same
composition as Kuiper belt objects, given that both populations were extracted
from the same primordial planetesimal disk. The recent data collected on the
satellite Phoebe by the Cassini mission argue in this direction
\cite{phoebe}\cite{JohnsonLunine}.

The new LHB scenario also has important implications for aspects of Solar
System formation and primordial evolution not discussed in this chapter.  The
formation of the terrestrial planets should be revisited, accounting for giant
planets on more compact, circular orbits, as required in \cite{gomesLHB}.
Similarly, the evolution of the asteroid belt should also be re-assessed.  As
mentioned before, the belt should have suffered two phases of dynamical
depletion and excitation. The first one during the formation of the
terrestrial planets, and the second one during the LHB. Therefore, during the
600~My period between the end of terrestrial planet accretion and the LHB, the
asteroid belt should have remained about 10--20 times more massive than the
current belt, in a dynamically excited state. The collisional evolution during
this period should have been very important, and the current size distribution
in the main belt should be a fossil of the one that was developed during this
phase.  A study similar to \cite{Bill} should be done, but taking into account
this two-stage evolution of the belt.

Finally, the LHB scenario constrains the orbital architecture of the giant
planets at the end of the gas disk phase. Future simulations of the formation
of these planets and of their interactions with the nebula will have to meet
these constraints. In particular, the compact configuration of the planetary
orbits and the presence of a massive disk of planetesimals outside the orbit
of the outermost planet constrain the maximum range of radial migration that
the giant planets could suffer during the gas phase. For instance, if Jupiter
had formed, say, at 30~AU and migrated down to 5~AU during the gas-disk
lifetime, the outer planetesimal disk required to trigger the LHB would have
been destroyed.  Most probably, the cores of all giant planets formed within
10--15~AU from the Sun \cite{thommes} and, for some reason not yet totally
clear, never migrated substantially.

\newpage
{\bf Acknowledgments:}

I wish to thank N. Thomas and W. Benz for their invitation to present a series
of lectures at the 35th Saas-Fee advanced course, from which this chapter has
been derived. I also thank R. Jedicke and D. Jewitt for their invitation to
re-present the same lectures at the Institute for Astronomy of the University
of Hawaii. I am grateful to all the colleagues with whom I had stimulating
discussions on comet dynamics and the formation of the cometary reservoirs, in
particular L. Dones and H. Levison. I also acknowledge that I re-cycled pieces
of text originally written by M. Brown and Ph. Claeys in papers that we
made together on the Kuiper belt and on the Late Heavy Bombardment. I'm in
debt to all those who read carefully the draft of this manuscript and gave
valuable suggestions for improvements: D. O'Brien, W. Benz. Finally, I wish
to devote this chapter to the memory of Michel Festou. This great French
expert of comets was particularly aware of the importance of dynamics for
understanding the role of these icy objects in the framework of Solar System
formation. My discussions with him greatly motivated me --originally an
asteroid person-- to know more about the primordial evolution of the outer
Solar System.

\end{document}